\documentclass[%
 reprint,
 amsmath,amssymb,
 aps,
floatfix
]{revtex4-2}

\usepackage{color}
\usepackage{graphicx}
\usepackage{dcolumn}
\usepackage{bm}

\def\ket#1{|#1\rangle}

\def\proj#1{|#1\rangle\langle #1|}

\newcommand{\rB}{{\mathrm{B}}}

\newcommand{\rI}{{\mathrm{I}}}

\newcommand{\rP}{{\mathrm{P}}}

\newcommand{\rS}{{\mathrm{S}}}
\newcommand{\rT}{{\mathrm{T}}}

\newcommand{\rf}{{\mathrm{f}}}

\newcommand{\ri}{{\mathrm{i}}}

\newcommand{\LR}{{\mathrm{LR}}}
\newcommand{\ETH}{{\mathrm{ETH}}}
\newcommand{\DE}{{\mathrm{DE}}}
\newcommand{\MC}{{\mathrm{MC}}}
\newcommand{\sh}{{\mathrm{sh}}}
\newcommand{\dG}{{\delta G}}
\newcommand{\can}{{\mathrm{can}}}
\newcommand{\tr}{{\mathrm{tr}}}

\newcommand{\mhyphen}{\mathrm{\mathchar`-}}

\begin{document}

\title{Eigenstate Fluctuation Theorem in the Short and Long Time Regimes}

\author{Eiki Iyoda$^{1}$, Kazuya Kaneko$^{2}$, Takahiro Sagawa$^{2,3}$}
\affiliation{
$^{1}$Department of Physics, Tokai University, 4-1-1 Kitakaname, Hiratsuka-shi, Kanagawa 259-1292, Japan \\
$^{2}$Department of Applied Physics, The University of Tokyo,
7-3-1 Hongo, Bunkyo-ku, Tokyo 113-8656, Japan \\
$^{3}$Quantum-Phase Electronics Center (QPEC), The University of Tokyo, 7-3-1 Hongo, Bunkyo-ku, Tokyo 113-8656, Japan}

\begin{abstract} 
The canonical ensemble plays a crucial role in statistical mechanics in and out of equilibrium.
For example, the standard derivation of the fluctuation theorem relies on the assumption that the initial state of the heat bath is the canonical ensemble.
On the other hand, the recent progress in the foundation of statistical mechanics has revealed that a thermal equilibrium state is not necessarily described by the canonical ensemble but can be a quantum pure state or even a single energy eigenstate, as formulated by the eigenstate thermalization hypothesis (ETH).
Then, a question raised is how these two pictures, the canonical ensemble and a single energy eigenstate as a thermal equilibrium state, are compatible in the fluctuation theorem. 
In this paper, we theoretically and numerically show that the fluctuation theorem holds in both of the long and short-time regimes, even when the initial state of the bath is a single energy eigenstate of a many-body system.
Our proof of the fluctuation theorem in the long-time regime is based on the ETH, while it was previously shown in the short-time regime on the basis of the Lieb-Robinson bound and the ETH [Phys. Rev. Lett. {\bf 119}, 100601 (2017)].
The proofs for these time regimes are theoretically independent and complementary, implying the fluctuation theorem in the entire time domain.
We also perform a systematic numerical simulation of hard-core bosons by exact diagonalization and verify the fluctuation theorem in both of the time regimes by focusing on the finite-size scaling.
Our results contribute to the understanding of the mechanism that the fluctuation theorem emerges from unitary dynamics of quantum many-body systems, and can be tested by experiments with, e.g., ultracold atoms.
\end{abstract}

\maketitle

\section{Introduction}
\label{sec:introduction}

Conventional statistical mechanics relies on the concept of ensembles, such as the microcanonical ensemble and the canonical ensemble. These are characterized as the maximum entropy states under certain energy constraints~\cite{Jaynes1957} and thus given by statistical mixtures of enormous energy eigenstates. 
The canonical ensemble plays a crucial role even beyond equilibrium situations. 
For example, we can prove the second law of thermodynamics and the fluctuation theorem~\cite{Jarzynski1997,Jarzynski2000,Crooks2000,Kurchan2000,Tasaki2000_0009244,Esposito2009,Campisi2011,Sagawa2012,Alhambra2016,Dorner2013,An2014,Xiong2018} out of equilibrium of the system, by relying on the assumption that the initial state of the heat bath is in the canonical ensemble. 
Here, the fluctuation theorem is a universal relation that incorporates the role of fluctuations of the entropy production, represented as $\langle e^{-\sigma}\rangle=1$ where $\sigma$ is the stochastic entropy production.
The fluctuation theorem implies the second law at the average level, stating that the average entropy production is nonnegative:
\begin{align}
\label{eq:2ndlaw}
    \langle \sigma \rangle
    :=
   \Delta S_\rS-\beta Q\geq 0,
\end{align}
where $\Delta S_\rS$ is the change of the von Neumann entropy of the system, $\beta$ is the inverse temperature of the heat bath, and $Q$ is the average heat emitted from the bath.
We note that the fluctuation theorem reproduces the second law through the Jensen inequality $\langle e^{-\sigma}\rangle\geq e^{-\langle \sigma\rangle}$.

On the other hand, in recent years, it has been established that most states, not limited to those ensembles (maximum entropy states), can represent thermal equilibrium states~\cite{Polkovnikov2011,Eisert2015,Gogolin2016,DAlessio2016,Mori2018,Trotzky2012,Clos2016,Kaufman2016,Neill2016,Gross2017,Parsons2015,Mello2019,Edabi2020}. 
The extreme case opposite to the canonical ensemble 
is zero entropy states such as a single energy eigenstate; The eigenstate thermalization hypothesis (ETH)~\cite{Deutsch1991,Srednicki1994,Rigol2008,Biroli2010,Steingeweg2013,Kim2014,Beugeling2014,Beugeling2015,Fratus2016,Mondaini2016,Mondaini2017,Yoshizawa2018,Garrison2018,Dymarsky2018,Khaymovich2019,Brenes2020,Kaneko2020} states that a single energy eigenstate can represent a thermal equilibrium state. 
The ETH is a sufficient condition for thermalization in isolated quantum many-body systems~\cite{Deutsch1991,Srednicki1994,Rigol2008} and numerically shown to be valid in various non-integrable quantum many-body systems~\cite{Rigol2008,Biroli2010,Steingeweg2013,Kim2014,Beugeling2014,Beugeling2015,Fratus2016,Mondaini2016,Mondaini2017,Yoshizawa2018,Garrison2018,Dymarsky2018,Khaymovich2019,Brenes2020,Kaneko2020}.
Furthermore, in recent experiments such as cold atoms~\cite{Trotzky2012,Kaufman2016,Gross2017,Parsons2015,Mello2019,Edabi2020}, trapped ions~\cite{Clos2016}, and superconducting qubits~\cite{Neill2016}, it has been observed that thermal equilibrium states are not necessarily canonical ensembles but can be pure states.

A critical question here is how universally the thermodynamic properties appear without canonical ensembles, even in non-equilibrium situations. 
In particular, it is interesting to investigate the validity of the fluctuation theorem: the question focuses on whether it is possible to understand how the fluctuation theorem and the second law emerge from quantum mechanics without assuming the conventional statistical ensembles. 

A partial understanding of the above question has been addressed in recent papers. 
Specifically, Ref.~\cite{Iyoda2017} theoretically showed the fluctuation theorem (and the second law) in the short-time regime, which is based on the ETH and the Lieb-Robinson bound~\cite{Lieb1972,Hastings2006,Haah2018}.
We note that in the numerical simulation of Ref.~\cite{Iyoda2017}, the initial state of the system was chosen to be a pure state, and the fluctuation theorem was apparently broken in the long-time regime.
This apparent breakdown originates from so-called absolute irreversibility~\cite{Murashita2014,Funo2015} induced by the pure initial state, and as will be shown in this paper, the fluctuation theorem still holds in the long-time regime if absolute irreversibility is properly treated.
Moreover, Ref.~\cite{Kaneko2017} theoretically showed the second law at the average level in the long-time regime on the basis of the ETH.
Also in previous papers~\cite{Jin2016,Schmidtke2018}, it has been numerically suggested that the fluctuation theorem holds even in the long-time regime.
Given these researches, it is desirable to make a comprehensive understanding of the validity of the fluctuation theorem in the entire time domain.

The main result of this paper is to show, analytically and numerically, that the fluctuation theorem holds with the non-canonical bath in both of the short and long-time regimes.
Specifically, the initial state of the bath is supposed to be a single energy eigenstate sampled from an energy shell. 
This establishes that the universal property of non-equilibrium fluctuations in entropy production emerges even without the canonical ensemble. 
In particular, since we prove the fluctuation theorem for the extreme case with a single energy eigenstate, the fluctuation theorem holds when the initial state of the bath is any mixture of energy eigenstates in the energy shell.

We discuss the entire time domain by considering the long and short-time regimes. 
We theoretically show that the long-time average of the deviation from the fluctuation theorem vanishes in the thermodynamic limit of the heat bath.
To confirm the theory, we perform numerical simulations and show that the deviation decreases with the bath size $N$. 
On the other hand, Ref.~\cite{Iyoda2017} theoretically showed the fluctuation theorem for the short-time regime in the thermodynamic limit on the basis of the Lieb-Robinson bound and the ETH.
While the theory of Ref.~\cite{Iyoda2017} is valid in the thermodynamic limit, the fluctuation theorem in the short-time regime  for the numerically-accessible system size was not fully established in Ref.~\cite{Iyoda2017} because of the large finite-size effect.
In the present paper, we perform systematic numerical calculations to clarify whether the theoretical scenario is indeed relevant for the numerically-accessible system size.
In particular, we numerically investigate the bath-size dependence of the error term of the fluctuation theorem, which eventually excludes other scenarios than the theoretical scenario of Ref.~\cite{Iyoda2017}.

The rest of this paper is organized as follows. 
In Sec.~\ref{sec:setup}, we introduce the setup of this study.
In Sec.~\ref{sec:SOR}, we overview the main results of this paper without going into details.
In Sec.~\ref{sec:LT}, we theoretically derive the fluctuation theorem in the long-time regime and show our numerical results to support the theory.
In Sec.~\ref{sec:ET}, we discuss the fluctuation theorem in the short-time regime and show the corresponding numerical results.
In Sec.~\ref{sec:Discussion}, we summarize the results and make some remarks.
In Appendix~\ref{sec:app:AI}, we introduce the concept of absolute irreversibility~\cite{Murashita2014,Funo2015}.
In Appendix~\ref{sec:app:error}, we examine the details of the interaction-induced error of the fluctuation theorem.
In Appendix~\ref{sec:app:RWA}, we show that the interaction-induced error vanishes under the rotating wave approximation.
In Appendix~\ref{sec:app:AnotherAttempt}, we discuss another naive approach to show the fluctuation theorem.
In Appendix~\ref{sec:app:InitialRise}, we discuss the initial time dependence of the errors of the fluctuation theorem.
In Appendix~\ref{sec:app:proof}, we explain the details of the proof of the fluctuation theorem in the long-time regime.
In Appendix~\ref{sec:app:interaction_perturb}, we show the additional calculations for the interaction-induced error.
In Appendix~\ref{sec:app:numerical}, we show the supplementary numerical results.

\section{Setup}
\label{sec:setup}

In this section, we introduce the setup of the study.
In Sec.~\ref{sec:setup:FT:can}, we explain the conventional setup for the fluctuation theorem with the canonical bath and briefly overview the fluctuation theorem~\cite{Jarzynski1997,Jarzynski2000,Crooks2000,Kurchan2000,Tasaki2000_0009244,Esposito2009,Campisi2011,Sagawa2012,Alhambra2016,Dorner2013,An2014,Xiong2018}.
In Sec.~\ref{sec:setup:FT:ee}, we explain the setup of the present study with the energy eigenstate bath.
In Sec.~\ref{sec:setup:Hamiltonian}, we explain the setup of our numerical simulation.

\subsection{Fluctuation theorem for the canonical bath}
\label{sec:setup:FT:can}

As a preliminary, we consider the conventional setup of the fluctuation theorem.
The total system is composed of the system S and the heat bath B.
The Hamiltonian of the total system is written as 
\begin{align}
H:=H_\rS+H_\rI+H_\rB,
\end{align}
where $H_\rS$, $H_\rB$ are the Hamiltonian of system S and bath B respectively, and $H_\rI (\neq 0)$ describes the interaction between S and B. 
The initial state of the total system SB is assumed to be a product state:
\begin{align}
\label{eq:initialstate}
\rho(0):=\rho_\rS(0)\otimes\rho_\rB(0).
\end{align}
We also define the canonical ensembles of system S and bath B as $\rho_\rS^\can:=e^{-\beta H_\rS}/Z_\rS$ with $Z_\rS:=\tr_\rS[e^{-\beta H_\rS}]$ and $\rho_\rB^\can:=e^{-\beta H_\rB}/Z_\rB$ with $Z_\rB:=\tr_\rB[e^{-\beta H_\rB}]$ respectively.

The time evolution of the total system is given by $ \rho(t)=U\rho_{\rS}(0)\otimes \rho_{\rB}(0)U^\dag $, where $U(t):=e^{-iHt/\hbar}$ is the unitary time evolution operator. 
We write the reduced density operator of system S and bath B at time $t$ as $ \rho_{\rS}(t):=\tr_\rB[\rho(t)], \rho_{\rB}(t):=\tr_\rS[\rho(t)] $, respectively.

In this setup, we explain the concept of the stochastic entropy production $\sigma$~\cite{Esposito2009}.
To introduce it,
let us consider an operator $\sigma (t) :=-\ln\rho_\rS(t)+\beta H_\rB$ (see \cite{Sagawa2012,Manzano2018,Landi2021} for the same approach to the stochastic entropy production). 
The first term on the right-hand side is the informational contribution, whose average is the von Neumann entropy of the system $S_\rS(t):=-\tr_\rS[\rho_\rS(t)\ln\rho_\rS(t)]$.
The second term of $\sigma(t)$ is the thermal contribution that gives the heat term.

While $\sigma(t)$ is not an ordinary physical observable because it depends on the state $\rho_\rS(t)$, 
the above definition of $\sigma(t)$ enables us to treat various physically-relevant situations in a unified manner. One of typical situations is that the initial state of system S is the canonical ensemble, in which the measurement of $\sigma(t)$ is equivalent to the measurement of energy, because $-\ln\rho_\rS^\can=\beta H_\rS+\ln Z_\rS$.
If we do not have any prior knowledge about the state of the system, we can measure $- \ln \rho_\rS(t)$ by performing quantum state tomography of $\rho_S(t)$ beforehand~\cite{Vogel1989}. In fact, by quantum state tomography, we obtain the information about the eigenbasis of $\rho_S(t)$ and its eigenvalues. Then, we can perform the projection measurement of the eigenbasis and associate an eigenvalue with the outcome.

We consider the projection measurement of $\sigma (t)$ at time $0$ and $t$. 
Let the measurement outcomes be
$\sigma_\ri$ at $t=0$ and $\sigma_\rf$ at $t$. 
We then define the stochastic entropy production $\sigma$ as $\sigma:=\sigma_\rf-\sigma_\ri$, whose average $\langle\sigma\rangle=\Delta S_\rS-\beta Q$ is the average entropy production of the total system. 
Here, $Q$ is the heat emitted from bath B to system S, defined as $Q:=-\tr_\rB[H_\rB(\rho_\rB(t)-\rho_\rB(0))].$

For the conventional fluctuation theorem, the initial state of bath B is assumed to be the canonical ensemble $\rho_\rB(0)=\rho_\rB^\can$.
The fluctuation theorem states that
\begin{align}
\langle e^{-\sigma}\rangle=1.
\label{eq:IFT}
\end{align}
It is straightforward to show~\cite{Esposito2009} that $\langle e^{-\sigma}\rangle$ can be rewritten as
\begin{align}
\langle e^{-\sigma}\rangle
=
\mathrm{tr}[
e^{-\beta H_\mathrm{B}}
U
e^{\beta H_\mathrm{B}}
\rho_\mathrm{B}^\mathrm{can}
U^\dag
\rho^\mathrm{can}_\mathrm{S}(t)
].
\label{eq:IFT:rep}
\end{align}
Substituting $\rho_\mathrm{B}^\mathrm{can}=e^{-\beta H_\rB}/Z_\rB$, we obtain
\begin{align}
\langle e^{-\sigma}\rangle
&=
\mathrm{tr}\left[
e^{-\beta H_\mathrm{B}}
U
e^{\beta H_\mathrm{B}}
\frac{e^{-\beta H_\rB}}{Z_\rB}
U^\dag
\rho^\mathrm{can}_\mathrm{S}(t)
\right]
\\
&=
\mathrm{tr}\left[
\frac{
e^{-\beta H_\mathrm{B}}
}
{Z_\rB}
U
U^\dag
\rho^\mathrm{can}_\mathrm{S}(t)
\right]
\\
&=
\mathrm{tr}\left[
\rho^\mathrm{can}_\mathrm{S}(t)\otimes \rho_\rB^\can
\right]
\\
&=
1.
\end{align}

We note that if absolute irreversibility~\cite{Murashita2014,Funo2015} occurs, the fluctuation theorem is modified (see details in Appendix~\ref{sec:app:AI}).
Absolute irreversibility is an apparent violation of the fluctuation theorem that occurs when $\rho_S(0)$ does not have the full rank (e.g., $\rho_S(0)$ is a pure state).
The modification term appears in the form of
\begin{align}
    \label{eq:IFT:AI}
    \langle e^{-\sigma}\rangle = 1-\lambda(t).
\end{align}
See Appendix~\ref{sec:app:AI} for details. 
Absolute irreversibility can be understood as a kind of singularity about the initial state of system S. We emphasize that this is not a purely mathematical problem, but can be observed experimentally~\cite{Masuyama2018}.

\subsection{Fluctuation theorem for the energy eigenstate bath}
\label{sec:setup:FT:ee}

\begin{figure}[t]
\begin{center}
\includegraphics[width=0.71\linewidth]{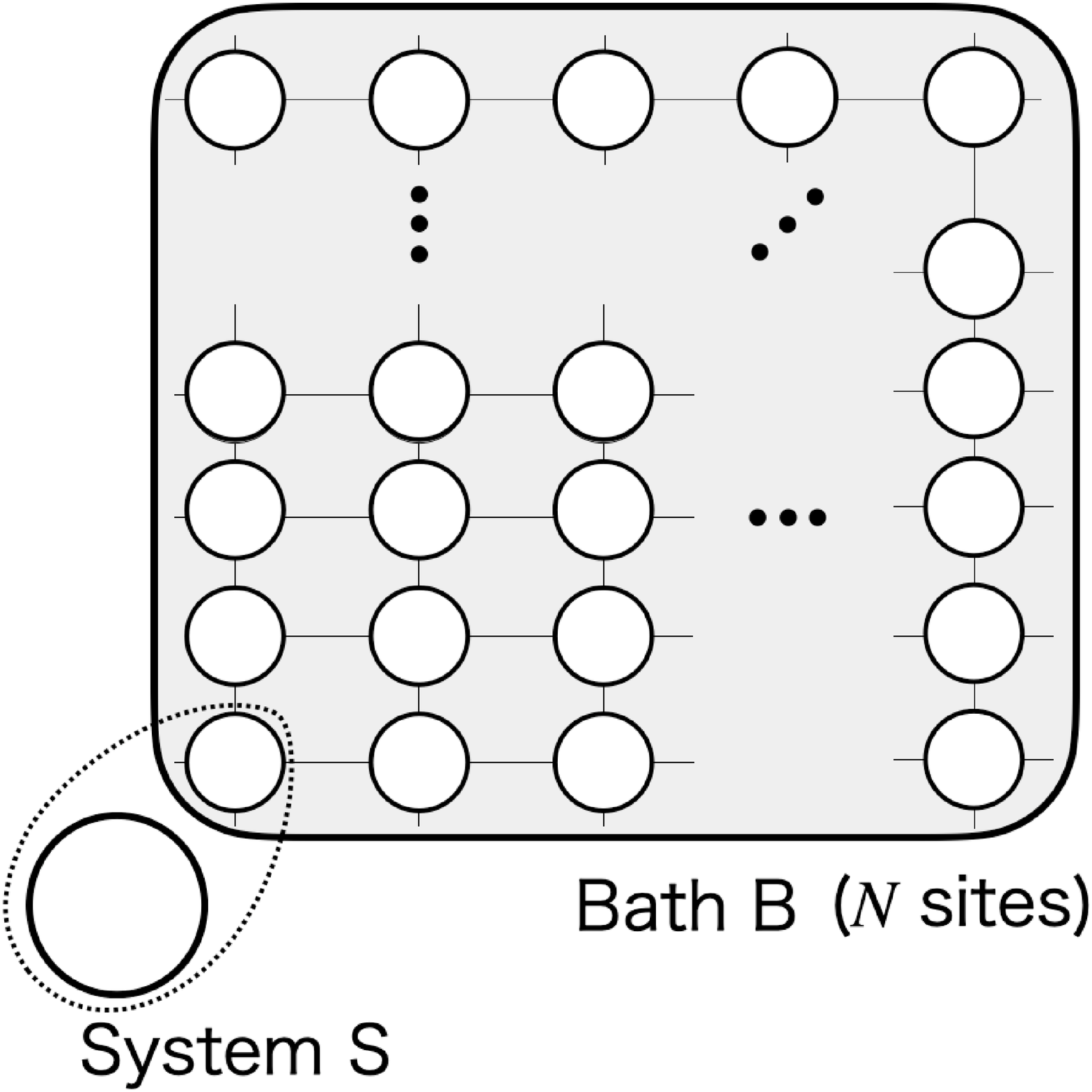}
\end{center}
\caption{The setup of our study. The total system consists of the system S and the heat bath B. The system S interacts with a local (bounded) region of B, and the interaction in B is also local.
The bath B is defined on a $d$-dimensional lattice.
We denote the number of sites of B by $N$ and the dimension of the Hilbert space of B by $D_\rB$.  We also consider an energy shell of B, whose Hilbert-space dimension is $D$.
See the inset of Fig.~\ref{fig:G_tdep} for the specific lattice used in our numerical simulation.
}
\label{fig:system}
\end{figure}
In this subsection, we introduce the setup of the fluctuation theorem for the energy eigenstate bath, which will be shown as the main result of the present paper.

The total system is composed of system S and bath B in the same manner as the setup for the conventional fluctuation theorem, while in the present setup, bath B is defined on a lattice as shown in Fig.~\ref{fig:system}.
We assume that the Hamiltonian of bath B is local and system S interacts with a local region of B by the interaction $H_\rI$. 
Let $N$ be the number of sites in bath B, and $d=1,2,3,\cdots$ be the spatial dimension of it. 
The size of system S and the support of $H_\rI$ are fixed and do not depend on $N$.
We refer to the dimensions of the Hilbert space of S, B, and the total system SB as $D_\rS$, $D_\rB$, and $D_{\rS\rB}$, respectively.
We denote an eigenstate of $H$ with eigenenergy $E_a$ as $|E_a\rangle$.
In the same manner, we denote eigenstates of $H_\rS,H_\rB$ as $|E_i^\rS\rangle,|E_\alpha^\rB\rangle$, respectively. 
We denote a matrix element of an operator with respect to the eigenstates of $H$ as $(\cdots)_{ab}:=\langle E_a|\cdots|E_b\rangle$. 
We also define an operator of system S as $q_\rS^{ij}:=|E_i^\rS\rangle\langle E_j^\rS|$.

In the present paper, we consider the thermodynamic limit that means the large-bath limit~($N\rightarrow\infty$)  without changing system S. 
To describe asymptotic behaviors of this limit, we use the following asymptotic notations:
\begin{align}
    &
    f(D)=\Theta(g(D))
    \Leftrightarrow
    0<
    \lim_{D\rightarrow \infty}
    \left|
    \frac{f(D)}{g(D)}
    \right|
    <\infty,
    \\
    &
    f(D)=\mathcal{O}(g(D))
    \Leftrightarrow
    \lim_{D\rightarrow \infty}
    \left|
    \frac{f(D)}{g(D)}
    \right|
    <\infty,
    \\
    &
    f(D)=o(g(D))
    \Leftrightarrow
    \lim_{D\rightarrow \infty}
    \left|
    \frac{f(D)}{g(D)}
    \right|
    =0.
\end{align}
We define the operator norm $\|X\|$ as the largest singular value of $X$ and the trace norm $\|X\|_1$ as the sum of the singular values of $X$.
For the sake of simplicity, we use $A\simeq B$ for operators $A$ and $B$ to represent that $\|A-B\|=o(1)$ or $\|A-B\|_1=o(1)$ depending on the context.

The initial state of SB is assumed to be a product state as in (\ref{eq:initialstate}).
We also assume that the initial state of S is diagonal with respect to $H_\rS$, because it simplifies the discussion of the time dependence in Secs.~\ref{sec:SOR} and \ref{sec:ET}.

Now, a crucial assumption is that the initial state of bath B is a single energy eigenstate, which is sampled from the energy shell $[E-\Delta E, E]$.
The energy $E$ is given by $E:=\tr_\rB[\rho_\rB^\can H_\rB]$ with the canonical ensemble at inverse temperature $\beta$.
The width $\Delta E$ can be chosen as
$\Theta(1)\leq\Delta E\leq\Theta(N)$, which gives the normal thermodynamic scaling $D=\exp(s N)$ with the entropy density $s$~\cite{RuelleTextBook}.

\begin{table*}[t]
\caption{The assumptions regarding the ETH.}
 \centering
  \begin{tabular}{|c||l|l|}
  \hline
    Time regime (section) & Long-time regime  (Sec.~\ref{sec:LT}) & Short-time regime (Sec.~\ref{sec:ET}) \\
    \hline
    Which ETH &  The diagonal and off-diagonal ETH of $H$ &  The diagonal ETH of $H_\rB$ \\
    \hline
    For which eigenstates & All the eigenstates 
&    Only for the given initial state of B \\
\hline
  \end{tabular}
\label{table:ETH}
\end{table*}

We next summarize the assumptions regarding the ETH, as summarized in Table~\ref{table:ETH}.
For the proof of the fluctuation theorem in the long-time regime~(Sec.~\ref{sec:LT}), we assume that $H$ satisfies the ETH  for all the eigenstates in the energy shell (the ``strong'' ETH) with respect to all observables of system S and $H_\rI$.
Here, we can introduce the energy shell of $H$, which is determined by the energy and energy width of the initial state~(\ref{eq:initialstate})~(see details in Appendix~\ref{sec:app:proof}).
Specifically, we assume the (strong) diagonal and off-diagonal ETH~\cite{Deutsch1991,Srednicki1994,Rigol2008,Mori2018} for all observables of system S and $H_\rI$, written as $O_{\rS\rI}$, stating that the following relations hold for all the eigenstates in the energy shell:
\begin{align}
&
\label{eq:diag_ETH_Setup}
\langle E_a|O_{\rS\rI} | E_a\rangle 
=
\langle O_{\rS\rI}\rangle_\MC
+
\mathcal{O}({D^\prime}^{-1/2}),
\\
&
\label{eq:offdiag_ETH_Setup}
|\langle E_a|O_{\rS\rI}|E_b\rangle|
=
\mathcal{O}({D^\prime}^{-1/2})
~~~~~(a\neq b),
\end{align}
where
$\langle \cdot\rangle_\MC$ is the microcanonical average of the energy shell of $H$.
Since the size of system S is $\Theta(1)$, the dimension of the energy shell $D^\prime$~$(<D_{\rS\rB})$ satisfies $D^\prime=\Theta(D)$.

In Sec.~\ref{sec:ET}, we discuss the fluctuation theorem in the short-time regime, which has been theoretically proved in Ref.~\cite{Iyoda2017}.
Here, we assume the diagonal ETH of $H_\rB$:
\begin{align}
&
\label{eq:diag_ETH_Setup:HB}
\langle E_a^\rB|L_{\rB_0} | E_a^\rB\rangle 
=
\langle L_{\rB_0}\rangle_{\MC,\rB}
+
\mathcal{O}(D^{-1/2}),
\end{align}
where
$\langle \cdot\rangle_{\MC,\rB}$ is the microcanonical average of the energy shell $[E-\Delta E,E]$, $D~(<D_\rB)$ is the dimension of the Hilbert space of the energy shell of $H_\rB$, and $L_{\rB_0}$ is a quasi-local operator defined in Sec.~\ref{sec:ET}. 
We note that Eq.~(\ref{eq:diag_ETH_Setup:HB}) is assumed only for the {\textit{given}} initial state $|E_a^\rB\rangle$ of B.

We remark on  the validity of the ETH. 
It has been numerically shown that the strong ETH holds in various non-integrable quantum many-body systems~\cite{Rigol2008,Biroli2010,Steingeweg2013,Kim2014,Beugeling2014,Beugeling2015,Fratus2016,Mondaini2016,Mondaini2017,Yoshizawa2018,Garrison2018,Dymarsky2018,Khaymovich2019,Brenes2020,Kaneko2020}, while it does not hold in the presence of quantum many-body scar~\cite{Turner2018} and in integrable systems~\cite{Kim2014}. On the other hand, the ETH for a given single energy eigenstate can be valid even in integrable systems~\cite{Biroli2010,Iyoda2017}.
Specifically, it has been shown that almost all energy eigenstates in the energy shell satisfy the ETH in integrable systems~\cite{Iyoda2017,Ogata2010,Mori2016} and quantum many-body scars~\cite{Turner2018}.
However, even the ETH for almost all energy eigenstates does not hold in many-body localized (MBL) systems~\cite{Pal2010,Nandkishore2015}. 

Meanwhile, in the following sections, we sometimes consider the following condition for the interaction Hamiltonian:
\begin{align}
\label{eq:com_int}
[H_\rS+H_\rB,H_\rI]=0,
\end{align}
which simplifies the derivation of the fluctuation theorem.
Under this condition, the sum of the energies of system S and bath B does not change, and we can rewrite the heat $Q$ by the energy change in system S.
We note that the interaction itself (i.e., $\|H_\rI\|$) is not necessarily small for the condition (\ref{eq:com_int}) to hold. 
While the condition~(\ref{eq:com_int}) is satisfied in simple models as the Jaynes-Cummings model at the resonant condition~\cite{BreuerTextBook}, 
it is not necessarily satisfied in generic quantum many-body systems.

When the condition (\ref{eq:com_int}) is not satisfied, an error term induced by the interaction can appear in the fluctuation theorem.
It is noteworthy, however, that Eq.~(\ref{eq:com_int})
is approximately satisfied under the rotating wave approximation~\cite{BreuerTextBook}, which holds well even for many-body systems in a long-time regime where high-frequency oscillations can be neglected.
In Appendix~\ref{sec:app:RWA}, we show that 
\begin{align}
\label{eq:com_int_norm}
\|[H_\rS+H_\rB,H_\rI]\|=o(1)
\end{align}
holds by using the rotating wave approximation and the off-diagonal ETH.
Intuitively, the rotating wave approximation ensures the energy conservation without including the interaction energy, i.e., prohibits transitions between eigenstates of  $H_\rS+H_\rB$, implying that  $H_\rI$ becomes commutable with  $H_\rS+H_\rB$. 

\subsection{Hamiltonian for numerical simulation}
\label{sec:setup:Hamiltonian}

In this subsection, we explain the setup of our numerical simulation discussed in Section \ref{sec:SOR}, \ref{sec:LT}, and \ref{sec:ET}.

We perform numerical calculations of hard-core bosons with nearest-neighbor repulsion using numerically exact diagonalization. 
System S is a single site, and bath B is on a two-dimensional lattice. 
The Hamiltonian is given by
\begin{align}
H_\rS
&:=
\omega n_0,
\\
H_\rI
&:=
-\gamma^\prime
\sum_{\langle 0,j\rangle}
(c_0^\dag c_j + c_j^\dag c_0),
\\
H_\rB
&:=
\omega
\sum_{i=1}^N  n_i
-\gamma
\sum_{\langle i,j\rangle}
(c_i^\dag c_j+c_j^\dag c_i)
+
g
\sum_{\langle i,j\rangle}
n_i n_j,
\end{align}
where $\omega$ is the onsite potential, $-\gamma$ is the hopping in bath B, $-\gamma^\prime$ is the hopping between system S and bath B, and $g>0$ represents repulsion between hard core bosons. 
We note that $\langle i,j\rangle$ means the sum over the nearest-neighbor sites.
Hard core bosons cannot exist simultaneously on a single site, and their annihilation (creation) operator $c_i$ ($c_i^\dag$) satisfies the following commutation relations $[c_i,c_j]=[c^\dag_i,c^\dag_j]=[c_i,c^\dag_j]=0$ for $i\neq j$,  $\{c_i,c_i\}=\{c^\dag_i,c^\dag_i\}=0$, and $\{c_i,c^\dag_i\}=1$. 
The occupation number operator is defined as $n_i:=c^\dag_i c_i$. 
The site of $i=0$ corresponds to system S. 
We set bath B as a two-dimensional square lattice with the open boundary condition, and the bath size $N$ is written as $N=L_x\times L_y$. 
The operator of the particle number in bath B is written as $n_\rB:=\sum_{i=1}^N n_i$.
The Hamiltonian $H$ is non-integrable when $g\neq 0, \gamma\neq 0$ and $\gamma^\prime\neq 0$.

We write the initial state of system S as
\begin{align}
\rho_\rS(0):=p\proj{1}+(1-p)\proj{0},
\end{align}
where $\ket{n_\mathrm{S}}$ is the eigenstate of $n_0$ and satisfies $n_0\ket{n_\mathrm{S}}=n_\mathrm{S}\ket{n_\mathrm{S}}$ $(n_\mathrm{S}=1,0)$. 
In order to avoid absolute irreversibility, we choose $\rho_\rS(0)$ to be a mixed state by setting $p=0.99$ in our numerical simulation.
See Appendices~\ref{sec:app:AI} and \ref{sec:app:proof:AI} for the case with absolute irreversibility.
We define the change of the particle number in S as
\begin{align}
    \label{eq:dnS}
    \delta n_\rS:=\tr_\rS[n_0(\rho_\rS(t)-\rho_\rS(0))].
\end{align}

The initial state of bath B is an energy eigenstate of $H_\rB$ with a particle number $N_\rP$, which samples from the energy shell $[E-\Delta E,E]$. 
We write the width of the energy shell as $\Delta E=N\delta_E$ with $\delta_E$ be a positive constant and set $\delta_E=0.02$ in our numerical simulation. 
We perform the calculation of the fluctuation theorem for each energy eigenstate in the energy shell and investigate the dependence of the error of the fluctuation theorem on the bath size and the initial state of bath B.
We define the inverse temperature of $|E_\alpha^\rB\rangle$ as $\beta_\alpha$ satisfying $E_\alpha^\rB=\tr_\rB[\rho_\rB^\can(\beta_\alpha) H_\rB]$, where we explicitly write the $\beta$-dependence of the canonical ensemble.
The onsite potential $\omega$ is determined such that the canonical expectation of the particle number of the bath equals $N_\rP$: $\tr_\rB[n_\rB\rho_\rB^\can]=N_\rP$. 
We set $\omega$ as above, because the ETH is satisfied only within each particle number sector, and $N_\rP$ should be close to the canonical expectation number.
To investigate the bath size dependence, we set $L_x=3$, $L_y=3,4,5$ and $N_\rP=N/3$.

\section{Overview of the Results}
\label{sec:SOR}

In this section, we give an overview of the main results of this paper: the fluctuation theorem holds in both the long and short-time regimes (as defined below) in the setting of the previous section. 
The proof of the fluctuation theorem in the long-time regime is based on the ETH of $H$ (Sec.~\ref{sec:LT}).
Also, the fluctuation theorem in the short-time regime has been shown on the basis of the ETH of $H_\rB$ and the Lieb-Robinson bound~\cite{Iyoda2017} (Sec.~\ref{sec:ET}). 
The proofs of the fluctuation theorem in these time regimes are theoretically independent of each other and play complementary roles.
Our systematic numerical calculations about the bath size dependence support the theories in both of the long and short-time regimes.

This section is organized as follows.
In Sec.~\ref{sec:SOR:ErrFT}, we show the numerical result of the real-time dynamics and define the errors of the fluctuation theorem.
In Secs.~\ref{sec:SOR:long} and \ref{sec:SOR:short}, we explain the results on the fluctuation theorem in the long-time and short-time regimes, respectively.
In Sec.~\ref{sec:SOR:regimes}, we comment on the two time regimes.

\begin{figure}[t]
\begin{center}
\includegraphics[width=0.9\linewidth]{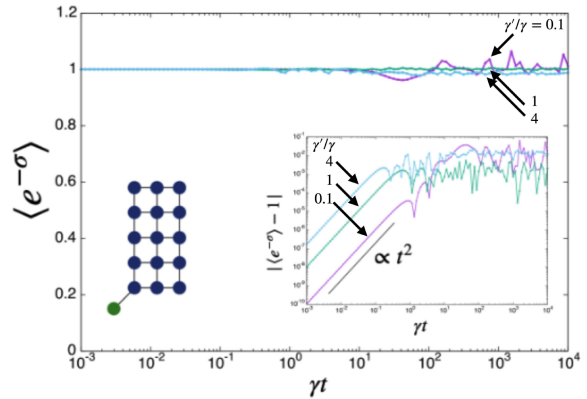}
\end{center}
\caption{
The time dependence of $\langle e^{-\sigma}\rangle$. The right inset shows the time dependence of the error of the fluctuation theorem $|\langle e^{-\sigma}\rangle-1|$, whose initial rise is proportionate to $t^2$. The left inset is a schematic of the total system used in numerical calculations. Parameters: $p=0.99,g=0.1\gamma$, $\gamma^\prime=0.1\gamma$~(purple), $\gamma$~(green),  $4\gamma$~(blue). The initial state of the bath is the energy eigenstates of $H_\rB$, whose energy is maximum in the energy shell at $\beta=0.1$.
The onsite potential $\omega$ is determined by $\tr_\rB[n_\rB\rho_\rB^\can]=N_\rP$.
}
\label{fig:G_tdep}
\end{figure}

\subsection{Errors of the fluctuation theorem}
\label{sec:SOR:ErrFT}
We first demonstrate whether the fluctuation theorem (\ref{eq:IFT}) holds or not by numerically investigating the real-time dynamics.
Figure~\ref{fig:G_tdep} shows the time dependence of $\langle e^{-\sigma}\rangle$, which implies that $\langle e^{-\sigma}\rangle\simeq 1$ holds in the entire time domain. 
The right inset of Fig.~\ref{fig:G_tdep} shows the time dependence of the error of the fluctuation theorem $|\langle e^{-\sigma}\rangle-1|$.
The error increases until the relaxation time of the system S ($t\sim 1$) and then saturates.
Qualitatively the same behavior is seen with other interacting parameters~(see Appendix~\ref{sec:app:numerical}), and Fig.~\ref{fig:dG_tdep} is a schematic of the typical time dependence of $|\langle e^{-\sigma}\rangle-1|$. 
\begin{figure}[t]
\begin{center}
\includegraphics[width=0.9\linewidth]{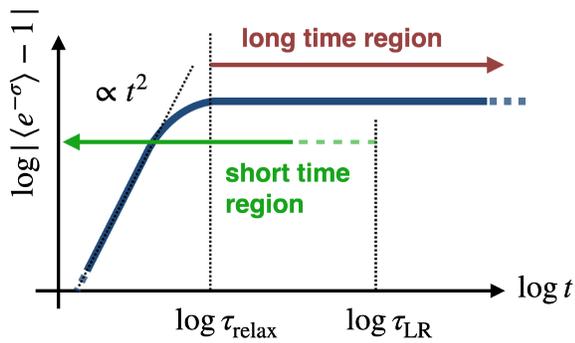}
\end{center}
\caption{
Sketch of a typical time dependence of $|\langle e^{-\sigma}\rangle-1|$, which initially rises in proportionate to $t^2$ and relaxes to the long-time average in $t\gtrsim\tau_\mathrm{relax}$. We refer to the time regime after the relaxation as the long-time regime. We also call the sufficiently shorter time regime than the Lieb-Robinson time $\tau_\LR$ (\ref{eq:tauLR:def}) as the short-time regime ($t\ll\tau_\LR$).
}
\label{fig:dG_tdep}
\end{figure}

We denote $\langle e^{-\sigma}\rangle$ when the initial state of bath B is $\rho_\rB(0)$~(resp. $\rho_\mathrm{B}^\mathrm{can}$) by $G$~(resp. $G^\can$).
We define $G$ and $G^\can$ as
\begin{align}
G
&:=
\mathrm{tr}[
e^{-\beta H_\mathrm{B}}
U
e^{\beta H_\mathrm{B}}
\rho_\mathrm{B}(0)
U^\dag
\rho_\mathrm{S}(t)
],
\\
G^\mathrm{can}
&:=
\mathrm{tr}[
e^{-\beta H_\mathrm{B}}
U
e^{\beta H_\mathrm{B}}
\rho_\mathrm{B}^\mathrm{can}
U^\dag
\rho^\mathrm{can}_\mathrm{S}(t)
]=1,
\end{align}
where $\rho_\mathrm{S}^\can(t)$ is the reduced density operator of system S defined as $
\rho^\mathrm{can}_\mathrm{S}(t)
:=\mathrm{tr}_\mathrm{B}[
U
\rho_\mathrm{S}(0)
\otimes
\rho_\mathrm{B}^\mathrm{can}
U^\dag
]
$.
Below, we focus on the error of the fluctuation theorem $G-G^\can$.

In the case where Eq.~(\ref{eq:com_int}) holds, $H_\rB$ in $G$ and $G^\can$ can be replaced by $-H_\rS$, and correspondingly we define
\begin{align}
\label{eq:GS:def}
G_\mathrm{S}
&:=\mathrm{tr}[
e^{\beta H_\mathrm{S}}
U
e^{-\beta H_\mathrm{S}}
\rho_\mathrm{B}(0)
U^\dag
\rho_\mathrm{S}(t)
],
\\
\label{eq:GScan:def}
G_\mathrm{S}^\mathrm{can}
&:=\mathrm{tr}[
e^{\beta H_\mathrm{S}}
U
e^{-\beta H_\mathrm{S}}
\rho_\mathrm{B}^\mathrm{can}
U^\dag
\rho^\mathrm{can}_\mathrm{S}(t)
].
\end{align}
Using the Baker-Campbell-Hausdorff formula for $e^{-\beta(H_\rS+H_\rB)}U e^{\beta(H_\rS+H_\rB)}$,
we can show $G=G_\rS$ and $G^\can=G_\rS^\can$ under Eq.~(\ref{eq:com_int}),
and the error of the fluctuation theorem is written as 
\begin{align}
G-G^\can
=
G_\rS-G_\rS^\can.
\end{align}

With the above argument, we define the error of the fluctuation theorem in the general case as
\begin{align}
\label{eq:decompose}
    G-G^\can = \delta G_\rS + \delta G_\rI,
\end{align}
where we define
\begin{align}
\delta G_\rS&:=
G_\rS-G_\rS^\can,
\label{eq:dGS_def}
\\
\delta G_\rI
&:=
G-G^\can-\delta G_\rS
\\
&=
G-G_\rS+G_\rS^\can-G^\can
\label{eq:dGI_def_4Gs}
\\
&=\dG_\rI^{(1)}+\dG_\rI^{(2)},
\label{eq:dGI_def_dGI1_dGI2}
\\
\dG_\rI^{(1)}
&:=
G-G_\rS,
\label{eq:dGI1_def}
\\
\dG_\rI^{(2)}
&:=
G_\rS^\can-G^\can.
\label{eq:dGI2_def}
\end{align}
We refer to $\delta G_\rI$ as the interaction-induced error.

As argued above, $\delta G_\rI=0$ holds under Eq.~(\ref{eq:com_int}).
Even if Eq.~(\ref{eq:com_int}) does not hold, the interaction-induced error $\delta G_\rI$ satisfies
\begin{align}
\label{eq:dGI_RWA}
|\delta G_\rI|=o(1)
\end{align}
under the rotating wave approximation and the off-diagonal ETH (see Appendix~\ref{sec:app:RWA}).
We also note that $\delta G_\rI=0$ holds without the rotating wave approximation when $\rho_\rB(0)=\rho_\rB^\can$, even if Eq.~(\ref{eq:com_int}) does not hold.
See also Appendix~\ref{sec:app:error} for the form of $\dG_\rI$.

We have decomposed the error of the fluctuation theorem into $\delta G_\rS$ and $\delta G_\rI$.
If we try to show the error without the decomposition, we cannot show that the error vanishes in the thermodynamic limit (see Appendix~\ref{sec:app:AnotherAttempt}).

\subsection{Long-time regime}
\label{sec:SOR:long}
We consider the fluctuation theorem in the long-time regime. 
In this paper, we prove that $\overline{\langle e^{-\sigma}\rangle}$, the long-time average of $\langle e^{-\sigma}\rangle$, nearly equals $1$ in the large-bath limit.
Note that we denote the long-time average of any quantity $O(t)$ as
\begin{align}
\overline{O(t)}
&:=
\lim_{T\rightarrow \infty}
\frac{1}{T}\int_0^T
O(t) dt.
\end{align}
In the special case that the condition
(\ref{eq:com_int}) is satisfied, we can prove that $| \overline{\langle e^{-\sigma}\rangle} - 1 | = o(1)$ holds, where the right-hand side represents the asymptotic behavior with respect to the bath size $N$.

If Eq.~(\ref{eq:com_int}) is not strictly satisfied, the interaction-induced error $\delta G_\rI$ defined in Sec.~\ref{sec:SOR:ErrFT} can generally appear in the fluctuation theorem.
However, the interaction-induced error vanishes in the thermodynamic limit under the rotating wave approximation, which holds well in the long-time regime (see Appendix~\ref{sec:app:RWA} for details).

In the absence of localization~\cite{Pal2010,Nandkishore2015} or persistent oscillations~\cite{Turner2018},  $\langle e^{-\sigma}\rangle$ relaxes to the long-time average $\overline{\langle e^{-\sigma}\rangle}$ after the finite relaxation time $\tau_\mathrm{relax}$.
We refer to the regime of $t$ satisfying $\tau_\mathrm{relax}\lesssim t$ as the long-time regime.
We argue that the relaxation time satisfies $\tau_\mathrm{relax}=\Theta(1)$ with respect to the bath size $N$ for the following reason.
The relaxation of $\langle e^{-\sigma}\rangle$ can be associated with observables of system S, because under the condition~(\ref{eq:com_int}), the heat $Q$ is replaced by the energy change of system S.
Under physically reasonable conditions, the relaxation time of observables of system S is $\Theta(1)$ when the initial state of bath B is the microcanonical ensemble~\cite{Garcia2017}.
We note that even when the condition~(\ref{eq:com_int}) does not strictly hold,
we can replace the heat Q under the rotating wave approximation, which again implies the relaxation time of  $\Theta(1)$.
We note that Ref.~\cite{Reimann2016} showed that the relaxation time is independent of the size of the total system if the Hamiltonian is random, and that numerical results consistent with the random Hamiltonians have been obtained with realistic models~\cite{Hetterich2015,Rigol2009}.
In summary, we argue that  $\tau_\mathrm{relax}=\Theta(1)$ holds in our setup, which is consistent with our own numerics as well.

We briefly remark on absolute irreversibility (see Appendices~\ref{sec:app:AI} and \ref{sec:app:proof:AI} for details).
As mentioned in the introduction, the numerical calculations in Ref.~\cite{Iyoda2017} showed that $\langle e^{-\sigma}\rangle$ is smaller than $1$ in the long-time regime, which was argued to be a violation of the fluctuation theorem. 
As a matter of fact, this is due to the numerical setup of Ref.~\cite{Iyoda2017} that the initial state of system S is a pure state, causing absolute irreversibility. 
That is, the reason why the fluctuation theorem in the long-time regime appeared to be broken in the numerical calculations of Ref.~\cite{Iyoda2017} is that the correction term $\lambda$ in Eq.~(\ref{eq:IFT:AI}) was not taken into account, while the numerical calculation itself is correct.
In the present paper, we prove and numerically confirm that the fluctuation theorem holds in the long-time regime with absolute irreversibility and the energy eigenstate bath, if the correction term $\lambda$ is taken into account. 
Finally, we remark that there is an alternative approach to make the fluctuation theorem hold by regularizing the pure initial state~\cite{HevelingPC}.

\subsection{Short-time regime}
\label{sec:SOR:short}
\begin{figure}[t]
\begin{center}
\includegraphics[width=0.55\linewidth]{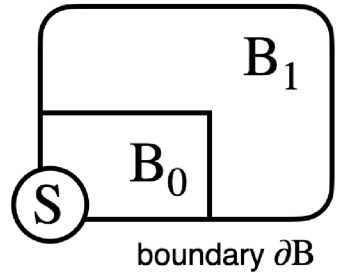}
\end{center}
\caption{Division of bath B used in the discussion of the short-time regime in Sec.~\ref{sec:ET}.
We divide bath B into the near part~($\rB_0$) and the far part~($\rB_1$).
The size of $\rB_0$ depends on $N$ as $\Theta(N^{\mu d})~(0<\mu<1/(2d))$. 
System S interacts with a part of $\rB_0$.
From the Lieb-Robinson bound, we introduce the Lieb-Robinson time $\tau_\LR=\Theta(N^\mu)$, at which the information of system S reaches $\rB_1$.
We note that the velocity of the information propagation does not depend on $N$.
}
\label{fig:systemLR}
\end{figure}

We next consider the short-time regime. 
To define the short-time regime in line with Ref.~\cite{Iyoda2017}, we divide bath B into $\rB_0$ and $\rB_1$, where $\rB_0$ is near S and $\rB_1$ is far from S~(see Fig.~\ref{fig:systemLR}).
We refer to the boundary between $\rB_0$ and $\rB_1$ as $\partial \rB$, and the size of $\rB_0$ is set to be $\Theta(N^{\mu d})~(0<\mu<1/(2d))$. 
Then, the short-time regime is defined as $t\ll \tau_\LR$. 
Here, $\tau_\LR$ is the Lieb-Robinson~(LR) time introduced by the Lieb-Robinson bound~\cite{Lieb1972,Hastings2006,Haah2018}, which, in the present case, represents a time scale that the information of S reaches $\rB_1$.
Specifically, under the above choice of the size of $\rB_0$, the LR time is given by
\begin{align}
    \label{eq:tauLR:def}
    \tau_\LR=\Theta(N^\mu).
\end{align}

Again in the special case that the condition (\ref{eq:com_int}) is satisfied,  $| \langle e^{-\sigma}\rangle - 1 | = o(1)$ holds in the short-time regime, as shown in Ref.~\cite{Iyoda2017} on the basis of the Lieb-Robinson bound and the ETH.
Even when the condition (\ref{eq:com_int}) does not hold, 
we show that the interaction-induced error vanishes within the perturbation theory at high temperature if the initial state of bath B satisfies the ETH (see Appendix~\ref{sec:app:interaction_perturb:ET}). 

In the present paper, to confirm the validity of the fluctuation theorem in the short-time regime, we perform systematic numerical calculations of the bath size dependence of the error of the fluctuation theorem. 
Our results in Sec.~\ref{sec:ET} show that the error decreases as $N$ increases, which supports the validity of the fluctuation theorem. 

We note that the error of the fluctuation theorem is $0$ at $t=0$, and initially increases in proportion to $t^2$ as shown in Fig.~\ref{fig:dG_tdep}. 
We theoretically show this time dependence in Appendix~\ref{sec:app:InitialRise}.
In Sec.~\ref{sec:ET}, assuming that the error of the fluctuation theorem initially grows in the form of $at^2$ with $a$ being a $t$-independent coefficient, we numerically confirm that this coefficient $a$ decreases as the bath size increases, implying $a=o(1)$. 

We note, as a side remark, that the time dependence of the change of the particle number in S~(\ref{eq:dnS}) also initially takes the form of $a_n t^2$ with $a_n$ being a $t$-independent coefficient.
In this case, we observe that the bath size dependence of $a_n$ is just $a_n=\Theta(1)$. 
The bath size independence of $a_n$ is contrastive to the error of the fluctuation theorem $a=o(1)$, implying that the ETH plays a crucial role only for the latter. 

\subsection{On the two time regimes}
\label{sec:SOR:regimes}
We emphasize that the short and long time regimes are defined and analyzed theoretically independently.
The short-time regime is defined by the Lieb-Robinson time, which increases at most linearly with the bath size, while the long-time regime is defined by coincidence with the long-time average, which covers a very long time scale characterized by the quantum recurrence time that increases doubly exponentially with the bath size~\cite{Bocchieri1957,Percival1961,Venuti2015}.
From the argument in Secs.~\ref{sec:SOR:long} and \ref{sec:SOR:short}, the short and long-time regimes have an overlap and thus cover the entire time domain as shown in Fig.~\ref{fig:dG_tdep}, when the bath size is sufficiently large (see also Sec.\ref{sec:ET:error}).

Suppose that the time evolution of the error is given as in Fig.~\ref{fig:dG_tdep}, i.e., the error of the fluctuation theorem initially increases in $t^2$ and relaxes monotonically to the long-time average after the $N$-independent relaxation time. Then, the fluctuation theorem in the long-time regime implies that in the short-time regime, and vice versa. 
In general, however, the fluctuation theorem in these time regimes are theoretically shown independently, and they together show the fluctuation theorem in the entire time domain. 
In fact, only from the fluctuation theorem in the long-time regime, we cannot exclude the situation that the error is not monotonic like Fig.~\ref{fig:dG_tdep}, but overshoots to a value larger than $o(1)$ after a time evolution of $t^2$ and then relaxes to the long-time average.
This possibility can be excluded from the fluctuation theorem in the short-time regime.

Let us discuss the time scales of real experimental setups.
For example,
in Ref.~\cite{Choi2016MBL}, real-time dynamics of ultracold atoms has been experimentally studied over about $300\tau$ with $\tau$ being the tunneling time.
On the other hand, the Lieb-Robinson time $\tau_\mathrm{LR}$ is estimated as $\tau_\LR\sim\ell\tau$, where $\ell=N^{1/4}$ is the distance between $\mathrm{S}$ and $\mathrm{B_1}$.
As the experiment is performed in the two-dimensional system where the site number is $N\sim 250$, the Lieb-Robinson time is evaluated as $4\tau$.
Thus, the Lieb-Robinson time is reasonably longer than the experimental time resolution, while it is much shorter than the experimentally tractable time regime.
In such an experimental setup, therefore, the theories for the short and long-time regimes are both necessary to address the entire time domain.
On the other hand, we note that the Lieb-Robinson time is too short  in our numerical setup of Sec.~\ref{sec:setup:Hamiltonian} for the short-time regime to be clearly visible, because our system size is small due to the limitation of numerical accessibility ($N\sim 15$).
Our theory for the short-time regime as well as for the long-time regime is more relevant to real-experimental setups with current or the near-future technologies, where the system sizes can be much bigger than numerics.

\section{Long-time regime}
\label{sec:LT}

In this section, we theoretically and numerically show the fluctuation theorem in the long-time regime.
In Sec.~\ref{sec:LT:average}, we discuss the long-time average and the relaxation time to the long-time average.
In Sec.~\ref{sec:LT:proof}, we outline the proof of the fluctuation theorem.
In Sec.~\ref{sec:LT:Numerical}, we show our numerical results to confirm our theory.

\subsection{Long-time average}
\label{sec:LT:average}

In this subsection, we consider the long-time average of $\dG_\rS=G_\rS-G_\rS^\can$ and discuss the relaxation times of $\dG_\rS$ and $\dG_\rI$.

To analyze the long-time average of $G_\rS$, we first write $G_\rS$ as follows:
\begin{align}
\label{eq:GSU4}
G_\mathrm{S}
&:=\mathrm{tr}[
e^{\beta H_\mathrm{S}}
U
e^{-\beta H_\mathrm{S}}
\rho_\mathrm{B}(0)
U^\dag
\tr_\rB[
U
\rho_\rS(0)\otimes \rho_\rB(0)
U^\dag
]
].
\end{align}
We note that Eq.~(\ref{eq:GSU4}) contains four unitary operators.

Then, under no degeneracy and the nonresonance condition (i.e., $E_a-E_b+E_c-E_d=0$ holds only when the pair of indexes $(a,c)$ equals $(b,d)$), the long-time average of $G_\mathrm{S}$ can be written as $\overline{G_\mathrm{S}}=\overline{G_\mathrm{S1}}+\overline{G_\mathrm{S2}}$. 
Here, $\overline{G_\mathrm{S1}}$ and $\overline{G_\mathrm{S2}}$ are defined as
\begin{align}
\nonumber
\overline{G_\mathrm{S1}}
:=
Z_\mathrm{S}
\sum_{a,c}
\mathrm{tr}_\mathrm{S}[
e^{\beta H_\mathrm{S}}
&
\mathrm{tr}_\mathrm{B}[
\pi_a
\rho_\mathrm{S}^\mathrm{can}
\otimes
\rho_\mathrm{B}(0)
\pi_a
]
\\
&
\mathrm{tr}_\mathrm{B}[
\pi_c
\rho_\mathrm{S}(0)\otimes\rho_\mathrm{B}(0)
\pi_c
]
],
\label{eq:dGS1}
\\
\nonumber
\overline{G_\mathrm{S2}}
:=
Z_\mathrm{S}
\sum_{\substack{a,b\\a\neq b}}
\mathrm{tr}_\mathrm{S}[
e^{\beta H_\mathrm{S}}
&
\mathrm{tr}_\mathrm{B}[
\pi_a
\rho_\mathrm{S}^\mathrm{can}
\otimes
\rho_\mathrm{B}(0)
\pi_b
]
\\
&
\mathrm{tr}_\mathrm{B}[
\pi_b
\rho_\mathrm{S}(0)\otimes\rho_\mathrm{B}(0)
\pi_a
]
],
\label{eq:dGS2}
\end{align}
where $\pi_a:=|E_a\rangle\langle E_a|$.
We refer to $\overline{G_\mathrm{S1}}$ as the diagonal term because it contains the diagonal ensemble~\cite{Reimann2008}, which is the long-time average of the density operator and will be defined in Sec.~\ref{sec:LT:proof}.
We also refer to $\overline{G_\mathrm{S2}}$ as the off-diagonal term, because $\overline{G_\mathrm{S2}}$ contains the off-diagonal matrix elements with respect to energy eigenstates of $H$. 
In a similar manner, we write $\overline{G_\mathrm{S}^\can}=\overline{G_\mathrm{S1}^\can}+\overline{G_\mathrm{S2}^\can}$.

In this paper, we prove 
\begin{align}
    |\overline{\dG_\rS}|=o(1)
    \label{eq:IFT_LT_dGS:result}.
\end{align}
From the foregoing discussion, the left-hand side is divided as 
$\overline{\dG_\rS}=(
\overline{G_\mathrm{S1}}
-
\overline{G_\mathrm{S1}^\mathrm{can}}
)+
(
\overline{G_\mathrm{S2}}
-
\overline{G_\mathrm{S2}^\mathrm{can}})
$.
In the next subsection, we will show that
\begin{align}
    |
\overline{G_\mathrm{S1}}
-
\overline{G_\mathrm{S1}^\mathrm{can}}
|&=o(1),
\label{eq:IFT_LT_dGS:result:diag}
\\
    |\overline{G_\mathrm{S2}}
|&=o(1),
\label{eq:IFT_LT_dGS:result:GS2}
\\
|\overline{G_\mathrm{S2}^\can}
|&=o(1),
\label{eq:IFT_LT_dGS:result:GS2can}
\end{align}
which imply Eq.~(\ref{eq:IFT_LT_dGS:result}).

We now discuss the relaxation time of $\dG_\rS$. 
From Eq.~(\ref{eq:GSU4}), we can write $G_\rS$ as
\begin{align}
    G_\rS
    =&
    \sum_{i,j,k,l}
    e^{\beta(E_i^\rS-E_j^\rS)}
    p_k^\rS
    \nonumber
    \\
    &
    \langle \psi_k(t)|
    q^{il}_\rS
    |\psi_k(t)\rangle
    \langle \phi_j(t)|
    q^{li}_\rS
    |\phi_j(t)\rangle,
    \label{eq:LT:GS:forRelaxTime}
    \\
    |\psi_k(0)\rangle
    :=&
    |p_k^\rS\rangle\otimes |E_\mathrm{ini}^\rB\rangle,
    \\
    |\phi_j(0)\rangle
    :=&
    |E_j^\rS\rangle\otimes |E_\mathrm{ini}^\rB\rangle,
\end{align}
where the spectral decomposition of $\rho_\rS(0)$ is $\rho_\rS(0)=\sum_k p_k^\rS |p_k^\rS\rangle\langle p_k^\rS|$.
As shown in Eq.~(\ref{eq:LT:GS:forRelaxTime}), $G_\rS$ is written as a combination of the expectation values of $q^{il}_\rS$. Besides, the same rewrite is possible for $G_\rS^\can$.
In the total system SB, if the initial state of bath B is the microcanonical ensemble, the relaxation time of an observable of system S to its long-time average is independent of the bath size under some conditions on the matrix elements of the operator and the initial state of SB~\cite{Garcia2017}.
Those conditions are satisfied if the amplitudes of the matrix elements obey the Gaussian distribution, which is numerically confirmed in various quantum many-body systems satisfying the off-diagonal ETH~\cite{Steingeweg2013,Beugeling2015}.
Then, we argue that the relaxation time of any system operator does not depend on the bath size also in our setup assuming the off-diagonal ETH.
Since $G_\rS$ and $G_\rS^\can$ are written as a combination of the expectation values of operators in S, the relaxation time of $\dG_\rS$ is also independent of the bath size.

We also discuss the relaxation time of $\dG_\rI$ to $\overline{\dG_\rI}$.
Since the interaction $H_\rI$ is local, we argue that the relaxation time to be $\Theta(1)$ by the same argument as above. 
\subsection{Outline of the proof}
\label{sec:LT:proof}

This subsection outlines the proof that the diagonal contribution $\left|
\overline{G_\mathrm{S1}}
-
\overline{G_\mathrm{S1}^\mathrm{can}}
\right|$ and the off-diagonal contributions 
$\left|
\overline{G_\mathrm{S2}}
\right|$ and
$\left|
\overline{G_\mathrm{S2}^\mathrm{can}}
\right|$ vanish in the thermodynamic limit.
See Appendix~\ref{sec:app:proof} for the complete proof.
We also note that the interaction-induced error $|\overline{\dG_\rI}|$ vanishes in the thermodynamic limit under the rotating wave approximation (see Appendix~\ref{sec:app:RWA} for the proof).

First, we show the outline of the proof of $\left|
\overline{G_\mathrm{S1}}
-
\overline{G_\mathrm{S1}^\mathrm{can}}
\right|=o(1)$.
We note that $\overline{G_\mathrm{S1}}$ and $\overline{G_\mathrm{S1}^\can}$ can be written as the expectation values of the observables of S. 
That is, the following relations hold:
\begin{align}
\overline{G_\mathrm{S1}}&=\tr[O_\rS\rho^\DE],
\label{eq:dGS1B}
\\
\overline{G_\mathrm{S1}^\can}&=\tr[O_\rS^\can\rho^{\can,\DE}],
\label{eq:dGS2B}
\end{align}
where $\rho^\DE:=\sum_a \pi_a \rho_\rS^\can\otimes \rho_\rB(0)\pi_a$ is the diagonal ensemble of $\rho_\rS^\can\otimes\rho_\rB(0)$ with respect to $H$ and $O_\rS:=Z_\rS\overline{\rho_\rS(t)}e^{\beta H_\rS}$ is an operator of system S. 
Similarly, 
$\rho^{\can,\DE}:=\sum_a \pi_a \rho_\rS^\can\otimes \rho_\rB^\can\pi_a$ is the diagonal ensemble of $\rho_\rS^\can\otimes\rho^\can_\rB$ with respect to $H$ and $O^\can_\rS:=Z_\rS\overline{\rho^\can_\rS(t)}e^{\beta H_\rS}$ is an operator of system S. 
It is clear that the energy width of $\rho_\rB(0)$ is $\Theta(1)$.
For the case of $\rho_\rB^\can$, the large-deviation type upper bound for the energy distribution has been obtained and the energy width of $\rho_\rB^\can$ is $\Theta(N^{1/2})$~\cite{Tasaki2018}.
Both of these energy widths are smaller than $\Delta=\Theta(N^\alpha)~(1/2<\alpha<1)$. 
Thus, we can define an energy shell with the energy width $\Delta$, which includes both of the supports of $\rho_\rB(0)$ and $\rho_\rB^\can$ and the contribution out of the shell is negligible in the thermodynamic limit.

When the energy eigenstates of $H$ satisfy the strong diagonal ETH for any observable of system S, $O_\rS\simeq O_\rS^\can$ and $\tr_\rB[\rho^\DE]\simeq\tr_\rB[\rho^{\can,\DE}]$ hold, implying that any observable of system S relaxes to the long-time average and the initial states of bath B cannot be distinguished by looking at system S alone. Then, we obtain
\begin{align}
\overline{G_\mathrm{S1}}
=
\tr[O_\rS\rho^\DE] 
\simeq
\tr[O_\rS^\can\rho^{\can,\DE}]
=
\overline{G_\mathrm{S1}^\can},
\label{eq:LT:dGS1:result}
\end{align}
which leads to $\left|
\overline{G_\mathrm{S1}}
-
\overline{G_\mathrm{S1}^\mathrm{can}}
\right|=o(1)$.

When the interaction between system S and bath B is weak, the above result can also be interpreted as follows.
We rewrite $O_\rS$ and $O_\rS^\can$ as $O_\rS=\overline{\rho_\rS(t)}(\rho_\rS^\can)^{-1}$ and $O^\can_\rS=\overline{\rho^\can_\rS(t)}(\rho_\rS^\can)^{-1}$, respectively. 
If the state of S relaxes to the canonical ensemble of $H_\rS$, i.e., $\overline{\rho_\rS(t)}\simeq\rho_\rS^\can$ and $\overline{\rho^\can_\rS(t)}\simeq\rho_\rS^\can$, $O_\rS\simeq 1_\rS$ and $O_\rS^\can\simeq 1_\rS$ hold, which leads to $\overline{G_\mathrm{S1}}\simeq \overline{G^\can_\mathrm{S1}}$.
That is, $\left|
\overline{G_\mathrm{S1}}
-
\overline{G_\mathrm{S1}^\mathrm{can}}
\right|=o(1)$ holds when the system-bath interaction is weak and the state of system S relaxes to the canonical ensemble.

We next prove 
$|\overline{G_\mathrm{S2}}|=o(1)$ and 
$|\overline{G_\mathrm{S2}^\can}|=o(1)$ by using the off-diagonal ETH for the energy eigenstates of $H$ and observables in S.
We first approximate $\overline{G_\mathrm{S2}}$ as
\begin{align}
\nonumber
\overline{G_\mathrm{S2}}
\simeq
Z_\mathrm{S}
\sum_{\substack{a,b\in\sh\\a\neq b\\i,j}}
&
(q_\rS^{ij})_{ab}
(q_\rS^{ji})_{ba}
\\
&
(\rho_\rS^\can\otimes\rho_\rB(0))_{ab}
(\rho_\rS(0)\otimes\rho_\rB(0))_{ba},
\label{eq:GS2_limit}
\end{align}
where the summation over the energy eigenstates of $H$ is restricted to the energy shell. 
If the off-diagonal ETH for $q_\rS^{ij}$ holds, the bath size dependence of the off-diagonal matrix elements $(q_\rS^{ij})_{ab}$ is written as
\begin{align}
(q_\rS^{ij})_{ab}=\mathcal{O}(1)/\sqrt{D^\prime}.
\label{eq:offETH}
\end{align}
In addition, by using the Cauchy-Schwartz inequality and the fact that the purity of any state $\rho$ is not larger than $1$, we obtain 
\begin{align}
\left|
\overline{G_\mathrm{S2}}
\right|
&\leq
\frac{\Theta(1)}{D^\prime}=o(1).
\label{eq:LT:dGS2:result}
\end{align}
See details in Appendix~\ref{sec:app:proof:S2}. 
In a similar manner, $\left|\overline{G^\can_\mathrm{S2}}\right|=o(1)$ is proved.

We next comment on the interaction-induced error $\dG_\rI$ defined in~(\ref{eq:dGI_def_4Gs}).
As mentioned in Sec.~\ref{sec:SOR:ErrFT}, the interaction-induced error is zero when the condition (\ref{eq:com_int}) is satisfied. 
Even when the condition (\ref{eq:com_int}) is not satisfied, we can prove that
\begin{align}
    |\overline{\dG_\rI}|=o(1)
    \label{eq:LT:dGI_expectation}
\end{align}
holds under the rotating wave approximation (see Appendix~\ref{sec:app:RWA}).

We here summarize the assumptions used in the foregoing proofs.
In the proof of
$\left|\overline{G_\mathrm{S1}}-\overline{G_\mathrm{S1}^\mathrm{can}}\right|=o(1)$~[Eq.~(\ref{eq:IFT_LT_dGS:result:diag})], we used the (strong) diagonal ETH~(\ref{eq:diag_ETH_Setup}) of $H$ for all observables of S, and the fact that the energy width of the canonical ensemble is narrower than $\Theta(N^a)~(1/2<a<1)$.
In the proof of $|\overline{G_\mathrm{S2}}|=o(1)$ and $|\overline{G_\mathrm{S2}^\can}|=o(1)$~[Eqs.~(\ref{eq:IFT_LT_dGS:result:GS2}) and (\ref{eq:IFT_LT_dGS:result:GS2can})], we used the (strong) off-diagonal ETH~(\ref{eq:offdiag_ETH_Setup}) of $H$ for all observables of S. 

\subsection{Numerical results}
\label{sec:LT:Numerical}

In this subsection, we numerically confirm that $|\overline{\dG_\rS}|=o(1)$~[Eq.~(\ref{eq:IFT_LT_dGS:result})] and $|\overline{\dG_\rI}|=o(1)$~[Eq.~(\ref{eq:LT:dGI_expectation})].
To numerically obtain $|\overline{\dG_\rS}|$ and $|\overline{\dG_\rI}|$,
we first analytically obtain the expressions like Eqs.~(\ref{eq:dGS1}) and (\ref{eq:dGS2}),
and then numerically evaluate these expressions.

\begin{figure}[t]
\begin{center}
\includegraphics[width=\linewidth]{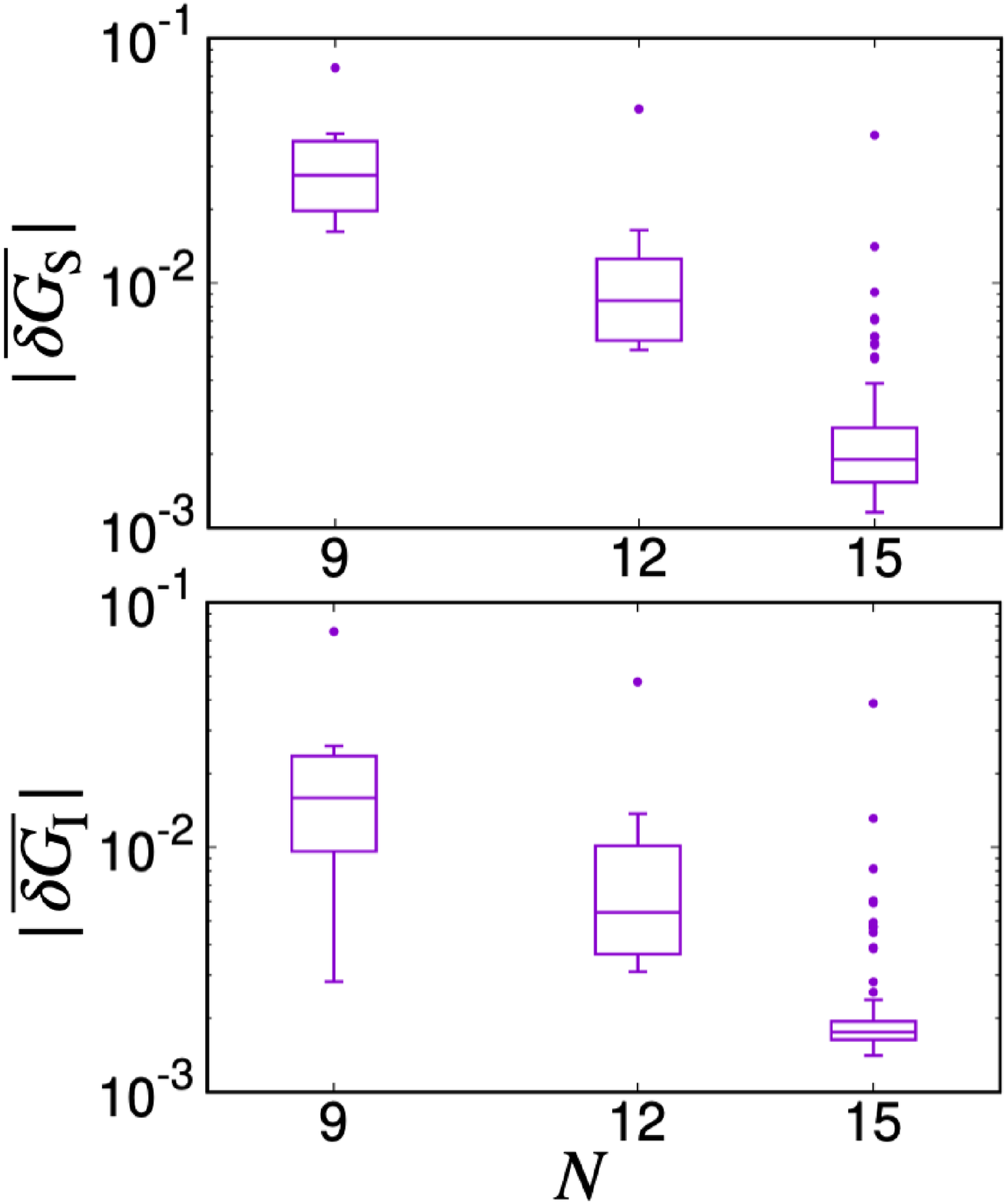}
\end{center}
\caption{
The boxplot of $|\overline{\dG_\rS}|$ and $|\overline{\dG_\rI}|$ to represent the dependence on the bath size and the initial state.
Each data in the boxplot corresponds to each initial eigenstate of bath B in the energy shell $[E-\Delta E,E]$.
Parameters : $p=0.99,g=0.1\gamma,\beta=0.1$, $\gamma^\prime=\gamma$.
Both $|\overline{\dG_\rS}|$ and $|\overline{\dG_\rI}|$ tend to decrease as $N$ increases.
}
\label{fig:dGS_Ndep}
\end{figure}

First, we show the numerical results about the long-time average of $\dG_\rS$ and $\dG_\rI$. 
We note that $|\overline{\dG_\rS}|$ and $|\overline{\dG_\rI}|$ are calculated for each initial eigenstate of bath B.
Since we consider $D$ eigenstates of B in the energy shell, $D$ data respectively for $|\overline{\dG_\rS}|$ and $|\overline{\dG_\rI}|$ are obtained in our numerical calculation.
Figure~\ref{fig:dGS_Ndep} shows the boxplot of the dependence of $|\overline{\dG_\rS}|$ and $|\overline{\dG_\rI}|$ on the bath size, which represents the distribution of $|\overline{\dG_\rS}|$ and $|\overline{\dG_\rI}|$.
Each data in the boxplot corresponds to each initial eigenstate of bath B in the energy shell $[E-\Delta E,E]$.
The middle line of the box represents the median and the top~(bottom) of the box represents the upper~(lower) quartile.
The half of the eigenstates in $[E-\Delta E,E]$ are included in the box. 
Figure~\ref{fig:dGS_Ndep} shows that $|\overline{\dG_\rS}|$ and $|\overline{\dG_\rI}|$ decreases as $N$ increases.

\begin{figure}[t]
\begin{center}
\includegraphics[width=\linewidth]{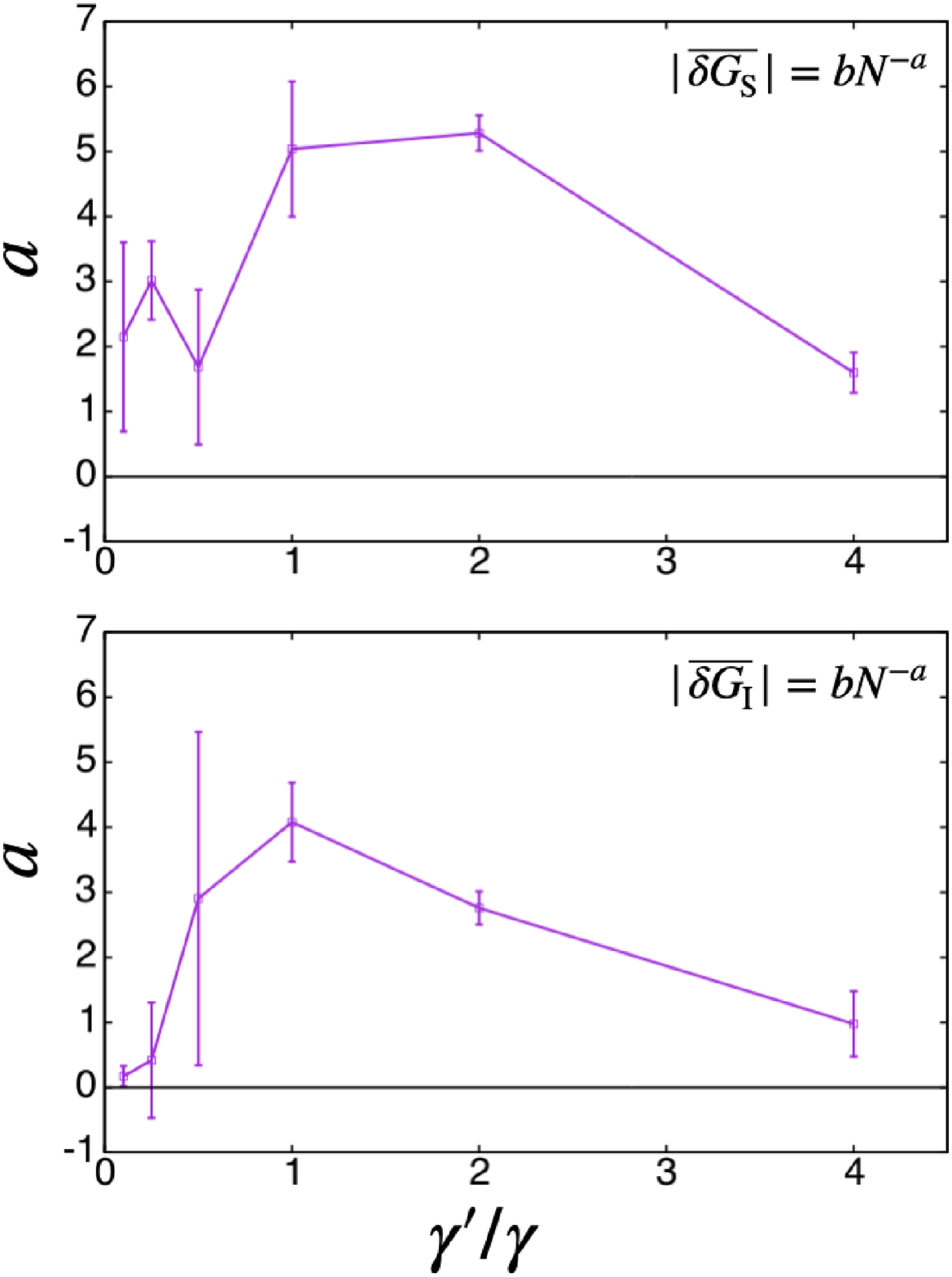}
\end{center}
\caption{
The $\gamma^\prime$-dependence of the exponent $a$, which is obtained by fitting the medians of $\log|\overline{\dG_\rS}|$ and $\log|\overline{\dG_\rI}|$ against $-a\log N+b$.
The fitting is performed with error and the error bars of $a$ comes from it.
Parameters : $p=0.99, g=0.1\gamma, \beta=0.1$.
For both $|\overline{\dG_\rS}|$ and $|\overline{\dG_\rI}|$, we obtain $a>0$, which supports the theories~(\ref{eq:IFT_LT_dGS:result}) and (\ref{eq:LT:dGI_expectation}).
When $\gamma^\prime$ is near $0$ or $\gamma^\prime$ is too large, the finite size effect is large and $a$ tends to be small.
}
\label{fig:a_dGS_gpdep}
\end{figure}

To investigate the $N$-dependence of $|\overline{\dG_\rS}|$ and $|\overline{\dG_\rI}|$ in more detail, we fit the numerical data of $\log|\overline{\dG_\rS}|$ and $\log|\overline{\dG_\rI}|$ against a fitting function $-a\log N+b$ with the fitting parameters $a$ and $b$. 
The positive $a$ implies $|\overline{\dG_\rS}|=o(1)$ and $|\overline{\dG_\rI}|=o(1)$. 
Figure~\ref{fig:a_dGS_gpdep} shows the $\gamma^\prime$-dependence of $a$, where we remind that $\gamma^\prime$ is the coupling strength of the interaction between system S and bath B.
Our result shows that $a$ for $|\overline{\dG_\rS}|$ is indeed positive for any $\gamma^\prime$. 
Thus, $|\overline{\dG_\rS}|=o(1)$ holds, which confirms the theory~(\ref{eq:IFT_LT_dGS:result}).

We note that system S is decoupled from bath B in the limit of $\gamma^\prime\rightarrow 0$. 
Also, system S and the support of $H_\rI$ are decoupled from the rest of the total system SB in the limit of $\gamma^\prime\rightarrow\infty$.
These decouplings imply that the ETH is no longer relevant to system S, and thus the fluctuation theorem does not hold.
Thus, it is reasonable that $a$ is small in the region of $\gamma^\prime/\gamma \simeq 0$ or $\gamma^\prime/\gamma\gg 1$ because of the large finite-size effect.
If the bath size is much larger than $N$ currently used, the finite-size effect is expected to become less significant, as a similar mechanism observed in Ref.~\cite{Brenes2020}.

For the interaction-induced error $|\overline{\dG_\rI}|$, the fitting parameter is positive within the margin of numerical errors as shown in Fig.~\ref{fig:a_dGS_gpdep},
which confirms (\ref{eq:LT:dGI_expectation})
, while the numerical errors are larger when $\gamma^\prime$ is small.
We consider this is because the finite-size effect of the ETH is larger there.

\section{Short-time regime}
\label{sec:ET}

In this section, we investigate the fluctuation theorem with the energy eigenstate bath in the short time regime.
While Ref.~\cite{Iyoda2017} theoretically showed that $|\dG_\rS|=o(1)$ holds on the basis of the Lieb-Robinson bound and the ETH, no quantitative numerical calculations of the bath size dependence of the error was made. 
In the present paper, we numerically investigate the dependence of the error of the fluctuation theorem on the bath size and the initial state.
In Sec.~\ref{sec:ET:error}, we define the error and briefly overview the theory developed in Ref.~\cite{Iyoda2017}.
In Sec.~\ref{sec:ET:numerical}, we show our numerical results, which support the theory.

\subsection{Definition of the error}
\label{sec:ET:error}

In this subsection, we discuss the error of the fluctuation theorem in the short-time regime. 
Specifically, we decompose $\dG_\rS$ introduced in Eq.~(\ref{eq:dGS_def}) into the term that can be evaluated by the Lieb-Robinson bound and the term that can be evaluated by the ETH.

From the Lieb-Robinson bound, when considering the dynamics of the short-time regime, the effect of the far side of bath B ($\rB_1$ in Fig.~\ref{fig:systemLR}) on system S is negligible.
Using $H_{\rB_0}$ which is given by restricting $H_\rB$ to $\rB_0$, we define the truncated Hamiltonian $H_\rT$ and the corresponding time evolution operator $U_\rT$ as $H_\rT:=H_\rS+H_\rI+H_{\rB_0}$ and $U_\rT:=\exp(-iH_\rT t/\hbar)$. 
Then, we define $G_\rS^\rT$ and $G_\mathrm{S}^{\can,\rT}$ as
\begin{align}
G_\rS^\rT
&:=\mathrm{tr}[
e^{\beta H_\mathrm{S}}
U_\rT
e^{-\beta H_\mathrm{S}}
\rho_\mathrm{B}(0)
U_\rT^\dag
\rho_\mathrm{S}(t)
],
\\
G_\mathrm{S}^{\can,\rT}
&:=\mathrm{tr}[
e^{\beta H_\mathrm{S}}
U_\rT
e^{-\beta H_\mathrm{S}}
\rho_\mathrm{B}^\mathrm{can}
U_\rT^\dag
\rho^\mathrm{can}_\mathrm{S}(t)
].
\end{align}
Due to the Lieb-Robinson bound, $G_\rS^\rT\simeq G_\rS$ and $G_\rS^{\can,\rT}\simeq G_\rS^\can$ hold in $t\ll\tau_\LR$. 
Using these arguments, we decompose $\dG_\rS$ as
\begin{align}
\dG_\rS = \dG_\LR + \dG_\ETH,
\end{align}
where we define
\begin{align}
\dG_\LR
&:=
\dG_\LR^{(1)}+\dG_\LR^{(2)},
\\
\dG_\LR^{(1)}
&:=
G_\rS-G_\rS^\rT,
\\
\dG_\LR^{(2)}
&:=
G_\rS^{\can,\rT}
-
G_\rS^{\can},
\\
\dG_\ETH
&:=
\delta G_\rS-\delta G_\LR
=
G_\rS^\mathrm{T}-G^{\can,\rT}_\rS.
\end{align}
We can evaluate these terms as follows.

Using the Lieb-Robinson bound~\cite{Lieb1972,Hastings2006}, we can show that
\begin{align}
    |\dG_\LR|=o(1)~~(t\ll\tau_\LR)
     \label{eq:LRB:dGLR}
\end{align}
holds. 
In this evaluation, the ETH is not used, while it is essential that $H_\rB$ and $H_\rI$ are local.
In Ref.~\cite{Iyoda2017}, it has been shown that
\begin{align}
    |\delta G_\mathrm{LR}^{(1,2)}|\leq C e^{-\kappa\ell}t^2
    \label{eq:LRB:IKS}
\end{align}
holds in $t\ll\tau_\LR$, where $\ell:=\mathrm{dist}(\tilde{\rS},\partial \mathrm{B})$ is the distance between $\tilde{\rS}$ and $\partial \rB$, and $\tilde{\rS}$ is the area that consists of system S and the support of $H_\rI$.
The positive constants $C$ and $\kappa$ are independent of $N$.
Since $\ell$ is increasing as the bath size increases, Eq.~(\ref{eq:LRB:IKS}) implies Eq.~(\ref{eq:LRB:dGLR}).
Furthermore, using the improved Lieb-Robinson bound~\cite{Haah2018}, we can show that
\begin{align}
    |\delta G_\mathrm{LR}^{(1,2)}|&\leq C^\prime
    \frac{t^\ell}{\ell!}
    \simeq
    C^\prime e^{-\ell(\log\ell-1)}
    t^\ell
    \label{eq:LRB:improved}
\end{align}
holds in $t\ll\tau_\LR$ with $C^\prime$ being a positive constant independent of $N$.
The evaluation of (\ref{eq:LRB:improved}) is tighter than (\ref{eq:LRB:IKS}), where the power of $t$ depends on $\ell$.

We next discuss $\dG_\ETH$.
Under the assumption that the initial state $|E_a^\rB\rangle$ of bath B satisfies the ETH of $H_\rB$ for observables in $\rB_0$, we show that
\begin{align}
    |\dG_\ETH| 
    &\simeq
    \left|
    \mathrm{tr}_{\rB_0}
    \left[
    L_{\rB_0}
    \tr_{\rB_1}[
    \rho_\mathrm{B}
    -
    \rho_\mathrm{B}^\can
    ]
    \right]\right|
    \\
    &=o(1),
    \label{eq:ETH:ET}
\end{align}
where 
$L_{\rB_0}
:=
\mathrm{tr}_{\rS}
[
U_\rT^\dag\rho_\mathrm{S}(t)
e^{\beta H_\mathrm{S}}
U_\rT
e^{-\beta H_\mathrm{S}}
]
$.
In Eq.~(\ref{eq:ETH:ET}), we use the ETH for $L_{\rB_0}$.
The initial state of bath B must be indistinguishable from the canonical ensemble, because the time evolution is restricted to $U_\rT$ and the far part $\rB_1$ can be neglected.

If the bath size is small (specifically, $\mathcal{O}(1)$), $\tau_\LR$ and $\tau_\mathrm{relax}$ becomes comparable (both $\mathcal{O}(1)$), which cannot be distinguished in practice.
Therefore, in our numerical calculations in the next subsection, we only focus on the time regime $t\lesssim\tau_\mathrm{relax}$ in which the error of the fluctuation theorem is proportionate to $t^2$.
We calculate the coefficients 
in our numerical calculations, $a_\ETH$, $a_\rI$ and $a_\LR$ defined as
\begin{align}
    \label{eq:def_aETH}
   |\dG_\ETH|&=a_\ETH t^2,
   \\
    \label{eq:def_aI}
   |\dG_\rI|&=a_\rI t^2,
   \\
    \label{eq:def_aLR}
   |\dG_\LR|&=a_\LR t^b,
\end{align}
where $b$ is a fitting parameter determined by numerical calculation.
We will show the $t^2$-dependence in Eqs.~(\ref{eq:def_aETH}) and (\ref{eq:def_aI}) in Appendix~\ref{sec:app:InitialRise}.
Since Eq.~(\ref{eq:LRB:improved}) is an inequality, $b$ does not necessarily equal $\ell$. We note that $b$ can depend on the bath size.
Thus, we find that the initial $t^2$ behavior in $\delta G_{\rm S}$ originates from $\dG_\ETH$, not from $\dG_\LR$.

In order for the time dependence (\ref{eq:def_aETH}) and (\ref{eq:def_aLR}) to continue until the relaxation time of $\Theta(1)$ and for Eqs.~(\ref{eq:LRB:dGLR}) and (\ref{eq:ETH:ET}) to hold, 
\begin{align}
    \label{eq:aLR_o1}
    a_\LR&=o(1),
    \\
    \label{eq:aETH_o1}
    a_\ETH&=o(1)
\end{align}
must be satisfied. 
We will numerically confirm these equations in the next subsection.

Using the perturbation theory and the off-diagonal ETH (see Appendix~\ref{sec:app:interaction_perturb:ET}), we can show that 
\begin{align}
    |\dG_\rI|=o(1)
    \label{eq:dGI_LRB_ETH}.
\end{align}
holds in $t\ll\tau_\LR$.
In order for the time dependence (\ref{eq:def_aI}) to continue until the relaxation time of $\Theta(1)$ and for Eq.~(\ref{eq:dGI_LRB_ETH}) to hold,
\begin{align}
    \label{eq:aI_o1}
    a_\rI&=o(1)
\end{align}
must be satisfied.
This will be numerically confirmed in the next subsection.

We here remark on the two separate time scales appeared above.
The time interval satisfying $\tau_\mathrm{relax}\lesssim t\ll\tau_\LR$ clearly exists for a sufficiently large bath size $N$,
while  such a large bath size is not accessible by our numerical calculations but may be achieved by real experiments with ultracold atoms.
In our numerical simulation (see Fig.~\ref{fig:App:FT_error_t2} of Appendix~\ref{sec:app:numerical}),
only $\tau_\mathrm{relax}\lesssim t \lesssim \tau_\LR$ can be guaranteed for the bath size $N=15$.
We should further remark that $\tau_\mathrm{relax}$ can depend on the system-bath coupling $\gamma^\prime$, and the bath size $N$ required for the existence of the time interval $\tau_\mathrm{relax}\lesssim t \lesssim \tau_\LR$ increases as $\gamma^\prime$ decreases.
In our numerical simulation,
such a time interval exists for $\gamma^\prime/\gamma=1,1.5,2,3,4$, while does not clearly exist for $\gamma^\prime/\gamma=0.05, 0.1, 0.4$ where our $N$ is not sufficiently large.

\subsection{Numerical results}
\label{sec:ET:numerical}

\begin{figure}[t]
\begin{center}
\includegraphics[width=0.4\linewidth]{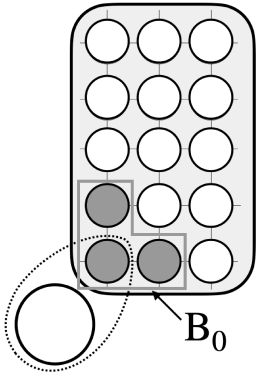}
\end{center}
\caption{$\rB_0$ used in our numerical simulation in Sec~\ref{sec:ET:numerical}.
}
\label{fig:systemB0}
\end{figure}
In this subsection, we numerically confirm Eqs.~(\ref{eq:aLR_o1}), (\ref{eq:aETH_o1}), and (\ref{eq:aI_o1}).
We choose three sites close to system S as $\rB_0$ as shown in Fig.~\ref{fig:systemB0}.

\begin{figure}[t]
\begin{center}
\includegraphics[width=\linewidth]{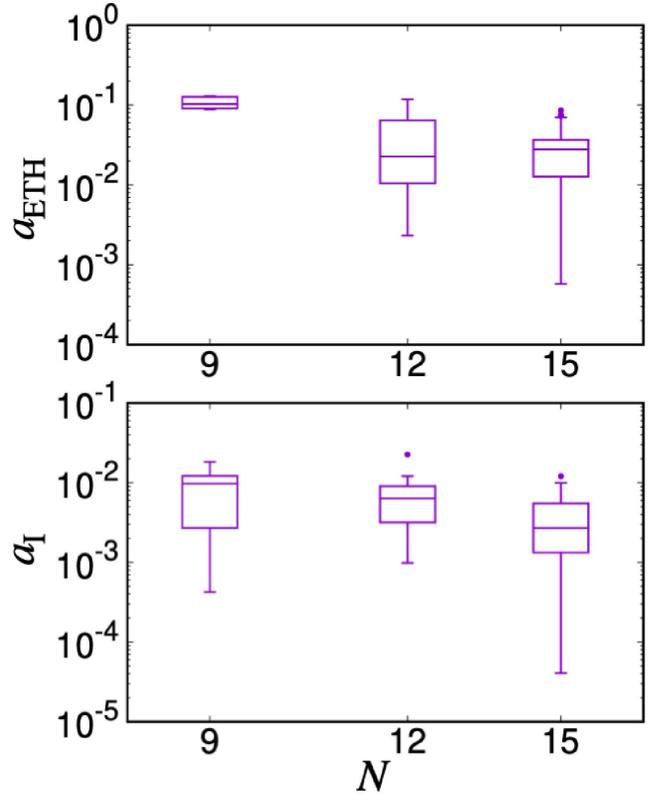}
\end{center}
\caption{
The dependence of $a_\ETH$ and $a_\rI$ on the bath size and the initial state.
Each data in the boxplot corresponds to each initial eigenstate of bath B in the energy shell $[E-\Delta E,E]$.
Parameters : $p=0.99,g=0.1\gamma, \gamma^\prime=\gamma, \beta=0.1$.
Both $a_\ETH$ and $a_\rI$ tend to decrease as $N$ increases.
}
\label{fig:aETH}
\end{figure}

Using the fitting functions $|\dG_\ETH|=a_\ETH t^{b_\ETH}$ and $|\dG_\rI|=a_\rI t^{b_\rI}$ and numerical data in $10^{-3}<\gamma t<10^{-2}$, we determine the fitting parameters $a_\ETH,b_\ETH,a_\rI$ and $b_\rI$.
Because the effect of the sub-leading terms are negligible in this time region, we choose this time region for simplicity.
Regardless of the parameters of the Hamiltonian, we obtain $b_\ETH= 2$ and $b_\rI= 2$ with negligibly small numerical errors.
Figure~\ref{fig:aETH} is a boxplot showing the dependence of $a_\ETH$ and $a_\rI$ on the bath size and the initial state.
Each data in the boxplot corresponds to each energy eigenstate in the energy shell $[E-\Delta E,E]$.
Both $a_\ETH$ and $a_\rI$ tend to decrease as the bath size increases.

To investigate the $N$-dependence of $a_\ETH$ and $a_\rI$, we fit the medians of $\log a_\ETH$ and $\log a_\rI$ against the fitting functions $\log a_\ETH= c_\ETH - \eta_\ETH \log N$ and $\log a_\rI=c_\rI-\eta_\rI \log N$. As a result, we obtain $\eta_\ETH=1.4\pm 0.8$, $\eta_\rI=2.0\pm 0.8$ for $\gamma^\prime=\gamma$, $g=0.1\gamma$ and $\eta_\ETH=1.8\pm 0.2$, $\eta_\rI=2.1\pm 0.1$ for $\gamma^\prime=\gamma$, $g=0.4\gamma$. 
These results support the theories based on the ETH~(\ref{eq:aETH_o1}) and (\ref{eq:aI_o1}).
We note that these values are almost independent of $\gamma^\prime$. 
This is understood from the fact that $a_\ETH=o(1)$ is brought by the ETH of $H_\rB$, which is independent of the interaction between system S and bath B.

\begin{figure}[t]
\begin{center}
\includegraphics[width=\linewidth]{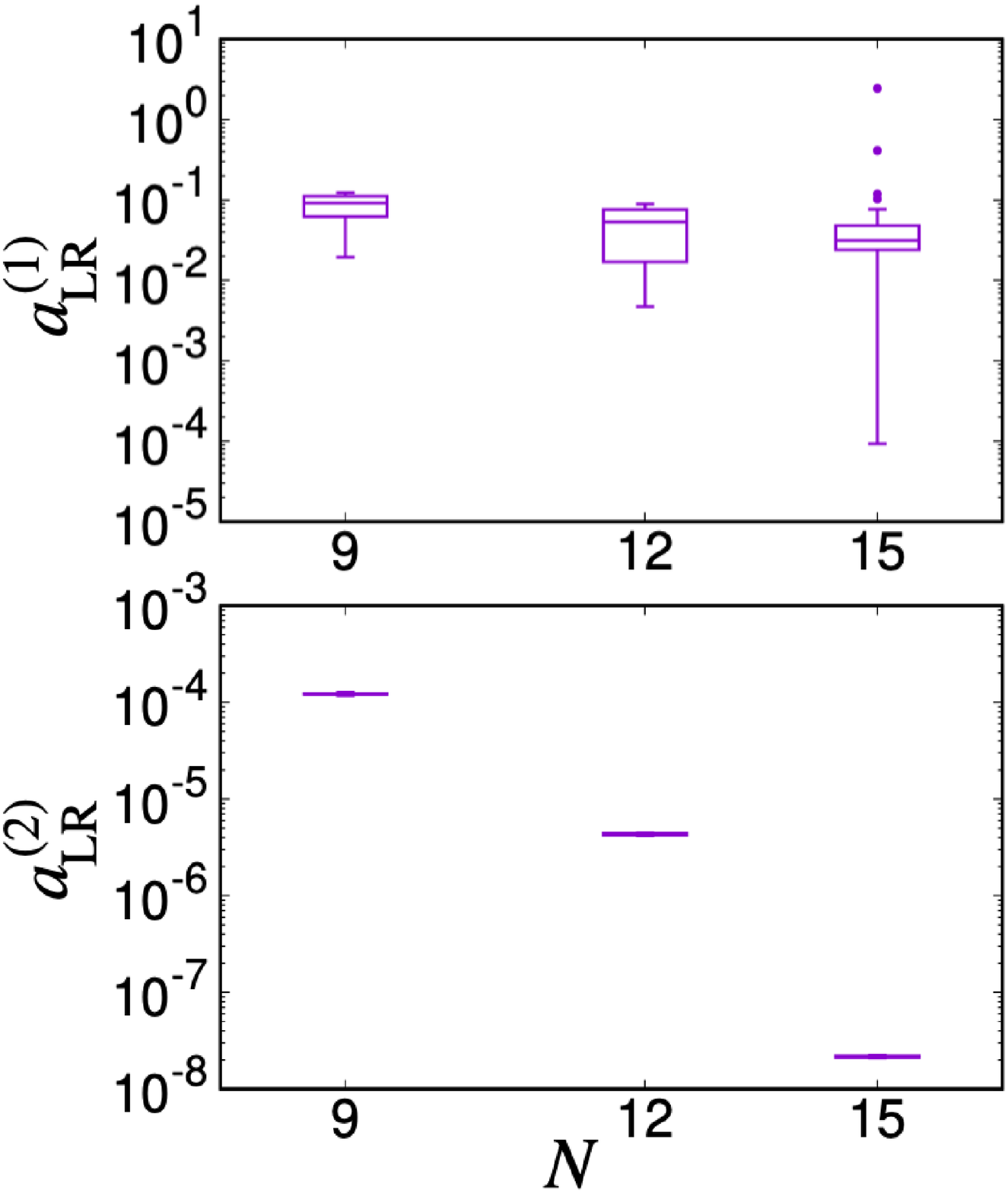}
\end{center}
\caption{
The bath size dependence of $a_\LR^{(1)}$ and $a_\LR^{(2)}$, which relate to the Lieb-Robinson bound.
Each data in the boxplot corresponds to each initial eigenstate of bath B in the energy shell $[E-\Delta E,E]$.
Parameters: $p=0.99,g=0.1\gamma$,$\gamma^\prime=\gamma$,$\beta=0.1$.
Both $a_\LR^{(1)}$ and $a_\LR^{(2)}$ tend to decrease as $N$ increases.
For $a_\LR^{(1)}$ at $N=15$, we note that the vertical range of the graph is widened because there are outliers on the smaller error side. 
The initial state dependence is small for $a_\LR^{(2)}$, implying that the change of $a_\LR^{(2)}$ with respect to $\beta$ is small when the initial state of bath B is the canonical ensemble.
}
\label{fig:aLR}
\end{figure}
We next show our numerical results on the errors related to the Lieb-Robinson bound.
We fit the median of $|\dG_\LR^{(1)}|$ against the function $a_\LR^{(1)}  t^{b_\LR^{(1)}}$.
Because $\dG_\LR^{(2)}$ cannot be fitted by the fitting function $a_\LR^{(2)}  t^{b_\LR^{(2)}}$, we numerically find $a_\LR^{(2)}  t^{b_\LR^{(2)}}$ only satisfying
$|\dG_\LR^{(2)}|\leq a_\LR^{(2)}t^{b_\LR^{(2)}}$ in $10^{-3}<\gamma t<10^{-2}$.
As a result, we obtain $b_\LR^{(1,2)}=4$, which is different from $b=2$ in Eq.~(\ref{eq:LRB:dGLR}).
Figure~\ref{fig:aLR} is a boxplot showing the dependence of $a_\LR^{(1)}$ and $a_\LR^{(2)}$ on the bath size and the initial state.
Both $a_\LR^{(1)}$ and $a_\LR^{(2)}$ tend to decrease as $N$ increases.

\begin{figure}[t]
\begin{center}
\includegraphics[width=0.8\linewidth]{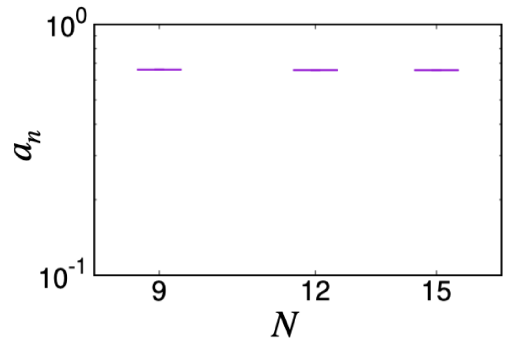}
\end{center}
\caption{
The bath size dependence of $a_n$.
Each data in the boxplot corresponds to each initial eigenstate of bath B in the energy shell $[E-\Delta E,E]$.
We note that the boxes are squashed because the initial-state dependence of $a_n$ is quite small.
Parameters : $p=0.99,g=0.1\gamma$, $\gamma^\prime=\gamma$, $\beta=0.1$.
$a_n$ is independent of the bath size $N$, which is reasonable because the size of system S does not depend on the bath size $N$.
}
\label{fig:an}
\end{figure}
Finally, we compare the time dependence of the errors of the fluctuation theorem with that of an ordinary observable of system S.
As mentioned in Sec.~\ref{sec:SOR}, the change of the occupation number in system S~(\ref{eq:dnS}) initially rises in the form of $a_n t^2$, which is apparently similar to the case of the fluctuation theorem~(\ref{eq:total_t2}).
Figure~\ref{fig:an} shows the dependence of $a_n$ on the bath size and the initial state of B.
We observe that $a_n$ does not depend on the bath size, i.e., $a_n=\Theta(1)$.
This bath size independence is reasonable because there is no physical mechanism for the change of the occupation number in system S to be suppressed in the thermodynamic limit.
This result $a_n=\Theta(1)$ is contrastive to $a_\ETH=o(1)$ and $a_\LR=o(1)$ about the fluctuation theorem.
Then, while the initial rise of the error of the fluctuation theorem is also proportionate to $t^2$, this is due to a different physical mechanism from the case of ordinary observables such as the occupation number. 
Thus, the above numerical result and discussion again support the scenario that the fluctuation theorem in the short-time regime holds with the nontrivial thermal mechanism based on the Lieb-Robinson bound and the ETH.

\section{Summary and discussion}
\label{sec:Discussion}

In this study, we have investigated the fluctuation theorem in the long and short-time regimes, when the initial state of bath B is an energy eigenstate and the time evolution of the total system is unitary. 
Our results theoretically and numerically show that the fluctuation theorem holds in the entire time domain.

In the long-time regime, we have considered the long-time average of $\langle e^{-\sigma}\rangle$. 
We theoretically showed that the error of the fluctuation theorem introduced in Eq.~(\ref{eq:dGS_def}) vanishes in the thermodynamic limit of the heat bath ~(Sec.~\ref{sec:LT}). 
The main assumptions used are the diagonal and off-diagonal ETH~(\ref{eq:diag_ETH_Setup})(\ref{eq:offdiag_ETH_Setup}).
We have also shown that the interaction-induced error (\ref{eq:dGI_def_4Gs}) decreases with increasing the bath size by using the rotating wave approximation (\ref{eq:RWA}).
We numerically investigated the dependence of the error of the fluctuation theorem on the bath size and the initial state.
Figures \ref{fig:dGS_Ndep} and \ref{fig:a_dGS_gpdep} support the above theory.

We remark that the fluctuation theorem in the long-time regime can also be derived from other assumptions than ours.
For example, Ref.~\cite{HevelingPC} is based on some natural assumptions 
on transition probabilities.
While those assumptions are not equivalent to our assumptions such as the ETH of $H$ discussed in Sec.~\ref{sec:LT}, their result and ours are both correct and would play complementary roles.

In the short-time regime, the fluctuation theorem has been theoretically shown in Ref.~\cite{Iyoda2017} on the basis of the ETH and the Lieb-Robinson bound. 
We performed systematic numerical calculations to confirm the validity of the fluctuation theorem in the short-time regime~(Sec.~\ref{sec:ET}). 
In particular, we focus on the dependence of the errors of the fluctuation theorem on the bath size, which establishes that the validity of the fluctuation theorem is due to the theoretically-proposed scenario based on the ETH and the Lieb-Robinson bound~\cite{Iyoda2017}, rather than a trivial scenario argued in the last paragraph of Sec.~\ref{sec:ET}.
Figures~\ref{fig:aETH}, \ref{fig:aLR}, and \ref{fig:an} show that our numerical results support this theoretical scenario.

The two time regimes play key roles in this study: the long and short-time regimes.
The long-time regime is defined by that $\langle e^{-\sigma}\rangle$ nearly equals the long-time average $\overline{\langle e^{-\sigma}\rangle}$, which is independent of the bath size as discussed in Sec.~\ref{sec:SOR:long}.
The short-time regime is defined by that the system information does not reach the far part of bath B, which becomes longer as the bath size increases. 
Therefore, the long and short-time regimes overlap and cover the entire time domain when the bath size is large enough. 
We again emphasize, however, that the fluctuation theorem has been shown independently in these time regimes.

We remark that our result highlights the connection between information and thermodynamics~\cite{SagawaProg,EspositoPRX,Parrondo2015}. The informational entropy (the von Neumann entropy) and the thermodynamic quantity (heat) are quantitatively connected to each other in the second law of thermodynamics~(\ref{eq:2ndlaw}), as historically demonstrated by the Szilard engine~\cite{Szilard1929} and the Landauer principle~\cite{Landauer1961}.
While in the conventional theory~\cite{Sagawa2012} this connection between the informational entropy and heat relies on the assumption that the initial state of the bath is canonical and thus has the maximum entropy, the present work shows that the same connection emerges even when the initial state of the bath is a single energy eigenstate.
Therefore, our result serves as a theoretical foundation of thermodynamics of information beyond the conventional canonical setup.

Meanwhile, in our numerical calculation, we used a specific model of the two-dimensional system introduced in Sec.~\ref{sec:setup:Hamiltonian}. 
However, 
our theory ensures that the fluctuation theorem should hold for a much broader class of models that satisfy the ETH (Table I). 
Since it has been numerically confirmed that the ETH holds for various non-integrable systems~\cite{Rigol2008,Biroli2010,Steingeweg2013,Kim2014,Beugeling2014,Beugeling2015,Fratus2016,Mondaini2016,Mondaini2017,Yoshizawa2018,Garrison2018,Dymarsky2018,Khaymovich2019,Brenes2020}, the fluctuation theorem holds for these systems.

We consider that our theory can be  experimentally verified. 
In particular, the dynamics of isolated non-integrable many-body systems have been investigated using artificial quantum systems such as cold atoms~\cite{Trotzky2012,Kaufman2016,Gross2017,Parsons2015,Mello2019,Edabi2020}, trapped ions~\cite{Clos2016}, and superconducting qubits~\cite{Neill2016}.
Not only local physical quantities but also informational entropy are experimentally measurable~\cite{Kaufman2016,Neill2016}.
Because our numerical calculation is performed with a small bath size of $15$ sites using numerically exact diagonalization, we cannot observe the separation between the two time scales, the Lieb-Robinson time $\tau_\LR$ and the relaxation time $\tau_\mathrm{relax}$, as mentioned in Sec.~\ref{sec:ET}.
However, experiments with around $100$ to $400$ sites are currently accessible~\cite{Parsons2015,Mello2019,Edabi2020}. 
Therefore, the separation of the above time scales would be experimentally observable.
It is an interesting future issue to directly verify the theory of the long and short-time regimes by such real experiments, which would open up the experimental investigation of the emergence of thermodynamics from quantum mechanics.


\textbf{Note added.}
We have been informed by J. Gemmer \textit{et al.} about their results~\cite{HevelingPC}, which are closely related to the present work and appeared on arXiv on the same day as this manuscript.
We recommend reading their submission.

\textbf{Acknowledgment.}
The authors are grateful to 
Krzysztof Ptaszy\'nski,
Takashi Mori, Jochen Gemmer,
and
Naoaki Kato
for valuable discussions.
E.I. and T.S. are supported by JSPS KAKENHI Grant No. JP16H02211. E.I. is supported by JSPS KAKENHI Grant No. JP19K14609. T.S. is supported by JSPS KAKENHI Grant No. JP19H05796. T.S. is also supported by Institute of AI and Beyond of the University of Tokyo.

\appendix

\section{Absolute irreversibility}
\label{sec:app:AI}

Absolute irreversibility means a violation of the fluctuation theorem, which occurs when the initial state of system S is singular in the sense that it does not have the full support in the Hilbert space (e.g., the initial state is pure)~\cite{Murashita2014,Funo2015}. 
This is because the final state of the reverse process is not necessarily in the support of the initial state of the forward process.
The violation of the fluctuation theorem can be described by a correction term as described later.
In this Appendix, we consider the generalized fluctuation theorem including the correction term in the presence of absolute irreversibility.

First, we consider the conventional case where the initial state of bath B is the canonical ensemble. 
We denote the projection operator onto the subspace which supports $\rho_\mathrm{S}(0)$ by $P_{\mathrm{ini}}^\rS$. 
When $D_\rS>\mathrm{rank}[P_\mathrm{ini}^\rS]$, the fluctuation theorem is accompanied by the correction term due to absolute irreversibility. 
In particular, if $\rho_\mathrm{S}(0)$ is pure, $\mathrm{rank}[P_\mathrm{ini}^\rS]=1<D_\rS$ holds. 
Then, it has been shown in Ref.~\cite{Murashita2014,Funo2015} that
\begin{align}
\label{eq:IFT:AI:can}
\langle e^{-\sigma}\rangle &= 1-\lambda^\can(t),
\\
\label{eq:IFT:AI:lambda:can}
\lambda^\can(t)
&:=\mathrm{tr}[
(1-P_{\mathrm{ini}}^\rS)
U^\dag \rho^\can_\mathrm{R}(0) U
],
\end{align}
where $\rho^\can_\mathrm{R}(0):=\rho^\can_\mathrm{S}(t)\otimes \rho^\can_\mathrm{B}$ is the initial state of the reverse process.
We note that one can measure $\langle e^{-\sigma}\rangle$ and $\lambda^\can(t)$ in the forward and the reverse processes independently.
In fact, Eq. (77) has been confirmed experimentally using projection measurements on the system~\cite{Masuyama2018}.

The correction term (\ref{eq:IFT:AI:lambda:can}) does not appear if the singularity is removed by regularizing the initial state of system S as follows:
\begin{align}
    \label{eq:rhoSini_epsilong}
    \rho_\rS^\mathrm{reg}(0)
    :=&
    \frac{\tilde{\rho}_\rS(0)}{
    \tr_\rS[\tilde{\rho}_\rS(0)]},
    \\
    \tilde{\rho}_\rS(0)
    :=&
    \rho_\rS(0)
    +
    \varepsilon (1-P_\mathrm{ini}^\rS)
    ~~~(\varepsilon>0),
\end{align}
which is essentially the same approach as adopted in Ref.~\cite{HevelingPC} (see also Ref.~\cite{Sagawa2012Reg}).
If we take the limit of $\varepsilon\rightarrow 0$ after calculating $\langle e^{-\sigma}\rangle$, then $\langle e^{-\sigma}\rangle=1$ holds.
On the other hand, if we take the limit $\varepsilon\rightarrow 0$ first and then calculate $\langle e^{-\sigma}\rangle$, Eq.~(\ref{eq:IFT:AI:can}) holds.
This is not just the problem of the (unphysical) order of the limits.
As mentioned above, however, Eq.~(\ref{eq:IFT:AI:can}) has been experimentally observed, and thus the effect of absolute irreversibility is physically relevant.
On the other hand, in the case where small noise is unavoidable in the initial state, the regularization approach would be relevant.
Therefore, these two approaches are both physically reasonable and play complementary roles.

We next consider the fluctuation theorem with the energy eigenstate bath in the long-time regime in the presence of absolute irreversibility. 
We define the correction term as
\begin{align}
\lambda(t)
&:=\mathrm{tr}[
(1-
P_{\mathrm{ini}}^\rS
)
U^\dag
\rho_\mathrm{R}(0)
U
],
\end{align}
where $\rho_\mathrm{R}(0):=\rho_\mathrm{S}(t)\otimes \rho_\mathrm{B}(0)$. 
We will prove in Appendix~\ref{sec:app:proof} that 
\begin{align}
\label{eq:IFT:AI:ee}
    \langle e^{-\sigma}\rangle =
    1-\lambda(t)
\end{align}
holds in the thermodynamic limit.
Because we have already discussed the long-time average of the left-hand side in Sec.~\ref{sec:LT:proof}, we only need to show
\begin{align}
|\overline{\lambda^\can(t)}-\overline{\lambda(t)}|=o(1).
\label{eq:app:AI:result}
\end{align}
We will prove Eq.~(\ref{eq:app:AI:result}) by using the ETH of the total system SB in Appendix~\ref{sec:app:proof}.

\begin{figure}[t]
\begin{center}
\includegraphics[width=\linewidth]{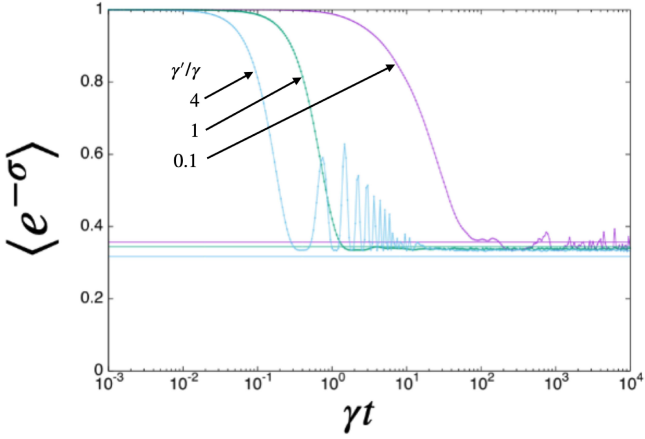}
\end{center}
\caption{
The time dependence of $\langle e^{-\sigma}\rangle$ when the initial state of system S is pure ($p=1$).
Parameters: $g=0.1\gamma$, $\gamma^\prime=0.1\gamma$~(purple), $\gamma$~(green), $4\gamma$~(blue).
The initial state of bath B is an energy eigenstates of $H_\rB$, whose energy is maximum in the energy shell at $\beta=0.1$.
The onsite potential $\omega$ is determined by $\tr_\rB[n_\rB\rho_\rB^\can]=N_\rP$.
The horizontal lines represent the numerically obtained values of $1-\overline{\lambda(t)}$.
}
\label{fig:app:AI}
\end{figure}
We here show the validity of Eq.~(\ref{eq:IFT:AI:ee}) by numerical simulation.
Figure~\ref{fig:app:AI} shows the time dependence of $\langle e^{-\sigma}\rangle $ when the initial state of system S is a pure state~($p=1$). 
In the long-time regime, $\langle e^{-\sigma}\rangle $ deviates from $1$ because absolute irreversibility occurs.
The convergent value nearly equals $1-\overline{\lambda(t)}$, implying the fluctuation theorem with the correction term (\ref{eq:IFT:AI:ee}) holds. 
We note that in the numerics of Ref.~\cite{Iyoda2017} ($L_x=L_y=4$, $N_\rP=4$), we confirm that Eq.~(\ref{eq:IFT:AI:ee}) holds by performing the same analysis as in this paper.

\section{Supplement on the interaction-induced error}
\label{sec:app:error}

In this Appendix, we discuss the form of the interaction-induced error.
Specifically, we compare $\dG_\rI$ defined in Eq.~(\ref{eq:dGI_def_4Gs}) and the interaction-induced error defined in Ref.~\cite{Iyoda2017}.
We show that $\dG_\rI$ reflects how thermal the initial state of bath B is.

First, we consider the characteristic function of the entropy production:
\begin{align}
\nonumber
\mathcal{G}(u;\rho_\rB)
:=
\tr[
&
U
e^{-iu\beta H_{\rB}}
e^{iu\ln \rho_\rS(0)}
\\
&
\rho_\rS(0)\otimes
\rho_\rB
U^\dag
e^{-iu\ln\rho_\rS(t)}
e^{iu\beta H_{\rB}}
],
\label{eq:GG:def}
\end{align}
which is the Fourier transformation of the probability distribution of the stochastic entropy production $\sigma$. We note that $\langle e^{-\sigma}\rangle
    =
    \mathcal{G}(i;\rho_\rB)$ holds.
Let us next consider the following function by replacing $H_\rB$ by $-H_\rS$, as in $G_S$ defined in Eq.~(\ref{eq:GS:def}):
\begin{align}
\nonumber
\mathcal{G}_\rS(u;\rho_\rB)
:=
\tr[
&
U
e^{iu\beta H_{\rS}}e^{iu\ln \rho_\rS(0)}
\\
&
\rho_\rS(0)\otimes
\rho_\rB
U^\dag
e^{-iu\ln\rho_\rS(t)}
e^{-iu\beta H_{\rS}}
].
\label{eq:GGS:def}
\end{align}
The interaction-induced error comes from the difference between Eqs.~(\ref{eq:GG:def}) and (\ref{eq:GGS:def}).  By  noting that the fluctuation theorem holds if the initial state of B is canonical, we define the interaction-induced error as
\begin{align}
\label{eq:App:dGI_def}
\dG_\rI:=&
\mathcal{G}(i;\rho_\rB(0))-\mathcal{G}_\rS(i;\rho_\rB(0))
\nonumber
\\
+&
\mathcal{G}_\rS(i;\rho_\rB^\can)-\mathcal{G}(i;\rho_\rB^\can).
\end{align}
By definition, $\dG_\rI=0$ holds when the initial state of bath B is the canonical ensemble ($\rho_\rB(0)=\rho_\rB^\can$).

As an alternative definition, in the supplemental information of Ref.~\cite{Iyoda2017}, the interaction-induced error is introduced using the operator norm as
\begin{align}
\delta_\rI(u)
&:=
\|
e^{iu\beta H_\rB} U e^{-iu\beta H_\rB}
-
e^{-iu\beta H_\rS}
U
e^{iu\beta H_\rS}
\|.
\end{align}
Using $\delta_\rI(u)$, the error of the fluctuation theorem is evaluated as
\begin{align}
\label{eq:App:dGI:supp1}
|
\mathcal{G}(u;\rho_\rB(0))
-
\mathcal{G}_\rS(i;\rho_\rB(0))
|
\leq \delta_\rI(u) C(u),
\end{align}
where $C(u)$ is an $N$-independent constant and written as
\begin{align}
\label{eq:App:dGI:supp2}
C(u)
=
\begin{cases}
 (p_b(0))^{-|u_\rI-1|}>1 & (1< u_\rI) \\
1 & (0\leq u_\rI\leq 1) \\    
 (p_b(t))^{-|u_\rI|}>1 & (u_\rI< 0), 
\end{cases}
\end{align}
where $p_b(t)$ is the minimum eigenvalue of $\rho_\rS(t)$ and $u_\rI:=\mathrm{Im}[u]$.
In the supplemental information of Ref.~\cite{Iyoda2017}, 
$C(u)$ in the right-hand side of (\ref{eq:App:dGI:supp1}) was considered to be $1$, while the
modification as above is required for $1<u_\rI$ and $u_\rI<0$.
We note that only the case of $u_\rI=1$ was used in the proof of the fluctuation theorem of the form~(\ref{eq:IFT}) in Ref.~\cite{Iyoda2017}, and therefore the proof of it is correct as it is.

In the above evaluation, $\delta_\rI(u)$ is independent of the initial state of bath B. 
On the other hand, $\dG_\rI=0$ holds when the initial state of bath B is the canonical ensemble. 
Therefore, $\dG_\rI$ is a more proper measure of the error, because it can reflect how thermal the initial state of bath B is. 

\section{Rotating wave approximation and the interaction-induced error}
\label{sec:app:RWA}

In this Appendix, we show that Eqs.~(\ref{eq:com_int_norm}) and (\ref{eq:dGI_RWA}) hold under the rotating wave approximation and the off-diagonal ETH.
We first introduce the rotating wave approximation  and show Eq.~(\ref{eq:com_int_norm}) in Appendix~\ref{sec:app:RWA1}.
Then, we show Eq.~(\ref{eq:dGI_RWA}) in Appendix~\ref{sec:app:RWA2}.

\subsection{Rotating wave approximation and the proof of Eq.~(\ref{eq:com_int_norm})}
\label{sec:app:RWA1}

We write the interaction Hamiltonian as
\begin{align}
H_\rI=
\sum_{ab}
(H_\rI)_{ab}
 | E_a\rangle 
 \langle E_b |.
\end{align}
Let $\Omega$ be the cutoff frequency for the rotating wave approximation, which is assumed to be independent of the system size.
We then introduce the rotating wave approximation of $H_\rI$ as
\begin{align}
\label{eq:RWA}
\tilde{H}_\rI
=
{\sum_{ab}}'
(H_\rI)_{ab}
 | E_a\rangle 
 \langle E_b |,
\end{align}
where the sum is taken over $(a,b)$ satisfying $|E_a-E_b|<\Omega$.
The rotating wave approximation holds in a long-time regime with $t\gg \Omega^{-1}$.

To show Eq.~(\ref{eq:com_int_norm}), we assume the off-diagonal ETH in the following form~\cite{Srednicki1994,Mondaini2017}:
\begin{align}
\label{eq:offdiag_ETH_random}
(\tilde{H}_\rI)_{ab}
=
\frac{1}{\sqrt{D^\prime}}
g_{ab}r_{ab},
\end{align}
where $g_{ab}$ characterizes the dependence of $(\tilde{H}_\rI)_{ab}$ on $E_a$ and $E_b$, and $r_{ab}$ is a random variable whose mean is $0$ and variance is $1$.
We note that 
Eq.~(\ref{eq:offdiag_ETH_random}) is stronger than the off-diagonal ETH in the form of Eq.~(\ref{eq:offdiag_ETH_Setup}).
We use Eq.~(\ref{eq:offdiag_ETH_random}) only in this Appendix.

Under the rotating wave approximation and the off-diagonal ETH,
the commutator in Eq.~(\ref{eq:com_int_norm}) is written as
\begin{align}
\label{eq:com_int_RWA_ETH}
[H_\rS+H_\rB, \tilde{H}_\rI]
=
{\sum_{ab}}'
|E_a\rangle\langle E_b|
\frac{R_{ab}}{\sqrt{D^\prime}},
\end{align}
where $R_{ab}:=(E_a-E_b)g_{ab}r_{ab}$. We note that $R_{ab}$ is the matrix elements of the band random matrix and the variance of $R_{ab}$ is less than $\Omega^2 \|\tilde{H}_\rI\|^2$.
From the Wigner semicircle law,
the maximum eigenvalue of a random matrix, whose dimension is $D$ and variance of the matrix elements is $\sigma^2$, is about  $2\sqrt{D}\sigma$.
Similarly, the maximum eigenvalue of  a band random matrix with band width $W$ is evaluated to be $2\sqrt{W}\sigma$, which is mathematically shown for some band random matrix~\cite{Sodin2010}.
We evaluate the band width of the band random matrix $R$ as $D^\prime \Omega/\|\tilde{H}\|$, where we defined $\tilde{H}:=H_\rS+H_\rB+\tilde{H}_\rI$.
Then, the operator norm of Eq.~(\ref{eq:com_int_RWA_ETH}) is evaluated as
\begin{align}
\|[H_\rS+H_\rB, \tilde{H}_\rI]\|
&\leq 
\frac{c\Omega\|\tilde{H}_\rI\|}
{\sqrt{N}},
\label{eq:RWA_OpNorm}
\end{align}
where we used $\|\tilde{H}\|=\Theta(N)$ and introduced $c:=\Theta(1)$.
Thus, Eq.~(\ref{eq:com_int_norm}) is proved.

In the same manner, we can show that
\begin{align}
\label{eq:RWA_OpNorm_f}
\|
f_n
\|
&\leq
\frac{c}{\sqrt{N}}\|\tilde{H}_\rI\|\Omega^n,
\\
\label{eq:RWA_OpNorm_g}
\|
g^a_n
\|
&\leq
\frac{ca}{\sqrt{N}}\|\tilde{H}_\rI\|^a\Omega^n,
\end{align}
where we inductively define
\begin{align}
f_{n+1}
&:=
[\tilde{H}, f_n],~~f_0=A,
\\
g^a_{n+1}
&:=
[\tilde{H}, g^a_n],~~g^a_0=\tilde{H}_\rI^a,
\\
A
&:=
[\tilde{H}_\rI, \tilde{U}^\dag],
\end{align}
where $\tilde{U}:=\exp(-i\tilde{H}t)$.
We note that Eq.~(\ref{eq:RWA_OpNorm}) is a special case ($a=n=1$) of Eq.~(\ref{eq:RWA_OpNorm_g}).

\subsection{Proof of Eq.~(\ref{eq:dGI_RWA})}
\label{sec:app:RWA2}

In this subsection, we show Eq.~(\ref{eq:dGI_RWA}). 
We first focus on $\delta G^{(1)}_\rI$:
\begin{align}
\delta G^{(1)}_\rI
=
\mathrm{tr}
[
\delta \tilde{U}_\beta
\rho
 \tilde{U}^\dag \Lambda_\rS(t)
],
\end{align}
where 
$\delta \tilde{U}_\beta:=e^{-\beta H_0} \tilde{U} e^{\beta H_0}-\tilde{U}$,
$H_0:=H_\rS+H_\rB$, $\rho:=\rho_\rS^\can\otimes\rho_\rB(0)$,
and $\Lambda_\rS(t):=\rho_\rS(t)(\rho_\rS^\can)^{-1}$.
Using the Cauchy-Schwartz inequality, we bound $(\delta G^{(1)}_\rI)^2$ from above as
\begin{align}
\nonumber
(\delta G^{(1)}_\rI)^2
\leq&
\tr[\tilde{U}\rho \tilde{U}^\dag \Lambda_\rS(t) \Lambda_\rS^\dag(t)]
\tr[
\delta \tilde{U}_\beta
\rho
\delta \tilde{U}_\beta^\dag
].
\end{align}
Since $|\tr[\tilde{U}\rho \tilde{U}^\dag \Lambda_\rS(t) \Lambda_\rS^\dag(t)]|=\Theta(1)$, we focus on 
\begin{align}
\nonumber
\tr[
\delta \tilde{U}_\beta
\rho
\delta \tilde{U}_\beta^\dag
]
=&
\tr[\delta \tilde{U}_{2\beta}^\dag \tilde{U} \rho]
-
2
\tr[\delta \tilde{U}_{\beta}^\dag \tilde{U} \rho]
\\
=:&
\sum_{n=2} \beta^n B_n.
\end{align}
In the first line, we used $[H_0,\rho]=0$.
In the second line, we used the Baker-Campbell-Hausdorff formula respectively for $\delta \tilde{U}_{2\beta}^\dag$ and $\delta \tilde{U}_{\beta}^\dag$.
We will show that $|B_n|=o(1)$ in the following.

We first consider $|B_2|$. We can write $B_2$ as
\begin{align}
\nonumber
B_2
&=
\tr[
[H_0, [H_0, \tilde{U}^\dag]]\tilde{U}\rho]
\\
&=
-
\tr[
(f_1-\tilde{H}_\rI f_0+f_0 \tilde{H}_\rI)
U\rho
].
\end{align}
Then, we obtain
\begin{align}
\nonumber
|B_2|
&\leq
(\|f_1\|+2\|\tilde{H}_\rI\| \|f_0\|
)
\|\tilde{U}\|
\|\rho\|_1
\\
&\leq
\nonumber
\frac{c}{\sqrt{N}}
\|\tilde{H}_\rI\|
(\Omega+\|\tilde{H}_\rI\|)
\\
&=
o(1),
\end{align}
where we used Eq.~(\ref{eq:RWA_OpNorm_f}) in the second line.

We next consider the case of $n\geq 3$.
$B_n$ is written as a linear combination of terms consisting of the product of $f_m, g_m^a, \tilde{H}_\rI$ ($m\leq n$) and each term contains a single $f_n$.
Since Eqs.~(\ref{eq:RWA_OpNorm_f}) and (\ref{eq:RWA_OpNorm_g}), the leading terms with respect to  $N$ do not contain $g_n^a$. Then, from the straight forward calculation, we obtain
\begin{align}
|B_n|
&\leq
\frac{1}{\sqrt{N}}
\frac{\|\tilde{H}_\rI\|}{y}
\frac{(6\beta y)^n}{n!}
+o\left(\frac{1}{\sqrt{N}}\right),
\\
y&:=\max(\Omega,\|\tilde{H}_\rI\|).
\end{align}
Summing $\beta^n B_n$ over $n$, we obtain
\begin{align}
\left|
\sum_{n=2}\beta^n B_n
\right|
\leq
\frac{\|\tilde{H}_\rI\|}{y}
\left(
e^{6\beta y}-6\beta y-1
\right)
\frac{c}{\sqrt{N}}
+o\left(\frac{1}{\sqrt{N}}\right).
\end{align}

From the foregoing arguments, $|\delta G_\rI^{(1)}|=o(1)$ is shown. In the same manner, $|\delta G_\rI^{(2)}|=o(1)$ is also shown. Thus, Eq.~(\ref{eq:dGI_RWA}) is shown under the rotating wave approximation (\ref{eq:RWA}) and the off-diagonal ETH (\ref{eq:offdiag_ETH_random}).
Furthermore, Eq.~(\ref{eq:LT:dGI_expectation}) is  shown since $\Omega=\Theta(1)$.

\section{Naive approach to evaluate the error of the fluctuation theorem}
\label{sec:app:AnotherAttempt}

In this Appendix, we discuss another naive approach to evaluate the error of the fluctuation without Eq.~(\ref{eq:decompose}).
In this approach, however, we cannot show that the error of the fluctuation theorem vanishes in the thermodynamic limit.

We note that $G-G^\can$ can be written as the difference between the expectation values of $O_\rB(t)$, which is non-local for $t>0$, as follows:
\begin{align}
\label{eq:dG_nonlocal}
    G-G^\can
=&
\mathrm{tr}_\mathrm{B}
\left[
O_\mathrm{B}(t)
(\rho_\mathrm{B}-\rho_\mathrm{B}^\mathrm{MC})
\right],
\nonumber
\\
+&
\mathrm{tr}_\mathrm{B}
\left[
O_\mathrm{B}(t)
(\rho_\mathrm{B}^\mathrm{MC}-\rho_\mathrm{B}^\mathrm{can})
\right]
\\
O_\mathrm{B}(t)
:=&
Z_\mathrm{B}
\mathrm{tr}_\mathrm{S}
\left[
U^\dag \rho_\mathrm{S}(t)\otimes \rho_\mathrm{B}^\mathrm{can}
U
\right]
e^{\beta H_\mathrm{B}}.
\end{align}
The first and second terms on the right-hand side of Eq.~(\ref{eq:dG_nonlocal}) are regarded as the errors associated with the ETH and the equivalence of ensembles for $O_\rB(t)$, respectively. Below, we discuss them in detail.

The equivalence of ensembles has been theoretically shown for (quasi-)local operators, whose support is at most $N^{1/2}$ and operator norm does not increase with respect to $N$~\cite{Tasaki2018}. The error of the equivalence of ensembles only polynomially decreases with respect to $N$. If we assume that the equivalence of ensembles in the same form as \cite{Tasaki2018} holds for $O_\rB (t)$,  we cannot show the error of the equivalence of ensembles vanishes in the thermodynamic limit due to the exponential increase of the norm of $O_\rB (t)$.

The ETH for highly non-local operators has been investigated theoretically~\cite{Hamazaki2018} and numerically~\cite{Kaneko2019}. In Ref.~\cite{Hamazaki2018}, the ETH is theoretically shown to hold with the error decreasing as $1/\sqrt{D}$ for typical many-body operators, whose operator norm does not grow exponentially. In Ref.~\cite{Kaneko2019}, the same scaling as~\cite{Hamazaki2018} was numerically observed. However, the operator norm of $O_\rB (t)$ exponentially increases and is bounded by  $D\sqrt{N}$~\cite{Tasaki2018} (see also Eq.~(\ref{eq:ZB_EOE})). Then, the error associated with the ETH is bounded by $\sqrt{DN}$ and we cannot show that the error of the fluctuation theorem associated with the ETH vanishes in the thermodynamic limit. 

For the above reasons, instead of the direct evaluation of $G-G_\can$, in the main text we have adopted the decomposition (\ref{eq:decompose}) and show the fluctuation theorem based on plausible assumptions such as the ETH and the equivalence of ensembles for (quasi-)local physical quantities.

\section{Initial rise of the error of the fluctuation theorem}
\label{sec:app:InitialRise}

In this Appendix, we show that the errors of the fluctuation theorem initially rises in proportionate to $t^2$ as mentioned in Secs.~\ref{sec:SOR} and \ref{sec:ET}.

The error of the fluctuation theorem $\langle e^{-\sigma}\rangle-1$ initially rises in proportionate to $t^2$, which is shown as
\begin{align}
    \nonumber
    &\langle e^{-\sigma}\rangle-1
    \\
    =&
    \tr[e^{-\beta H_\rB}U e^{\beta H_\rB}\rho_\rB(0) U^\dag \rho_\rS(t)]-1
    \nonumber
    \\
    =&
    \tr\left[
    e^{-\beta H_\rB}
    \left(-\frac{itH}{\hbar}\right)
    e^{\beta H_\rB} 
    \rho_\rS(t)\otimes \rho_\rB(0)\right]
    \nonumber
    \\
    +&
    \tr\left[
    \rho_\rB(0) 
    \left(\frac{itH}{\hbar}\right)
    \rho_\rS(t)\right]+\mathcal{O}(t^2)
    \nonumber
    \\
    =&
    \tr\left[
    \frac{-itH}{\hbar}
    \left(
    e^{\beta H_\rB}
    \rho_t
    e^{-\beta H_\rB}
    -
    \rho_t
    \right)
    \right]
    +\mathcal{O}(t^2)
    \nonumber
    \\
    =&
    \mathcal{O}(t^2),
    \label{eq:total_t2}
\end{align}
where we defined $\rho_t:=\rho_\rS(t)\otimes \rho_\rB(0)$.
In the second line, we expand the time evolution operator with respect to $t$.
In the last line, we used $[H_\rB,\rho(0)]=0$.
We note that Eq.~(\ref{eq:com_int}) is not assumed here.

Similarly, by using $[H_\rS,\rho(0)]=0$ in addition to $[H_\rB,\rho(0)]=0$, we show that $\dG_\rS$ and $\dG_\rI$ initially rise in $t^2$. In fact,
\begin{align}
    &
    \dG_\rS
    \nonumber
    \\
    =&
    \tr[e^{\beta H_\rS}U 
    e^{-\beta H_\rS}\rho_\rB(0) U^\dag \rho_\rS(t)]
    \nonumber
    \\
    -&\tr[e^{\beta H_\rS}U e^{-\beta H_\rS}\rho_\rB^\can U^\dag \rho_\rS^\can(t)]
    \nonumber
    \\
    =&
    \tr\left[
    \left(-\frac{itH}{\hbar}\right)
    (
    e^{-\beta H_\rS}
    \rho_0
    e^{\beta H_\rS}
    -\rho_0
    )
    \right]
    \nonumber
    \\
    -&
    \tr\left[
    \left(-\frac{itH}{\hbar}\right)
    (
    e^{-\beta H_\rS}
    \rho_0^\can
    e^{\beta H_\rS}
    -\rho_0^\can
    )
    \right]
    +
    \mathcal{O}(t^2)
    \nonumber
    \\
    =&
    \mathcal{O}(t^2),
    \label{eq:dGS_t2}
\end{align}
where $\rho_t^\can:=\rho_\rS(t)\otimes \rho_\rB^\can$.
Next, the interaction-induced error $\dG_\rI$ is decomposed into $\dG_\rI^{(1)}$ and $\dG_\rI^{(2)}$ as in Eq.~(\ref{eq:dGI_def_dGI1_dGI2}).
Then,
\begin{align}
    &
    \dG_\rI^{(1)}
    \nonumber
    \\
    =&
    \tr[e^{-\beta H_\rB}U 
    e^{\beta H_\rB}\rho_\rB(0) U^\dag \rho_\rS(t)]
    \nonumber
    \\
    -&\tr[e^{\beta H_\rS}U e^{-\beta H_\rS}\rho_\rB(0) U^\dag \rho_\rS(t)]
    \nonumber
    \\
    =&
    \tr\left[
    \left(-\frac{itH}{\hbar}\right)
    (
    e^{\beta H_\rB}
    \rho_t
    e^{-\beta H_\rB}
    -\rho_t
    )
    \right]
    \nonumber
    \\
    -&
    \tr\left[
    \left(-\frac{itH}{\hbar}\right)
    (
    e^{-\beta H_\rS}
    \rho_t
    e^{\beta H_\rS}
    -\rho_t
    )
    \right]
    +
    \mathcal{O}(t^2)
    \nonumber
    \\
    =&
    \mathcal{O}(t^2),
    \label{eq:dGI_t2}
\end{align}
where we used the fact that $\rho_0$ and $\rho_0^\can$ commute with both of $H_\rS$ and $H_\rB$.
In the same manner as in Eq.~(\ref{eq:dGI_t2}), it is shown that the initial rise of $\dG_\rI^{(2)}$ defined in Eq.~(\ref{eq:dGI2_def}) is also proportionate to $t^2$.

Furthermore, we can show that $\dG_\ETH$ defined in Sec.~\ref{sec:ET} initially rises as $t^2$ by replacing $U$ in Eq.~(\ref{eq:dGS_t2}) by $U_\rT$.
Thus, Eqs.~(\ref{eq:def_aETH}) and (\ref{eq:def_aI}) are confirmed.

\section{Proof in the long-time regime}
\label{sec:app:proof}

This Appendix shows the details of the proof of the fluctuation theorem in the long-time regime, which was discussed in Sec.~\ref{sec:LT}.
We show $|\overline{G_{\mathrm{S1}}}-\overline{G_{\mathrm{S1}}^\can}|=o(1)$~[Eq.~(\ref{eq:IFT_LT_dGS:result:diag})] in Appendix~\ref{sec:app:proof:S1} and $|\overline{G_{\mathrm{S2}}}|=o(1)$ and $|\overline{G_{\mathrm{S2}}^\can}|=o(1)$~[Eqs.~(\ref{eq:IFT_LT_dGS:result:GS2}) and (\ref{eq:IFT_LT_dGS:result:GS2can})] in Appendix~\ref{sec:app:proof:S2}.
In Appendix~\ref{sec:app:proof:AI}, we show that $|\overline{\lambda(t)}-\overline{\lambda^\can(t)}|=o(1)$~[Eq.~(\ref{eq:app:AI:result})] holds, which complements the discussion about absolute irreversibility discussed in Appendix~\ref{sec:app:AI}.
Finally, we show that the temporal fluctuation of the error of the fluctuation theorem in the long-time regime vanishes in the thermodynamic limit in Appendix~\ref{sec:app:proof:TempFluc}.

\subsection{Proof of Eq.~(\ref{eq:IFT_LT_dGS:result:diag}) in Sec.~\ref{sec:LT}}
\label{sec:app:proof:S1}

In this subsection, we show that $|\overline{G_{\mathrm{S1}}}-\overline{G_{\mathrm{S1}}^\can}|=o(1)$~[Eq.~(\ref{eq:IFT_LT_dGS:result:diag})] holds.
The assumption used here is the diagonal ETH of $H$ for the observable of system S~[Eq.~(\ref{eq:diag_ETH_Setup})].
We also assume Eqs.~(\ref{eq:assump_sh1}) and (\ref{eq:assump_sh2}) below, which state that the contribution from the outside of the energy shell is negligible in the thermodynamic limit.

We first define the energy shell of $H$ as $[E^\prime-\Delta,E^\prime+\Delta]$, where
$E^\prime=\tr[H \rho^\DE]$ and $\Delta=\Theta(N^a)~(1/2<a<1)$.
We denote by $\sum_{a\in\sh}$ the sum over $a$ such that $E^\prime-\Delta\leq E_a\leq E^\prime+\Delta$.
For the energy shell, we define  $\rho^{\DE\mhyphen\sh}:=\sum_{a\in\sh} \pi_a \rho^\DE\pi_a$  and $\rho^{\can,\DE\mhyphen\sh}:=\sum_{a\in\sh} \pi_a \rho^{\can,\DE}\pi_a$, where $\rho^{\DE\mhyphen\sh}$ and $\rho^{\can,\DE\mhyphen\sh}$ are not normalized.
Also, we define the microcanonical ensemble $\rho_\MC:=\sum_{a\in\sh}\pi_a/D^\prime$, where $D^\prime$ is the dimension of the energy shell.

From Eqs.~(\ref{eq:dGS1B}) and (\ref{eq:dGS2B}), we evaluate
$|\overline{G_{\mathrm{S1}}}-\overline{G_{\mathrm{S1}}^\can}|$ by using the triangle inequality as follows:
\begin{align}
    &
    |\overline{G_\mathrm{S1}}
    -
    \overline{G_\mathrm{S1}^\can}
    |
    \\
    =&
    |\tr[O_\rS\rho^\DE]
    -\tr[O^\can_\rS\rho^{\can,\DE}]|
    \\
    \leq&
    \nonumber
    |
    \tr[O_\rS\rho^\DE]
    -\tr[O_\rS\rho^{\DE\mhyphen\sh}]
    |
    \\
    +&
    \nonumber
    |
    \tr[O_\rS\rho^{\DE\mhyphen\sh}]
    -\tr[O_\rS\rho^{\MC}]
    |
    \\
    +&
    \nonumber
    |
    \tr[O_\rS\rho^\MC]
    -\tr[O_\rS^\can\rho^{\MC}]
    |
    \\
    +&
    \nonumber
    |
    \tr[O_\rS^\can\rho^\MC]
    -\tr[O_\rS^\can\rho^{\can,\DE\mhyphen\sh}]
    |
    \\
    +&
    |
    \tr[O_\rS^\can\rho^{\can,\DE\mhyphen\sh}]
    -\tr[O_\rS^\can\rho^{\can,\DE}]|.
    \label{eq:dGS1C}
\end{align}

The energy widths of $\rho_\rB(0)$ and $\rho_\rB^\can$ are $\Theta(1)$ and $\Theta(N^{1/2})$~\cite{Tasaki2018}, respectively. 
Since they are narrower than $\Delta$ for sufficiently large $N$, the truncation error, which originates when $\rho^{\DE}$ and $\rho^{\can,\DE}$ are restricted to the energy shell, are negligible in the thermodynamic limit~\cite{Kaneko2017}: 
\begin{align}
    \label{eq:assump_sh1}
    \|\rho^{\DE}-\rho^{\DE\mhyphen\sh}\|_1&=o(1),
    \\
    \label{eq:assump_sh2}
    \|\rho^{\can,\DE}-\rho^{\can,\DE\mhyphen\sh}\|_1&=o(1).
\end{align}
From Eqs.~(\ref{eq:assump_sh1}) and (\ref{eq:assump_sh2}), $|1-\tr[\rho^{\DE\mhyphen\sh}]|=o(1)$ and $|1-\tr[\rho^{\can,\DE\mhyphen\sh}]|=o(1)$ hold.
Then, the first and fifth terms on the right-hand side of Eq.~(\ref{eq:dGS1C}) are evaluated as
\begin{align}
    \nonumber
    &
    |
    \tr[O_\rS\rho^\DE]
    -\tr[O_\rS\rho^{\DE\mhyphen\sh}]
    |
    \\
    \leq&
    \|O_\rS\|
    \|\rho^{\DE}-\rho^{\DE\mhyphen\sh}\|_1
    =o(1),
    \\
    \nonumber
    &
    |
    \tr[O^\can_\rS\rho^{\can,\DE\mhyphen\sh}]
    -\tr[O^\can_\rS\rho^{\can,\DE}]
    |
    \\
    \leq&
    \|O_\rS^\can\|
    \|\rho^{\can,\DE\mhyphen\sh}-\rho^{\can,\DE}\|_1
    =o(1).
\end{align}

We next evaluate the second term on the right-hand side of Eq.~(\ref{eq:dGS1C}) by using the ETH of $H$ for $O_\rS$.
We write the spectral decomposition of $\rho^{\DE\mhyphen\sh}$ as $\rho^{\DE\mhyphen\sh}=\sum_{a\in\sh}p^{\DE}_a\pi_a$.
Then, we can evaluate the second term on the right-hand side of Eq.~(\ref{eq:dGS1C}) as follows:
\begin{align}
\nonumber
    &
    |\tr[O_\rS(\rho^{\DE\mhyphen\sh}-\rho^\MC)]|
    \\
    =&
    |
    \sum_{a\in\sh}p^\DE_a (O_\rS)_{aa}-\langle O_\rS\rangle_\MC
    |
    \\
    \leq &
    \max_{a\in\sh}
    |
    (O_\rS)_{aa}-\langle O_\rS\rangle_\MC
    |
    \sum_{a\in\sh} p_a^\DE
    \nonumber
    \\
    &+
    (1-\sum_{a\in\sh} p_a^\DE)
    |\langle O_\rS\rangle_\MC|
    \\
    =&
    o(1).
\end{align}
In the last line, we used the ETH of $H$ for $O_\rS$ and Eq.~(\ref{eq:assump_sh1}).
In the same manner, we show that the fourth term on the right-hand side of Eq.~(\ref{eq:dGS1C}) is $o(1)$ by using the ETH for $O_\rS^\can$.

The third term on the right-hand side of Eq.~(\ref{eq:dGS1C}) can be bounded from above by $\|\rho_\MC\|\|O_\rS-O_\rS^\can\|_1$. Besides, we can evaluate $\|O_\rS-O_\rS^\can\|_1$ as $\|O_\rS-O_\rS^\can\|_1\leq 
Z_\rS\|e^{\beta H_\rS}\|
\|
    \tr_\rB[
    \rho^\DE-\rho^{\can,\DE}
    ]
    \|_1$. 
Then, using the triangle inequality, we obtain
\begin{align}
    \nonumber
    \|
    \tr_\rB[
    \rho^\DE-\rho^{\can,\DE}
    ]
    \|_1
    \leq &
    \|
    \tr_\rB[
    \rho^\DE-\rho^{\DE\mhyphen\sh}
    ]
    \|_1
    \\
    \nonumber
    +&
    \|
    \tr_\rB[
    \rho^{\DE\mhyphen\sh}-\rho^{\MC}
    ]
    \|_1
    \\
    \nonumber
    +&
    \|
    \tr_\rB[
    \rho^\MC-\rho^{\can,\DE\mhyphen\sh}
    ]
    \|_1
    \\
    +&
    \|
    \tr_\rB[
    \rho^{\can,\DE\mhyphen\sh}-\rho^{\can,\DE}
    ]
    \|_1.
    \label{eq:dGS1D}
\end{align}
Using Eqs.~(\ref{eq:assump_sh1}) and (\ref{eq:assump_sh2}), we can show that the first and fourth terms on the right-hand side are both $o(1)$.
From the ETH for any operator of system S, we show that the second and third terms on the right-hand side are also $o(1)$.

By summing up the foregoing arguments, $|
\overline{G_{\mathrm{S1}}}
-
\overline{G_{\mathrm{S1}}^\can}
|=o(1)$ is proved.

\subsection{Proof of Eqs.~(\ref{eq:IFT_LT_dGS:result:GS2}) and (\ref{eq:IFT_LT_dGS:result:GS2can}) in Sec.~\ref{sec:LT}}
\label{sec:app:proof:S2}

This Appendix shows that Eqs.~(\ref{eq:IFT_LT_dGS:result:GS2}) and (\ref{eq:IFT_LT_dGS:result:GS2can}) hold by using the off-diagonal ETH for all operators of system S~(\ref{eq:offETH}) and the assumption on the initial energy distribution~(\ref{eq:assump_sh1}) and (\ref{eq:assump_sh2}).

From Eqs.~(\ref{eq:assump_sh1}) and (\ref{eq:assump_sh2}), we can neglect the contribution from the outside of the energy shell and obtain Eq.~(\ref{eq:GS2_limit}) from Eq.~(\ref{eq:dGS2}). Then, along with the off-diagonal ETH~(\ref{eq:offETH}), we show that 
\begin{align}
    |\overline{G_{\mathrm{S2}}}|
    \leq
    \frac{\Theta(1)}{D^\prime}
    \left|
    \sum_{\substack{a,b\in\sh\\a\neq b\\ i,j}}
    e^{\beta E_i^\rS}
    (\rho_1)_{ab}
    (\rho_2)_{ba}
    \right|
\end{align}
holds, where $\rho_1:=\rho_\rS^\can\otimes\rho_\rB(0)$ and $\rho_2:=\rho_\rS(0)\otimes\rho_\rB(0)$. Using the Cauchy-Schwartz inequality, we obtain
\begin{align}
    \nonumber&
    \left|
    \sum_{\substack{a,b\in\sh\\a\neq b}}
    (\rho_1)_{ab}
    (\rho_2)_{ba}
    \right|
    \leq
    \sqrt{
    \tr[\rho_1^2]
    \tr[\rho_2^2]
    }
    \leq
    1.
\end{align}
From the above, 
\begin{align}
    \left|
    \overline{G_{\mathrm{S2}}}
    \right|
    \leq
    \frac{\Theta(1)}{D^\prime}
    \sum_{i,j}
    e^{\beta E_i^\rS}
    =o(1)
\end{align}
is proved.
In the same manner, $\left|\overline{G_\mathrm{S2}^\can}\right|\leq o(1)$ is proved.

\subsection{Proof of Eq.~(\ref{eq:app:AI:result}) in Appendix~\ref{sec:app:AI}}
\label{sec:app:proof:AI}

In this Appendix, we show Eq.~(\ref{eq:app:AI:result})
in Appendix~\ref{sec:app:AI}.
As in Sec.~\ref{sec:LT},
the long-time averages of $\lambda(t)$ and $\lambda^\can(t)$ are written as the sums of the diagonal term and the off-diagonal term:
$\overline{\lambda(t)}=\overline{\lambda_1}+\overline{\lambda_2}$ and 
$\overline{\lambda^\can(t)}=\overline{\lambda^\can_1}+\overline{\lambda^\can_2}$.

First, we evaluate the diagonal term.
We define $Q^\rS:=1-P_\mathrm{ini}^\rS$. Then, $\overline{\lambda_1}$ is written as
\begin{align}
\overline{\lambda_1}
&=
\sum_a
\tr[
Q^\rS
|E_a\rangle\langle E_a|
\rho^\DE_\rS\otimes\rho_\rB(0)
|E_a\rangle\langle E_a|
]
\\
&=
\langle Q^\rS\rangle_\MC
+
\delta\lambda_1,
\\
\delta\lambda_1
&:=
\sum_a
(
(Q^\rS)_{aa}
-
\langle Q^\rS\rangle_\MC
)
(\rho^\DE_\rS\otimes\rho_\rB(0))_{aa},
\end{align}
where $\langle Q^\rS\rangle_\MC$ only depends on the energy shell and does not depend on the microscopic details of the initial state.
On the other hand, $\delta\lambda_1$ can depend on the details.
By neglecting the outside of the energy shell, we obtain
\begin{align}
  \left|\delta \lambda_1\right|
  &\simeq
  \left|
  \sum_{a\in\mathrm{sh}}
    (
    (Q^\rS)_{aa}
    -
    \langle Q^\rS\rangle_\MC
    )
    (\rho^\DE_\rS\otimes\rho_\rB(0))_{aa}
    \right|
  \\
  &\leq
  \max_{a\in\mathrm{sh}} |(Q^\rS)_{aa}-\langle Q^\rS\rangle_\MC|
  \sum_b (\rho^\DE_\rS\otimes\rho_\rB(0))_{bb}
  \\
  &=o(1),
\end{align}
where we used the ETH for $Q^\rS$.
We can show $|\overline{\lambda_1^\can}
-
\langle Q^\rS\rangle_\MC
|=o(1)$ in the same manner.
Thus, $\left|\overline{\lambda_1}-\overline{\lambda_1^\can}\right|=o(1)$ is proved.

The off-diagonal term $\overline{\lambda_2}$ can be evaluated as follows:
\begin{align}
    \overline{\lambda_2}
    &\simeq
    \sum_{\substack{a,b\in\mathrm{sh}\\a\neq b\\ i,j}}
    (Q^\rS)_{ba}
    (q_\rS^{ij})_{ab}
    (q_\rS^{ji})_{ba}
    (\rho(0))_{ab}
    \\
    &=
    \sum_{\substack{a,b\in\sh\\a\neq b}}
    \frac{\Theta(1)}{D^\prime}
    (q_\rS)_{ba}
    (\rho(0))_{ab},
\end{align}
where we used the off-diagonal ETH for $Q^\rS$ and $q_\rS^{ij}$ and defined $q_\rS:=\sum_{ji}q_\rS^{ji}$, which is a hermitian operator of system S.
Using the Cauchy-Schwartz inequality, we obtain
\begin{align}
    \left|\overline{\lambda_2}\right|
    &\leq
    \frac{\Theta(1)}{D^\prime}
    \sqrt{
        \left|
        \sum_{\substack{a,b\in\sh\\a\neq b}}
        (q_\rS)_{ba}
        (q_\rS)_{ab}
        \right|
        \left|
        \sum_{\substack{a,b\in\sh\\a\neq b}}
        (\rho(0))_{ab}
        (\rho(0))_{ba}
        \right|
    }.
\end{align}
The right-hand side of this inequality is evaluated as
\begin{align}
    \nonumber
    &
    \left|
    \sum_{\substack{a,b\in\sh\\a\neq b}}
    (q_\rS)_{ba}
    (q_\rS)_{ab}
    \right|
    \leq
    D^\prime
    \left|
    \tr_\rS[(q_\rS)^2]
    \right|
    =
    \Theta(D),
    \\
    &
    \sum_{\substack{a,b\in\sh\\a\neq b}}
    (\rho(0))_{ab}
    (\rho(0))_{ba}
    \leq
    \tr[(\rho(0))^2]
    \leq
    1.
\end{align}
Then,
\begin{align}
    \left|\overline{\lambda_2}\right|
    &\leq
    \frac{\Theta(1)}{\sqrt{D^\prime}}
    =o(1)
\end{align}
is proved. In the same manner, $\left|\overline{\lambda_2^\can}\right|=o(1)$ is proved.

From the foregoing argument, $|\overline{\lambda(t)}-\overline{\lambda^\can(t)}|=o(1)$ is proved. We note that if the ETH holds, the correction of absolute irreversibility in the long-time regime is given by $\langle Q^\rS\rangle_\MC=1-\langle P_\mathrm{ini}^\rS\rangle_\MC$ regardless of the initial state of bath B.

\subsection{Temporal fluctuation}
\label{sec:app:proof:TempFluc}

In this subsection, we show that the temporal fluctuation around the long-time average of the error of the fluctuation theorem vanishes in the thermodynamic limit.
Together with the fact that the long-time average of the error is $o(1)$, we show that the fluctuation theorem holds for almost all the times after the relaxation time, except for the effect of quantum recurrence~\cite{Bocchieri1957,Percival1961,Venuti2015}.

We define the temporal fluctuation around the long-time average $\overline{A}$ as
\begin{align}
    \Delta_T(A):=\overline{(A-\overline{A})^2}
\end{align}
and the cross correlation as
\begin{align}
    \Delta_T(A,B):=
    \overline{AB}-
    \overline{A}~\overline{B}.
\end{align}

The temporal fluctuations of $\delta G:=G-1$, $\dG_\rS$ and $\dG_\rI$ are written as
\begin{align}
\label{eq:Tfluc_DeltaTdG}
    \Delta_T(\dG)
    =&
    \Delta_T(G),
    \\
\label{eq:Tfluc_DeltaTdGS}
    \Delta_T(\dG_\rS)
    =&
    \Delta_T(G_\rS)+\Delta_T(G_\rS^\can)-2\Delta_T(G_\rS,G_\rS^\can),
    \\
    \nonumber
    \Delta_T(\dG_\rI)
    =&
    \Delta_T(\delta G_\rS)+
    \Delta_T(G)
    \\
    &-
    2\Delta_T(G,G_\rS)+
    2\Delta_T(G,G_\rS^\can),
\label{eq:Tfluc_DeltaTdGI}
\end{align}
respectively.
From the Cauchy-Schwartz inequality, $|\Delta_T(A,B)|^2\leq \Delta_T(A)\Delta_T(B)$ holds. Then, if all of the equations $\Delta_T(G_\rS)=o(1)$, $\Delta_T(G_\rS^\can)=o(1)$ and $\Delta_T(G)=o(1)$ hold, the temporal fluctuations (\ref{eq:Tfluc_DeltaTdG}),(\ref{eq:Tfluc_DeltaTdGS}) and (\ref{eq:Tfluc_DeltaTdGI}) vanish in the thermodynamic limit.
Below, we will show $\Delta_T(G_\rS)=o(1)$, $\Delta_T(G_\rS^\can)=o(1)$ and $\Delta_T(G)=o(1)$.

\begin{figure}[t]
\begin{center}
\includegraphics[width=\linewidth]{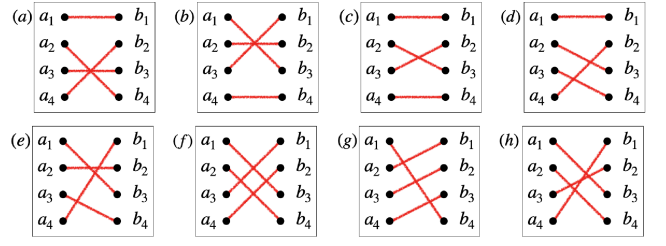}
\end{center}
\caption{
The terms that appear in the calculation of the temporal fluctuation. The indexes connected by the lines are equal. For example, (a) shows $a_1=b_1,a_2=b_4,a_3=b_3$, and $a_4=b_2$.
}
\label{fig:app:permutation}
\end{figure}

First, to show $\Delta_T(G_\rS)=o(1)$, we evaluate $\overline{G_\rS^2}$.
In the same manner as in Sec.~\ref{sec:LT:average}, we obtain 
\begin{align}
\nonumber
\overline{G_\rS^2}=
&
\sum_{\substack{(a_1,a_2,a_3,a_4)=(b_1,b_2,b_3,b_4)\\i,j,k,l}}
\rho_{a_1,b_1}
\rho^\prime_{a_2,b_2}
\rho_{a_3,b_3}
\rho^\prime_{a_4,b_4}
\\
&
(A^{ij}_\rS)_{b_1,a_1}
(B^{ij}_\rS)_{b_2,a_2}
(A^{kl}_\rS)_{b_3,a_3}
(B^{kl}_\rS)_{b_4,a_4},
\label{eq:GS2_perm}
\end{align}
where
$\rho:=\rho_\rS^\can\otimes\rho_\rB(0)$,
$\rho^\prime:=\rho_\rS(0)\otimes\rho_\rB(0)$,
$A^{ij}_\rS:=Z_\rS q_\rS^{ij} e^{\beta H_\rS}$ and
$B^{ij}_\rS:=Z_\rS q_\rS^{ji}$.
We note that the summation is taken over the case of $(a_1,a_2,a_3,a_4)=(b_1,b_2,b_3,b_4)$, i.e., a permutation of $(a_1,a_2,a_3,a_4)$ equals $(b_1,b_2,b_3,b_4)$. In the calculation of $(\overline{G_\mathrm{S}})^2$, only the combination of the form $(a_1,a_2)=(b_1,b_2)$ and $(a_3,a_4)=(b_3,b_4)$ appears in the sum.
Then, this combination does not appear in the temporal fluctuation $\overline{G_\rS^2}-(\overline{G_\mathrm{S}})^2$.
In Fig.~\ref{fig:app:permutation},
we show that combinations of $(a_1,a_2,a_3,a_4)$ and $(b_1,b_2,b_3,b_4)$ that should be evaluated.
The combinations that do not appear in Fig.~\ref{fig:app:permutation} are equivalent to some of the combinations in Fig.~\ref{fig:app:permutation}.
In the following evaluation, the diagonal matrix elements for the observable of system S are evaluated using the diagonal ETH as 
\begin{align}
\label{eq:diag_elem}
(A^{ij}_\rS)_{a,a}
 &\simeq
 \langle A^{ij}_\rS\rangle_\MC
 =\Theta(1).
\end{align}

In $\Delta_T(G_\rS)$, the term corresponding to Fig.~\ref{fig:app:permutation}(a) is evaluated as
\begin{align}
\nonumber
&\sum_{\substack{a_1,a_2,a_3,a_4\\i,j,k,l}}
\rho_{a_1,a_1}
\rho^\prime_{a_2,a_4}
\rho_{a_3,a_3}
\rho^\prime_{a_4,a_2}
\\&
(A^{ij}_\rS)_{a_1,a_1}
(B^{ij}_\rS)_{a_4,a_2}
(A^{kl}_\rS)_{a_3,a_3}
(B^{kl}_\rS)_{a_2,a_4}
\\
&=
\frac{\Theta(1)}{D^\prime}
\sum_{\substack{a_1,a_2,a_3,a_4\\i,j,k,l}}
\rho_{a_1,a_1}
\rho^\prime_{a_2,a_4}
\rho_{a_3,a_3}
\rho^\prime_{a_4,a_2}
\\
&=
\frac{\Theta(1)}{D^\prime}
(\tr[\rho])^2\tr[(\rho^\prime)^2]
\\
&=
o(1).
\end{align}
We used the off-diagonal ETH for $B^{ij}_\rS$
and
$B^{kl}_\rS$
and the diagonal ETH for $A^{ij}_\rS$
and
$A^{kl}_\rS$.
Because we can evaluate the other terms in the similar manner, $\Delta_T(G_\rS)=o(1)$ is proved.

In the evaluation of $\Delta_T(G_\rS)$, 
only $\tr[\rho]=1$ and $\tr[\rho^2]\leq 1$ are used as the properties of the density operator. Then, $\Delta_T(G_\rS^\can)=o(1)$ is proved by setting $\rho=\rho_\rS^\can\otimes\rho_\rB^\can$ and
$\rho^\prime=\rho_\rS(0)\otimes\rho_\rB^\can$.

We next evaluate $\Delta_T(G)$.
As in Eq.~(\ref{eq:GS2_perm}), we can write $\overline{G^2}$ as
\begin{align}
\nonumber
\overline{G^2}=
&
(D_\rS
Z_\rB
e^{\beta E^\mathrm{B}_\mathrm{ini}})^2
\\
&
\nonumber
\sum_{\substack{(a_1,a_2,a_3,a_4)=(b_1,b_2,b_3,b_4)\\i,j,k,l}}
\tilde{\rho}_{a_1,b_1}
\rho^\prime_{a_2,b_2}
\tilde{\rho}_{a_3,b_3}
\rho^\prime_{a_4,b_4}
\\
&
(\rho^{ij})_{b_1,a_1}
(B^{ij}_\rS)_{b_2,a_2}
(\rho^{kl})_{b_3,a_3}
(B^{kl}_\rS)_{b_4,a_4},
\label{eq:GS_perm}
\end{align}
where
$E^\mathrm{B}_\mathrm{ini}$ is the energy of the initial state of bath B,
$\tilde{\rho}:=(1_\rS/D_\rS)\otimes \rho_\rB(0)$ and
$\rho^{ij}:=q_\rS^{ij}\otimes \rho_\rB^\can$.

Regarding the partition function, the following relation holds~\cite{Tasaki2018}:
\begin{align}
\label{eq:ZB_EOE}
(Z_\rB
e^{\beta E^\mathrm{B}_\mathrm{ini}})^2\leq D^2 N.
\end{align}
We also have
\begin{align}
    \label{eq:Deff:rho2}
\tr[(\rho^{ij})^2]=\delta_{ij}/D_\mathrm{eff}[\rho_\rB^\can],
\end{align}
where $D_\mathrm{eff}[\rho_\rB^\can]$ is the effective dimension of $\rho_\rB^\can$ with respect to $H_\rB$. For simplicity, when there is no degeneracy in the eigenenergy and the state is diagonal with the energy eigenstates as $\rho=\sum_i p_i |E_i\rangle\langle E_i|$, the effective dimension is written as
\begin{align}
    \label{eq:Deff:def}
    D_\mathrm{eff}[\rho]
    :=\left(\sum_i |p_i|^2\right)^{-1}.
\end{align}
We note that $D_\mathrm{eff}[\rho_\rB^\can]$ increases exponentially with respect to the bath size $N$~\cite{Kaneko2017}.
Besides, we assume that the matrix elements of the density operator $\rho^\prime$ and $\tilde{\rho}$ are $\mathcal{O}({D^\prime}^{-1})$:
\begin{align}
    &
    \label{eq:DOp_elem_cond1}
    |(\rho^\prime)_{ab}|=\mathcal{O}({D^\prime}^{-1}),
    \\
    &
    \label{eq:DOp_elem_cond2}
    |(\tilde{\rho})_{ab}|=
    \mathcal{O}({D^\prime}^{-1}).
\end{align}
This is not satisfied when the density operators $\rho^\prime$ and $\tilde{\rho}$, which are product states as in Eq.~(\ref{eq:initialstate}), are localized to some specific eigenstates of $H$.
When the initial state is in the form of Eq.~(\ref{eq:initialstate}), we argue that $\mathcal{O}({D^\prime}^{-1})$ is naturally satisfied if the Hamiltonian $H$ is chaotic and mixes system S and bath B sufficiently.

In $\Delta_T(G_\rS)$, the term corresponding to Fig.~\ref{fig:app:permutation}(b) is evaluated as
\begin{align}
\nonumber
&
\left|
(D_\rS
Z_\rB
e^{\beta E^\mathrm{B}_\mathrm{ini}})^2
\right.
\\
\nonumber
&
\sum_{\substack{a_1,a_2,a_3,a_4\\i,j,k,l}}
\tilde{\rho}_{a_1,a_3}
\rho^\prime_{a_2,a_2}
\tilde{\rho}_{a_3,a_1}
\rho^\prime_{a_4,a_4}
\\
&
\left.
(\rho^{ij})_{a_3,a_1}
(B^{ij}_\rS)_{a_2,a_2}
(\rho^{kl})_{a_1,a_3}
(B^{kl}_\rS)_{a_4,a_4}
\right|
\\
\leq&
\Theta(N)
\left|
\sum_{\substack{a_1,a_2,a_3,a_4\\i,j,k,l}}
\rho^\prime_{a_2,a_2} \rho^\prime_{a_4,a_4}
(\rho^{ij})_{a_3,a_1} (\rho^{kl})_{a_1,a_3}
\right|
\\
\leq&
\Theta(N)
(\tr[\rho^\prime])^2
\sum_{i,j,k,l}
\sqrt{
\tr[(\rho^{ij})^2]
\tr[(\rho^{kl})^2]
}
\\
=&
\frac{\Theta(N)}{D_\mathrm{eff}[\rho_\rB^\can]}
=o(1).
\label{eq:app:proof1}
\end{align}
In the first inequality, we used Eqs.~(\ref{eq:diag_elem})(\ref{eq:ZB_EOE}) and (\ref{eq:DOp_elem_cond1}).
In the second inequality, we used the Cauchy-Schwartz inequality. In the last inequality, we used Eq.~(\ref{eq:Deff:rho2}).
Because we can evaluate the other terms in the same manner, $\Delta_T(G)=o(1)$ is proved.

\section{Perturbation theory on the interaction-induced error}
\label{sec:app:interaction_perturb}

In this Appendix, by using the perturbation theory, we show that the interaction-induced error does not grow significantly in the long-time regime and vanishes in the thermodynamic limit in the short-time regime.

\subsection{The long-time regime}
\label{sec:app:interaction_perturb:LT}

We show that $|\overline{\dG_\rI}|\leq\Theta(1)$ holds at high temperature using the perturbation theory.
The long-time average of the interaction-induced error (\ref{eq:dGI_def_4Gs}) is written as
\begin{align}
\overline{\dG_\rI^{(1)}}
&:=
\overline{G}-\overline{G_\rS},
\\
\overline{\dG_\rI^{(2)}}
&:=
\overline{G_\rS^\can}-\overline{G^\can}.
\end{align}

First, we divide $\overline{\dG_\rI^{(1)}}$ into the diagonal term $\overline{\dG_{\rI 1}^{(1)}}$ and the off-diagonal term $\overline{\dG_{\rI 2}^{(1)}}$.
The diagonal term is written as
\begin{align}
    \overline{\dG_{\rI1}^{(1)}}
    =
    \sum_a
    &
    \tr[
    e^{-\beta H_\rB}\pi_a e^{\beta H_\rB}\rho_\rB(0) \pi_a \rho_\rS^\DE
    ]
    \nonumber
    \\
    -\sum_a&
    \tr[
    e^{\beta H_\rS}\pi_a e^{-\beta H_\rS}\rho_\rB(0) \pi_a \rho_\rS^\DE
    ],
\end{align}
where $\rho_\rS^\DE:=\tr_\rB[\rho^\DE]$. By using the first order perturbation theory for the eigenstates, we obtain
\begin{align}
    \nonumber
    &\overline{\dG_{\rI1}^{(1)}}
    \\
    \nonumber
    \simeq&
    \sum_{i\alpha}{\sum_{j\beta}}^\prime{\sum_{k\gamma}}^\prime
    \frac{
    |\langle E_j^\rS E_\beta^\rB|H_\rI|E_i^\rS E_\alpha^\rB\rangle|^2
    }{
    (\delta E_{ji}^\rS+\delta E_{\beta\alpha}^\rB)
    (\delta E_{ki}^\rS+\delta E_{\beta\alpha}^\rB)
    }
    \delta_{\beta\gamma}
    \\
    \nonumber
    &\left[
        \langle E_j^\rS|\rho_\rS^\DE|E_k^\rS\rangle\delta_{\alpha,\mathrm{ini}}
        (
        e^{-\beta \delta E_{\beta\alpha}^\rB}
        -
        e^{-\beta \delta E_{ki}^\rS}
        )
    \right.
    \\
    &+\left.
\langle E_j^\rS|\rho_\rS^\DE|E_k^\rS\rangle\delta_{\alpha,\mathrm{ini}}
        (
        e^{-\beta \delta E_{\beta\alpha}^\rB}
        -
        e^{-\beta \delta E_{ki}^\rS}
        )
    \right],
    \label{eq:dGI1_perturb}
\end{align}
where the summation over $j,k,\beta,\gamma$ are restricted to $j\neq i$,$k\neq i$,$\beta\neq \alpha$,$\gamma\neq \alpha$ and $\delta_{\alpha,\mathrm{ini}}$ means $\alpha$ equals the index of the initial state of bath B. We also define $\delta E^\rS_{ij}:=E^\rS_i-E^\rS_j$ and $\delta E^\rB_{\beta\alpha}:=E^\rB_\beta-E^\rB_\alpha$.

Here, we assume the following relation as the off-diagonal ETH instead of Eq.~(\ref{eq:offETH}).
\begin{align}
    |\langle E_j^\rS E_\beta^\rB|H_\rI|E_i^\rS E_\alpha^\rB\rangle|^2
    \sim 
    \frac{\Theta(1)}{D}
    e^{-\nu |\delta E_{\beta\alpha}^\rB|},
    \label{eq:offETH:expdecay}
\end{align}
where $\nu$ is an $N$-independent constant.
The exponential decay with respect to energy difference $\delta E_{\beta\alpha}^\rB$ in Eq.~(\ref{eq:offETH:expdecay}) has been observed in numerical calculations of the off-diagonal ETH~\cite{Mondaini2017}. Besides, the exponential decay is theoretically shown when the Hamiltonian is local~\cite{Arad2016}.
Then, the terms that can diverge in Eq.~(\ref{eq:dGI1_perturb}) are written as
$e^{-(\nu\pm\beta)|\delta E_{\beta\alpha}^\rB|}$. At high temperature $\nu>\beta$, $\overline{\dG_{\rI1}^{(1)}}\leq \Theta(1)$ is shown.

We next show that the off-diagonal term vanishes as $N$ increases: $\left|\overline{\dG_{\rI 2}^{(1)}}\right|=o(1)$. Since $\left|\overline{G_{\rS 2}}\right|=o(1)$ is already shown, it is sufficient to show $|\overline{G_2}|=o(1)$, where $\overline{G_2}$ is the off-diagonal contribution of $\overline{G}$. In the same manner as Appendix~\ref{sec:app:proof}, we show $|\overline{G_2}|=o(1)$ by using the off-diagonal ETH (\ref{eq:offETH}) and the conditions on the density operators (\ref{eq:DOp_elem_cond1}) and (\ref{eq:DOp_elem_cond2}).

We divide $\overline{\dG_\rI^{(2)}}$ into the diagonal term $\overline{\dG_{\rI 1}^{(2)}}$ and the off-diagonal term $\overline{\dG_{\rI 2}^{(2)}}$. 
Since $\overline{G^\can_{\rS 2}}=o(1)$ is already shown, $\left|\overline{\dG_{\rI 2}^{(2)}}\right|=o(1)$ holds. By using the diagonal ETH for the operator of system S, the diagonal term $\overline{\dG_{\rI 1}^{(2)}}$ is evaluated as
\begin{align}
\overline{
    \dG_{\rI 1}^{(2)}
    }
    \simeq
    1-
    \langle \rho_\rS^\DE (\rho_\rS^\can)^{-1}
    \rangle_\MC.
\end{align}
If the density operator of system S in the long-time regime $\rho_\rS^\DE$ relaxes to the canonical ensemble of $H_\rS$, the second term on the right-hand side equals $1$. Then, we argue that $\overline{
\dG_{\rI 1}^{(2)}
}=o(1)$ holds when $\rho_\rS^\DE$ relaxes to the canonical ensemble.
On the other hand, 
$\rho_\rS^\DE$ does not necessarily relax to the canonical ensemble of $H_\rS$ for general interactions. Then, we assume $\overline{
\dG_{\rI 1}^{(2)}
}\leq\Theta(1)$ in general.

From the foregoing argument, $|\overline{\dG_\rI}|\leq\Theta(1)$ is shown.

\subsection{The short-time regime}
\label{sec:app:interaction_perturb:ET}

We show that $|\dG_\rI|=o(1)$ holds in the short-time regime~($t\ll\tau_\LR$) at high temperature using the perturbation theory.

The interaction-induced error is defined in Eq.~(\ref{eq:dGI_def_4Gs}). 
Because $\rho_\rS(t)\simeq \rho_\rS^\can(t)$ in $t\ll\tau_\LR$ holds due to the Lieb-Robinson bound and the ETH, we obtain
\begin{align}
    \dG_\rI
    \simeq&
    \tr[
    e^{-\beta H_\rB}
    U
    e^{\beta H_\rB}
    \delta \rho_\rB 
    U^\dag
    \rho_\rS(t)
    ]
    \nonumber
    \\
    -&
    \tr[
    e^{\beta H_\rS}
    U
    e^{-\beta H_\rS}
    \delta \rho_\rB 
    U^\dag
    \rho_\rS(t)
    ],
    \label{eq:dG_rI_ET_approx}
\end{align}
where $\delta \rho_\rB:=\rho_\rB(0)-\rho_\rB^\can$.
By using the Lieb-Robinson bound and the ETH again, the second line is shown to be $o(1)$. Below, we evaluate the first line using the perturbation calculation for the time evolution operator.

From the first order perturbation theory, we obtain
\begin{align}
    U
    \simeq&
    U_0+U_1,
    \\
    U_0
    :=
    &
    e^{-iH_0 t/\hbar},
    \\
    U_1
    :=&
    -i U_0\int_0^t d\tau
    e^{-iH_0\tau/\hbar}
    H_\rI
    e^{iH_0\tau/\hbar},
\end{align}
where we define $H_0:=H_\rS+H_\rB$. We also define
\begin{align}
    \delta G_\rI^\rB[U,V^\dag]
    :=&
    \tr[
    e^{-\beta H_\rB}
    U
    e^{\beta H_\rB}
    \delta \rho_\rB 
    V^\dag
    \rho_\rS(t)
    ],
    \\
    \delta G_{\rI,(i,j)}^\rB
    :=&
    \delta G_\rI^\rB[U_i,U_j^\dag].
\end{align}
Then the first line of Eq.~(\ref{eq:dG_rI_ET_approx}) is approximated as $\sum_{i=0}^1\sum_{j=0}^1\delta G_{\rI,(i,j)}^\rB$. 

We first evaluate $\delta G_{\rI,(0,0)}^\rB$ as
\begin{align}
    \delta G_{\rI,(0,0)}^\rB
    =&
    \tr[
    e^{-\beta H_\rB}U_0 e^{\beta H_\rB}
    \delta \rho_\rB U^\dag_0 \rho_\rS(t)
    ]
    \nonumber
    \\
    =&
    \tr[
    \rho_\rS(t)\otimes\delta\rho_\rB
    ]
    \nonumber
    \\
    =&
    0,
\end{align}
where we used $[U_0,H_\rB]=0$ and $[U_0,\delta\rho_\rB]=0$.
We next calculate $\delta G_{\rI,(0,1)}^\rB$ as
\begin{align}
    &
    |\delta G_{\rI,(0,1)}^\rB|
    \nonumber
    \\
    =&
    \left|
    \tr[
    e^{-\beta H_\rB} U_0 e^{\beta H_\rB}
    \delta \rho_\rB U^\dag_1 \rho_\rS(t)
    ]
    \right|
    \nonumber
    \\
    =&
    \left|
    \int_0^t
    d\tau
    \tr[
    U_0
    H_\rI 
    \rho_\rS(t,\tau)
    \otimes \delta \rho_\rB
    ]
    \right|
    \nonumber
    \\
    \leq&
    \int_0^t
    d\tau
    \|U_0\|
    \|
    H_\rI
    \rho_\rS(t,\tau)
    \otimes \delta \rho_\rB
    \|_1,
\end{align}
where we define $\rho_\rS(t,\tau):=e^{i H_\rS(t-\tau)}\rho_\rS(t)e^{-iH_\rS(t-\tau)}$. We note that $\|U_0\|=1$ holds and $\|
    H_\rI
    \rho_\rS(t,\tau)
    \otimes \delta \rho_\rB
    \|_1=\mathcal{O}(D^{-1/2})$ is satisfied from the ETH.
Then, in the short-time regime $t\ll\tau_\LR$, we obtain
\begin{align}
    |\delta G_{\rI,(0,1)}^\rB|
    \leq\frac{\tau_\LR}{\Theta(D^{1/2})}
    =
    \frac{\Theta(N^\mu)}{\Theta(D^{1/2})}
    =
    o(1).
\end{align}
Besides, $\delta G_{\rI,(1,0)}^\rB$ is evaluated as
\begin{align}
    \delta G_{\rI,(1,0)}^\rB
    =
    -i
    \int_0^t
    d\tau
    \tr[
    U_0
    \rho_\rS(t,\tau)
    \otimes\delta\rho
    H_\rI
    ].
\end{align}
Then, $\delta G_{\rI,(1,0)}^\rB=o(1)$ in $t\ll\tau_\LR$ is shown to hold in the same manner as $\delta G_{\rI,(0,1)}^\rB$.

We finally calculate $\delta G_{\rI,(1,1)}^\rB$ as
\begin{align}
    &
    \delta G_{\rI,(1,1)}^\rB
    \nonumber
    \\
    =&
    \int_0^t \int_0^t
    d\tau d\tau^\prime 
    \tr
    [
    e^{-\beta H_\rB} 
    e^{i H_0 (t-\tau)}
    H_\rI
    \nonumber
    \\
    &
    ~~~~~
    e^{-iH_0\tau}
    e^{\beta H_\rB}
    \delta\rho_\rB 
    e^{iH_0 \tau^\prime}
    H_\rI 
    e^{iH_0(t-\tau^\prime)}
    \rho_\rS(t)
    ]
    \nonumber
    \\
    \leq&
    t^2
    \left|
    \sum_{a,b,c}e^{-\beta \delta E_\rB^{a_\rB b_\rB}}
    (H_\rI)_{ab}
    (H_\rI)_{bc}
    (\delta\rho_\rB)_{bb}
    (\rho_\rS(t))_{ca}
    \right|
    \nonumber
    \\
    =&
    t^2
    \left|
    \sum_{a,b,c}
    M(a,b,c,t)
    \right|,
    \label{eq:app:dGI_ET:2nd}
\end{align}
where
\begin{align}
    M(a,b,c,t):=&e^{-\beta \delta E_\rB^{a_\rB b_\rB}}
    (H_\rI)_{ab}
    (H_\rI)_{bc}
    (\delta\rho_\rB)_{bb}
    (\rho_\rS(t))_{ca}.
\end{align}
We note that the index $a$ represents a pair of indexes of the eigenstates of system S and bath B and is written as $a=(a_\rS,a_\rB)$.
We write the corresponding eigenenergy as $E^a=E_\rS^{a_\rS}+E_\rB^{a_\rB}$.
The same applies to $b$ and $c$.
We write the matrix elements as
$(A)_{ab}=\langle E_\rS^{a_\rS}E_\rB^{a_\rB}|A|E_\rS^{b_\rS}E_\rB^{b_\rB}\rangle$, 
$(A)_{a_\rS b_\rS}=\langle E_\rS^{a_\rS}|A|E_\rS^{b_\rS}\rangle$
and
$(A)_{a_\rB b_\rB}=\langle E_\rB^{a_\rB}|A|E_\rB^{b_\rB}\rangle$.
We also write the energy change of B as $\delta E_\rB^{a_\rB b_\rB}$.

We define the energy ranges of bath B as follows.
First, we define $\Lambda_1:=[E-\Delta,E+\Delta]$, $E=\tr_\rB[H_\rB \rho_\rB(0)]=\tr_\rB[H_\rB \rho_\rB^\can]$, $\Delta=\Theta(N^a)$ $(1/2<a<1)$. 
We note that the energy widths of $\rho_\rB(0)$ and $\rho_\rB^\can$ are included in $\Lambda_1$. 
We also define $\Lambda_2:=[E-2\Delta,E+2\Delta]$ and denote the outside of $\Lambda_2$ by $\overline{\Lambda_2}$.
We write the sum range as $a\in\Lambda_1$, which means that $\sum_{a\in\Lambda_1}=\sum_{a_\rS}\sum_{a_\rB\in\Lambda_1}$.

We evaluate $\sum_{a,b,c} M(a,b,c,t)$ for the following three cases: (i) $E_a\in\overline{\Lambda_2}$, (ii) $E_a\in\Lambda_2, a_\rB\neq b_\rB$,  (iii) $E_a\in\Lambda_2, a_\rB=b_\rB$.
We note that $a_\rB=c_\rB$ always holds since $\rho_\rS(t)$ is the operator of system S.

First, we consider the case (i):
\begin{align}
    &
    \sum_{\substack{a\in{\overline{\Lambda_2}}\\ b\in        \Lambda_1\\c;c_\rB=a_\rB}}
    M(a,b,c,t)
    \nonumber
    \\
    &=
    \frac{\Theta(1)}{D}
    \sum_{\substack{a\in{\overline{\Lambda_2}}\\ b\in        \Lambda_1\\c;c_\rB=a_\rB}}
    (\delta\rho_\rB)_{bb}
    (\rho_\rS(t))_{ca}
    \nonumber
    \\
    &~~~~
    e^{-\nu |\delta E_\rB^{a_\rB b_\rB}|}
    e^{-\nu |\delta E_\rB^{b_\rB c_\rB}|}
    e^{-\beta \delta E_\rB^{a_\rB b_\rB}},
    \label{eq:app:dGI_ET:2nd:aout}
\end{align}
where we used the off-diagonal ETH for $H_\rI$.
We write $b_\rB$ maximizing
$
e^{-\nu |\delta E_\rB^{a_\rB b_\rB}|}
$ and $
e^{-\beta \delta E_\rB^{a_\rB b_\rB}}
$ as $b_{\rB}^\nu$ and $b_{\rB}^\beta$, respectively.
By using
$e^{-\nu |\delta E_\rB^{b_\rB c_\rB}|}\leq 1$ and
$|(\rho_\rS(t))_{ca}|\leq 1$,
The absolute value of the right-hand side of Eq.~(\ref{eq:app:dGI_ET:2nd:aout}) is bounded from above as follows:
\begin{align}
    &
    \frac{D_\rS}{D}
    \sum_{a\in{\overline{\Lambda_2}}}
    e^{-\nu |\delta E_\rB^{a_\rB b_\rB^\nu}|}
    e^{+\beta |\delta E_\rB^{a_\rB b_\rB^\beta}|}
    \sum_{b\in\Lambda}
    |\delta\rho_{bb}|
    \nonumber
    \\
    \leq&
    \frac{D_\rS D_{\overline{\Lambda_2}}}{D}
    \max_{a_\rB} (e^{-\nu |\delta E_\rB^{a_\rB b_\rB^\nu}|}
    e^{+\beta |\delta E_\rB^{a_\rB b_\rB^\beta}|}),
\end{align}
where we used $\sum_{b\in\Lambda}|\delta\rho_{bb}|\leq 1$ and $D_{\overline{\Lambda_2}}$ is the dimension of $\overline{\Lambda_2}$. The right-hand side exponentially decays as $e^{-(\nu-\beta)\Theta(\Delta)}$ if $\beta<\nu$. Then, the right-hand side of Eq.~(\ref{eq:app:dGI_ET:2nd}) in this case is evaluated as $\tau_\LR^2 e^{-(\nu-\beta)\Theta(\Delta)}=o(1)$.

Next, we consider the case (ii).
When $a_\rB\neq b_\rB$, $c_\rB\neq b_\rB$ holds from $a_\rB=c_\rB$.
Then, using the off-diagonal ETH,
we evaluate $\sum_{a,b,c} M(a,b,c,t)$ restricted to the case (ii) as
\begin{align}
    \frac{1}{\Theta(D)}
    \sum_{\substack{a\in{\overline{\Lambda_2}}\\ b\in        \Lambda_1;b_\rB\neq a_\rB\\c;c_\rB=a_\rB}}
    &
    (\delta\rho_\rB)_{bb}
        (\rho_\rS(t))_{ca}
        \nonumber
    \\
    &
    e^{-\beta \delta E_\rB^{a_\rB b_\rB}-\nu(|\delta E_\rB^{a_\rB b_\rB}|+|\delta E_\rB^{c_\rB b_\rB}|)},
\end{align}
which is shown to exponentially decay as $e^{-(\nu-\beta)\Theta(\Delta)}$ in $\beta<\nu$ using $\sum_{b\in\Lambda}|\delta\rho_{bb}|\leq 1$.
Then, the right-hand side of Eq.~(\ref{eq:app:dGI_ET:2nd}) in the case (ii) is $o(1)$.

Finally, we consider the case  (iii).
We evaluate $\sum_{a,b,c} M(a,b,c,t)$ restricted to this case as
\begin{align}
    &
    \sum_{\substack{
    b_\rB\in\Lambda_1^\prime\\
    a_\rS,b_\rS,c_\rS
    }}
    (J_\rI^{a_\rS  b_\rS})_{b_\rB b_\rB}
    (J_\rI^{b_\rS  c_\rS})_{b_\rB b_\rB}
    (\rho_\rS(t))_{c_\rS a_\rS}
    (\delta \rho_\rB)_{b_\rB b_\rB}
    \nonumber
    \\
    \simeq &
    \sum_{a_\rS b_\rS c_\rS}
    \langle J_\rI^{a_\rS b_\rS}\rangle_\MC
    \langle J_\rI^{b_\rS c_\rS}\rangle_\MC
    (\rho_\rS(t))_{c_\rS a_\rS}
    \sum_{b_\rB\in\Lambda_1}
    (\delta \rho_\rB)_{b_\rB b_\rB},
\end{align}
where we define $J_\rI^{a_\rS b_\rS}:=\langle E_\rS^{a_\rS}|H_\rI|E_\rS^{b_\rS}\rangle$ and use the ETH for $J_\rI^{a_\rS b_\rS}$.
Besides,
$|\sum_{b_\rB\in\Lambda_1}(\delta \rho_\rB)_{b_\rB b_\rB}|$
is the contribution of $\rho_\rB^\can$ from outside $\Lambda_1$, which decays as $e^{-\Theta(\Delta)}$. Therefore, the right-hand side of Eq.~(\ref{eq:app:dGI_ET:2nd}) in the case (iii) is also $o(1)$.

From the foregoing argument, $|\dG_\rI|=o(1)$ is shown in the short-time regime~($t\ll\tau_\LR$).

\section{Supplementary numerical results}
\label{sec:app:numerical}

\begin{figure}[!t]
\begin{center}
\includegraphics[width=\linewidth]{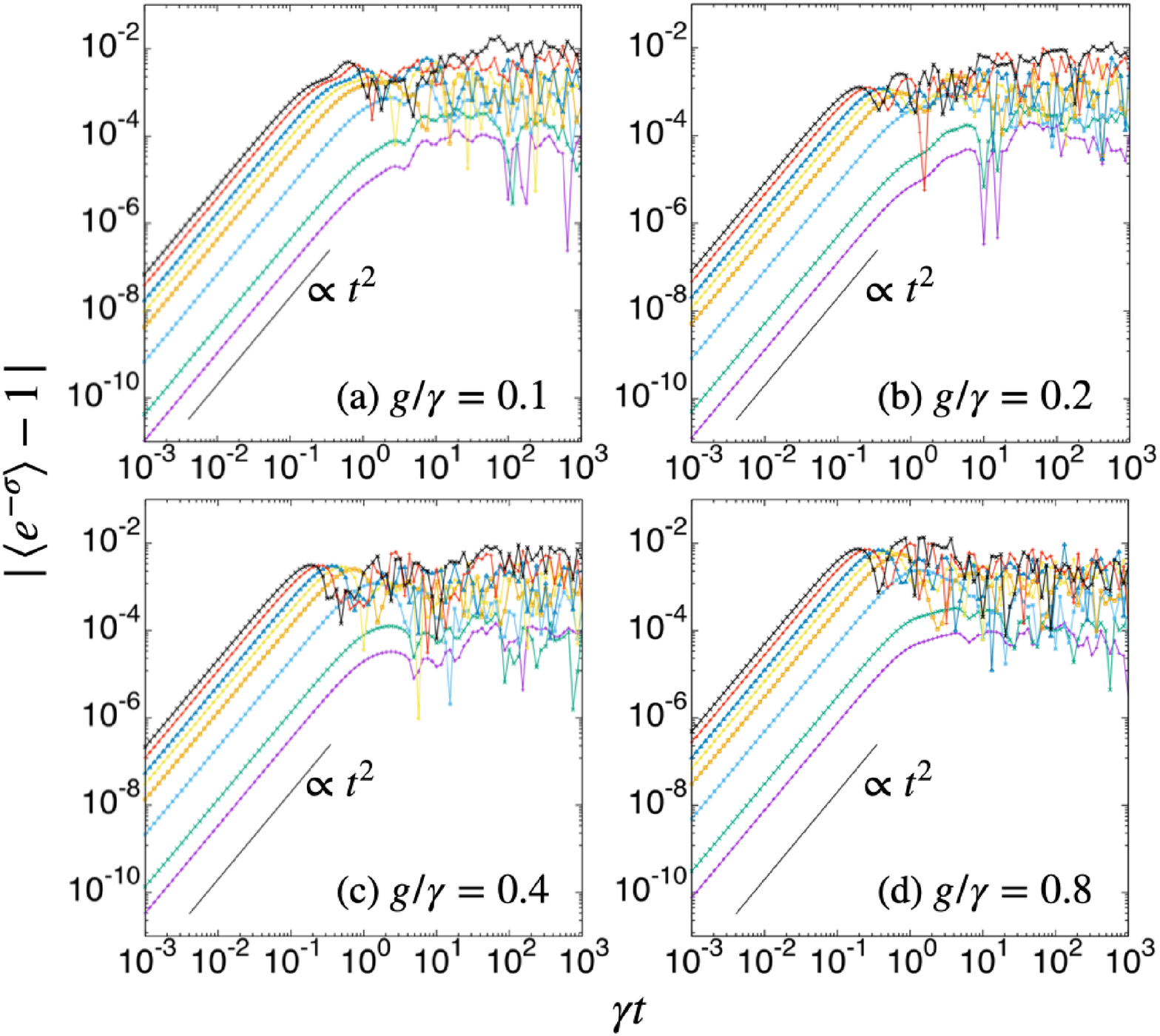}
\end{center}
\caption{
The time dependence of the error of the fluctuation theorem $|\langle e^{-\sigma}\rangle-1|$, whose initial rise is proportionate to $t^2$. The initial state of bath B is the energy eigenstates of $H_\rB$, whose energy is maximum in the energy shell at $\beta=0.1$. Parameter: $\gamma^\prime/\gamma=0.05, 0.1, 0.4, 1, 1.5, 2, 3, 4$ (from bottom to top) and $p=0.99$.
The onsite potential $\omega$ is determined by $\tr_\rB[n_\rB\rho_\rB^\can]=N_\rP$.
}
\label{fig:App:FT_error_t2}
\end{figure}
We first show the time dependence of the error of the fluctuation theorem $|\langle e^{-\sigma}\rangle-1|$ for various interaction parameters.
As in Fig.~\ref{fig:App:FT_error_t2},
we observe qualitatively the same behavior as the inset of Fig.~\ref{fig:G_tdep} as mentioned in Sec.~\ref{sec:SOR}.

\begin{figure}[!t]
\begin{center}
\includegraphics[width=\linewidth]{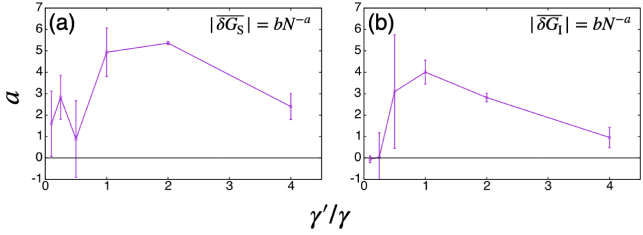}
\end{center}
\caption{
The $\gamma^\prime$-dependence of the exponent $a$, which is obtained by fitting 
$|\overline{\dG_\rS}|$ and $|\overline{\dG_\rI}|$.
Parameters: $p=0.9$,  $g=0.1\gamma,\beta=0.1$, $\gamma^\prime=\gamma$.
The left figure shows $|\overline{\dG_\rS}|$ and the right figure shows $|\overline{\dG_\rI}|$.
}
\label{fig:App:4}
\end{figure}
\begin{figure}[!t]
\begin{center}
\includegraphics[width=\linewidth]{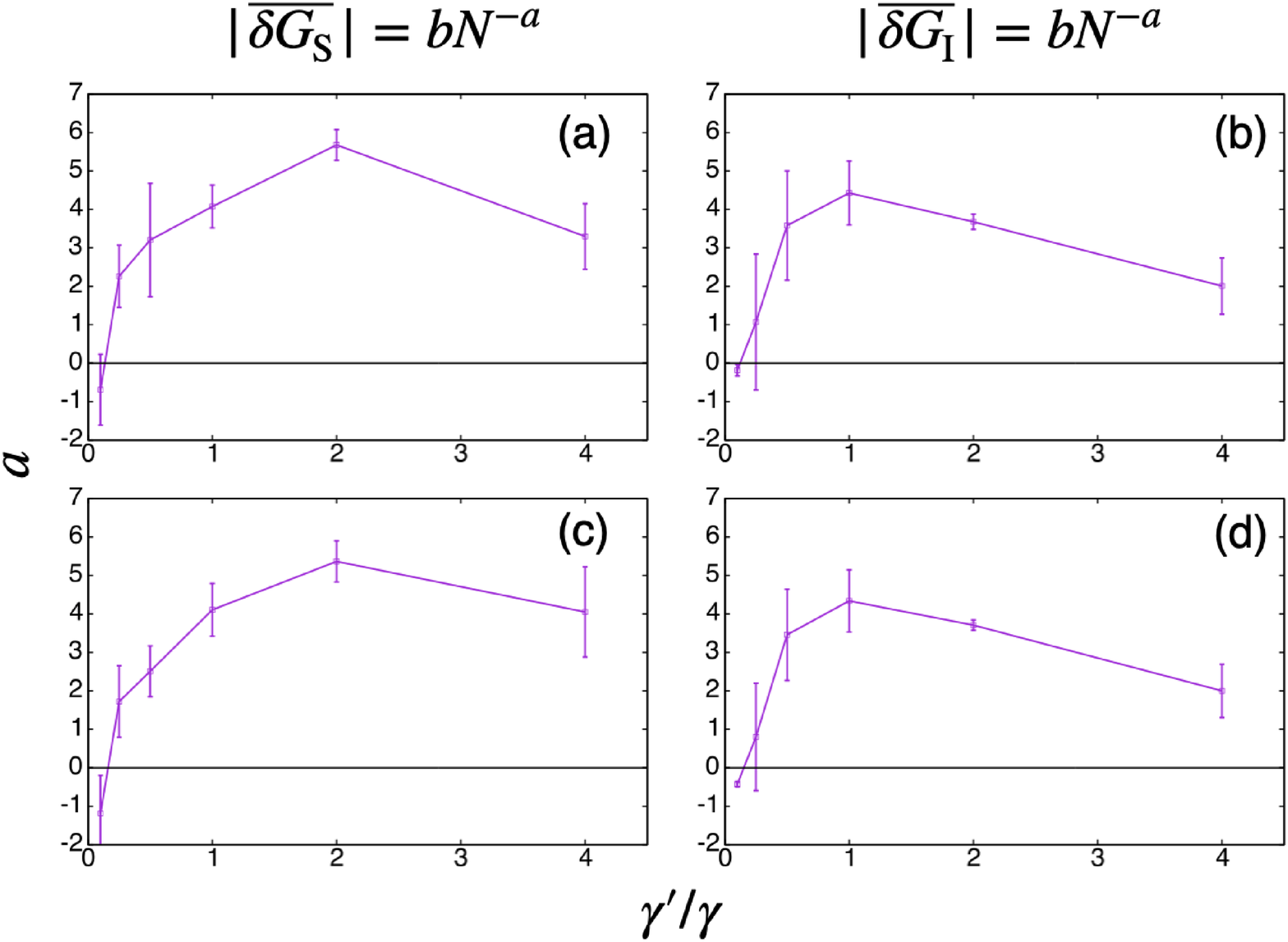}
\end{center}
\caption{
The $\gamma^\prime$-dependence of the exponent $a$, which is obtained by fitting 
$|\overline{\dG_\rS}|$ and $|\overline{\dG_\rI}|$.
Parameters: $g=0.4\gamma,\beta=0.1$, $\gamma^\prime=\gamma$.
$p=0.99$ (upper panels) and $p=0.9$ (lower panels).
The left panels show $|\overline{\dG_\rS}|$ and the right panels show $|\overline{\dG_\rI}|$.
}
\label{fig:App:5}
\end{figure}
We next show the supplemental data in the long-time regime.
Figure~\ref{fig:App:4} shows the $\gamma^\prime$-dependence of $|\overline{\dG_\rS}|$ and $|\overline{\dG_\rI}|$ with $p=0.9$.
Also, Fig.~\ref{fig:App:5} shows the $\gamma^\prime$-dependence of $|\overline{\dG_\rS}|$ and $|\overline{\dG_\rI}|$ with $g=0.4\gamma$, showing the similar result as in the main text.

\begin{figure}[t]
\begin{center}
\includegraphics[width=\linewidth]{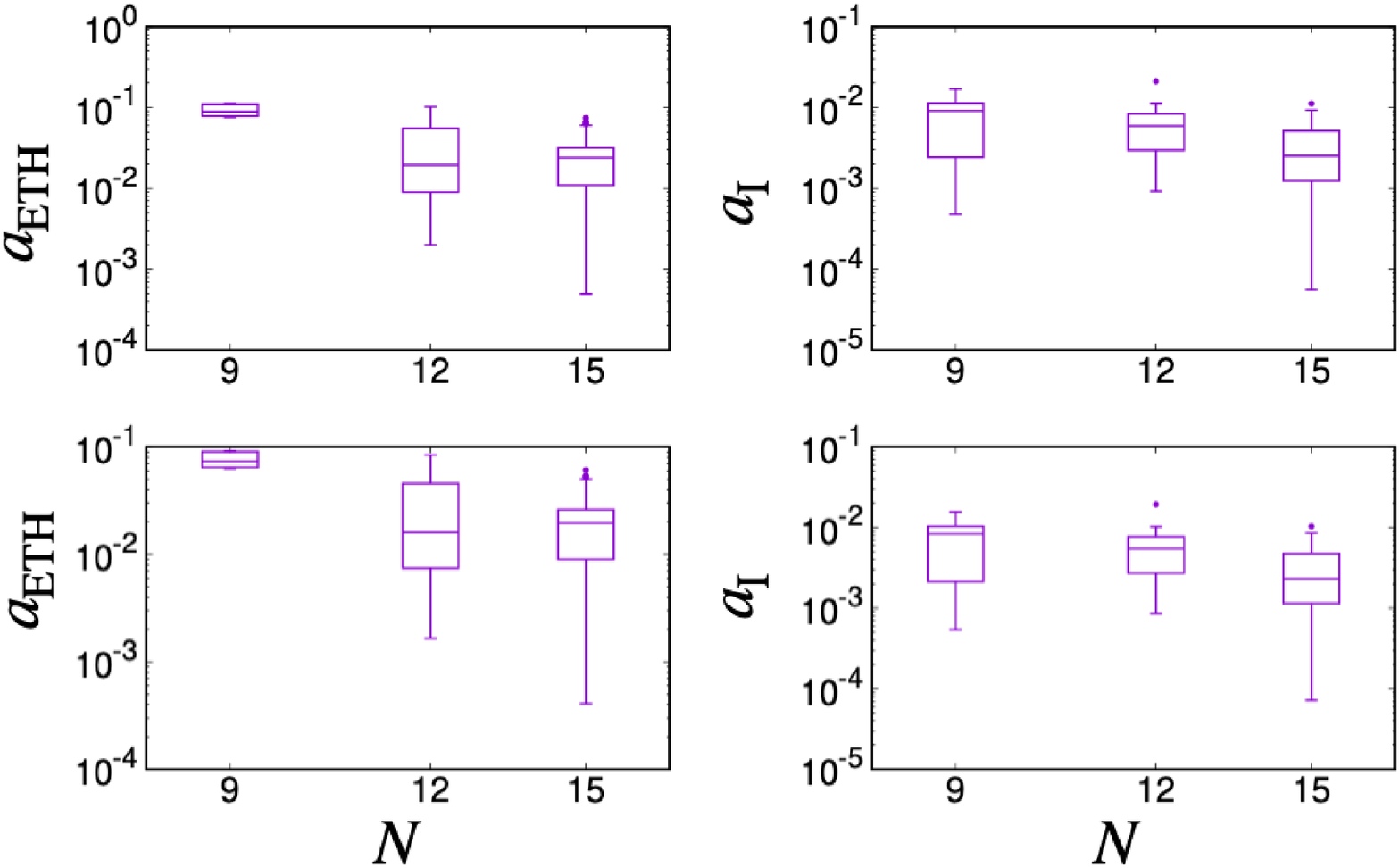}
\end{center}
\caption{
The dependence of $a_\ETH$ and $a_\rI$ on the bath size and the initial state.
Parameters: 
$p=0.99$,
$g=0.1\gamma$,
$\gamma^\prime=4\gamma$~(upper panels) and $0.1\gamma$~(lower panels), $\beta=0.1$.
Left panels are $a_\ETH$ and right panels are $a_\rI$.
}
\label{fig:App:7}
\end{figure}
\begin{figure}[t]
\begin{center}
\includegraphics[width=\linewidth]{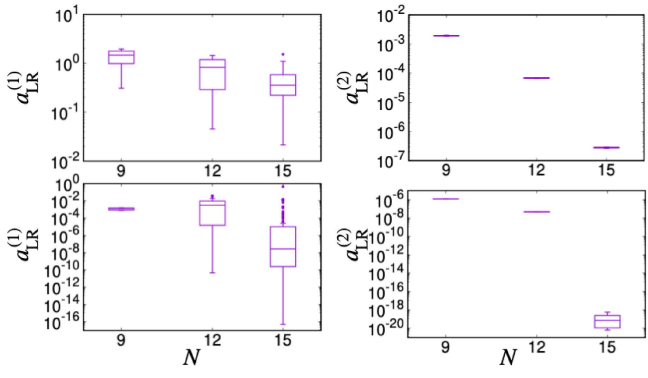}
\end{center}
\caption{
The dependence of $a_\LR^{(1)}$ and $a_\LR^{(2)}$ on the bath size and the initial state.
Parameters: 
$p=0.99$,
$g=0.1\gamma$,
$\gamma^\prime=4\gamma$~(upper panels) and $0.1\gamma$~(lower panels), $\beta=0.1$.
Left panels are $a_\LR^{(1)}$ and right panels are $a_\LR^{(2)}$.
}
\label{fig:App:8}
\end{figure}

Finally, we show the supplemental data in the short-time regime.
Figure~\ref{fig:App:7} shows the $N$-dependence of $a_\ETH$ and $a_\rI$ with $\gamma^\prime/\gamma=4$ and $0.1$. Both $a_\ETH$ and $a_\rI$ decrease as $N$ increases.
Figure \ref{fig:App:8} shows the $N$-dependence of $a_\LR^{(1,2)}$. We see that both $a_\LR^{(1)}$ and $a_\LR^{(2)}$ decrease in $N$.

\clearpage


\end{document}